**Version d'auteur de l'article :**

Nicolas PERREAUX, « *In nomine patris*. Éléments pour une sémantique de la paternité médiévale », dans Léo DUMONT, Octave JULIEN et Stéphane LAMASSE (dir.), *Histoires de mots. Saisir le passé grâce aux données textuelles*, Éditions de la Sorbonne, Paris, 2023, p. 127-159.

*

NB : Cette version contient des développements absents de la publication éditée, ainsi que de références bibliographiques complémentaires.

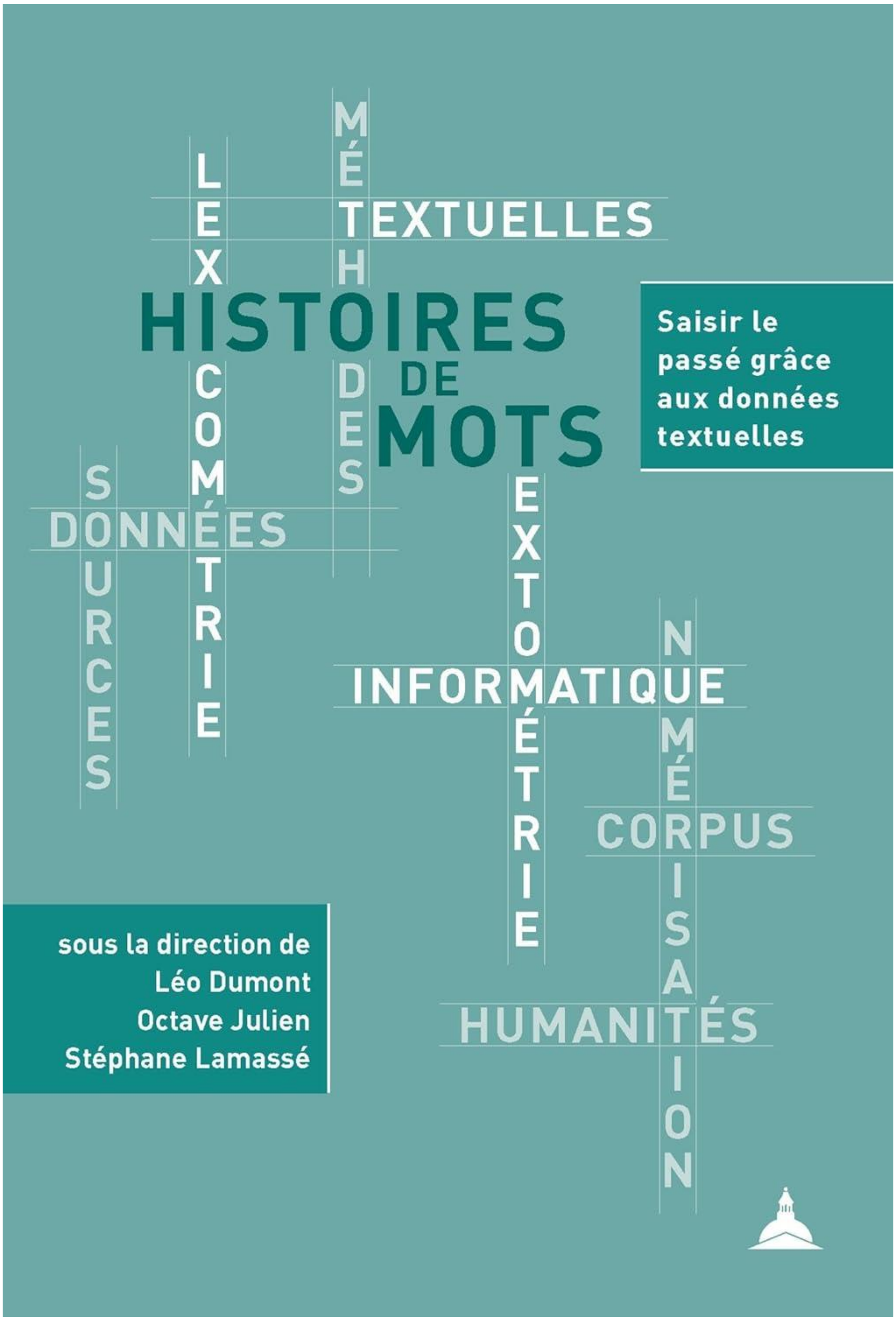

## *In nomine patris*

## Éléments pour une sémantique de la paternité médiévale[1]

Depuis maintenant plusieurs décennies, les historiens se sont intéressés aux rapports de parenté médiévale dans une perspective d'anthropologie historique[2]. Longtemps assez rares, les travaux sur les pères et la paternité font aujourd'hui partie de ceux dont l'historiographie est la plus dynamique[3]. Les enquêtes sur le sens des termes médiévaux désignant ces personnages (*pater*/*patres*) et cette relation ou condition (*paternitas, paternus*) demeurent toutefois encore minoritaires, constituant d'une certaine façon un angle mort de la recherche[4]. Or, elles sont d'autant plus nécessaires qu'on a longtemps insisté sur la continuité entre la paternité antique et médiévale[5], ce que certains chercheurs tendent actuellement à contester[6]. Parmi les éléments toujours en discussion, on relève par exemple le rôle du modèle trinitaire dans la sémantique des pères médiévaux, celui de la paternité biologique face aux paternités adoptives ou spirituelles, mais encore les liens entre père et fils, y compris dans leurs implications sociales[7]. Le premier objectif du présent article est ainsi d'aborder les textes numérisés de l'Europe médiolatine en tant que corpus, afin d'étudier les mentions des termes relatifs à la paternité et d'en dégager les principales tendances[8].

---

[1] Une première version de ce texte a été présentée lors du colloque que nous avons été coorganisé avec Anna Dorofeeva, Bernhard Jussen et Tim Geelhaar à Francfort en 2017 : « *Social Tensions Between Kinship and Community in the Middle Ages* ». Le texte actuel a été relu à différents stades par Jérôme Baschet, Charles de Miramon, Alain Guerreau, Anita Guerreau-Jalabert, Sylvie Joye, Eliana Magnani, Joseph Morsel, Alain Rauwel, Daniel Russo, Geneviève Bührer-Thierry : nous les en remercions chaleureusement. Enfin, il a bénéficié de discussions lors de plusieurs séances de la « pépinière » organisée par Joseph Morsel, en janvier et février 2021.

[2] Quelques jalons importants dans cette historiographie aujourd'hui très dense : Jack Goody, *L'évolution de la famille et du mariage en Europe*, Paris, Armand Colin, 1985 (première publication anglaise en 1983) ; Bernhard Jussen, *Patenschaft und Adoption im Frühen Mittelalter*, Göttingen, Vandenhoeck und Ruprecht, 1991 ; Anita Guerreau-Jalabert, « Parenté », dans Jacques Le Goff, Jean-Claude Schmitt (dir.), *Dictionnaire raisonné de l'Occident médiéval*, Paris, Fayard, 1999, p. 861-876 ; Régine Le Jan, *Famille et pouvoir dans le monde franc, VII^e^-X^e^ siècle. Essai d'anthropologie sociale*, Paris, Publications de la Sorbonne, 1995 ; Christiane Kalpisch-Zuber, *L'ombre des ancêtres. Essai sur l'imaginaire médiéval de la parenté*, Paris, Fayard, 2000 ; Chloé Maillet, *La parenté hagiographique : d'après Jacques de Voragine et les manuscrits enluminés de la Légende dorée (c. 1260-1490)*, Turnhout, Brepols, 2014 ; Hans Hummer, *Visions of kinship in medieval Europe*, Oxford, Oxford University Press, 2018.

[3] Didier Lett (dir.), *Être père à la fin du Moyen Âge*, numéro spécial des *Cahiers de Recherches Médiévales et Humanistes*, vol. 4, 1997 ; Robert J. Bast, *Honor Your Fathers: Catechisms and the Emergence of a Patriarchal Ideology in Germany 1400-1600*, Leiden, Brill, 1997 ; Jérôme Baschet, *Le sein du père. Abraham et la paternité dans l'Occident médiéval*, Paris, Gallimard, 2000 ; Jean Delumeau, Daniel Roche (dir.), *Histoire des pères et de la paternité*, Paris, Larousse, 2000 (2^e^ édition) ; Paul Payan, *Joseph : une image de la paternité dans l'Occident médiéval*, Paris, Aubier, 2006 ; Bernhard Jussen, Christopher H. Johnson, David Warren Sabean, Simon Teuscher (éd.), *Blood and Kinship. Matter for metaphor from Ancient Rome to the Present*, New York, Berghahn Books, 2013 ; Hannah Probert, *Fatherhood between Late Antiquity and the Early Middle Ages*, thèse de l'université de Sheffield, 2015 ; Aude-Marie Certin, « La cité des pères. Paternité, mémoire, société dans les villes méridionales de l'Empire du milieu du XIV^e^ siècle au milieu du XVI^e^ siècle (Nuremberg, Augsbourg, Francfort-sur-le-Main) » (résumé de la thèse de l'auteure), *L'Atelier du Centre de recherches historiques*, 2014, http://journals.openedition.org/acrh/6143, consultée le 14.04.2020 ; Id., « Überlegungen zu einer Geschichte der Vaterschaft in Westeuropa in der langen Dauer », *Historische Anthropologie*, vol. 28/1, 2020, p. 57-77 ; Aude-Marie Certin (dir.), *Formes et réformes de la paternité à la fin du Moyen Âge et au début de l'époque moderne*, Francfort, Peter Lang, 2016 ; Carole Avignon (dir.), *Bâtards et bâtardises dans l'Europe médiévale et moderne*, Rennes, Presses universitaires de Rennes, 2016 ; Sylvie Joye, *L'autorité paternelle en Occident à la fin de l'Antiquité et au haut Moyen Âge*, mémoire d'Habilitation à diriger des recherches, université de Paris 1 Panthéon-Sorbonne, 2016 (nous remercions vivement l'auteur de nous avoir confié une copie de son texte) ; Roxanne Louot, *La paternité à la fin du Moyen Âge. Des pères à travers les testaments d'Anjou (XIV^e^-XV^e^ siècles)*, mémoire de Master de l'université d'Angers, 2019.

[4] Parmi les exceptions les plus notables : Anita Guerreau-Jalabert, « La désignation des relations et des groupes de parenté en latin médiéval », *Archivum Latinitatis Medii Aevi*, vol. XLVI, 1988, p. 65-108 ; id., « Qu'est-ce que l'*adoptio* dans la societe chretienne medievale ? », *Médiévales*, vol. 35, 1998, p. 33-50 ; id., « *Nutritus*/*Oblatus* : parenté et circulation d'enfants au Moyen Âge », dans Mireille Corbier (dir.), *Adoption et fosterage*, Paris, De Boccard, 1999, p. 263-290 ; Sylvie Joye, *L'autorité paternelle en Occident [...]*, op.cit.

Sur quels ensembles documentaires mener une telle enquête ? Les historiens n'ont jamais produit autant de corpus digitaux, en particulier les médiévistes qui sont très actifs en la matière. Mais paradoxalement, ces ensembles documentaires demeurent sous-exploités. En dépit du rôle pionnier de Roberto Busa et de son *Corpus Thomisticum*, débuté en collaboration avec IBM dès 1949[9], aucune analyse numérique globale des textes de Thomas d'Aquin n'a été réalisée à ce jour. Ce constat n'est pas spécifique aux textes du « Docteur angélique » et pourrait être généralisé aux documents anciens en général[10]. Comment l'expliquer ? Une hypothèse revient à énoncer que la méthode historique, particulièrement efficace lorsqu'il s'agit d'analyser un groupe limité d'occurrences, se heurte à la nature même des corpus numériques. Ces derniers posent en effet la question documentaire d'une façon plutôt exogène à l'heuristique historienne traditionnelle, en proposant d'exploiter des ensembles textuels beaucoup plus larges[11]. Que faire par exemple des 500 000 mentions de *terra* que nous rencontrons dans les bases de données médiolatines, ou encore des presque 700 000 occurrences de *sanctus*[12] ?

La sémantique historique apparaît comme une réponse potentielle pour dépasser ces blocages en matière de parenté médiévale, ainsi que l'a démontré à différentes reprises Anita Guerreau-Jalabert. En plaçant le vocabulaire au cœur du questionnaire historien, elle place aussi indirectement le corpus textuel au centre des préoccupations. La méthode implique toutefois un changement de perspective, puisqu'elle présuppose que le sens des mots n'est pas universel et que c'est cette variation qui doit être reconstruite. Sans remonter à Humboldt ou Herder, il faut rappeler ici le rôle fondamental de Jost Trier, inventeur du concept de « champ sémantique », situé à l'articulation du « champ lexical » (*Wortfeld*) et du « champ de concepts » (*Bedeutungsfeld-Sinnfeld*)[13]. Son apport fut toutefois un semi-échec, puisqu'il échoua à expliciter la méthode sur laquelle il fondait son analyse : les champs sémantiques existaient bel et bien, mais leur modélisation se heurtait alors à des obstacles techniques quasi-insurmontables[14]. Après Pierre Guiraud et Georges Matoré[15], puis les historiens allemands de la *Begrifgeschichte*[16], la médiévistique doit à Alain Guerreau d'avoir émis une hypothèse fondamentale : les bases de données et les algorithmes pourraient être une réponse méthodologique pour la sémantique historique[17].

---

[5] Voir par exemple Geoffrey S. NATHAN, *The Family in Late Antiquity. The Rise of Christianity and the Endurance of Tradition*, Londres/ New York, Routledge, 2000.

[6] Dès 1997, Didier Lett soulignait que « l'histoire de la paternité médiévale s'est construite essentiellement à partir de sources normatives (coutumiers, traités de droit, etc.) », dans Didier Lett, « Pères modèles, pères souverains, pères réels », dans Didier Lett (dir.), *Être père à la fin du Moyen Âge,* op.cit., p. 7-14, ici p. 8). Cette approche partielle va toutefois au-delà de la question typologique, puisque c'est toute la dimension lexicale du problème qui a largement été sous-estimée (y compris dans les textes dits normatifs).

[7] L'historiographie de certaines de ces questions est présentée dans Hans Hummer, *Visions of kinship in medieval Europe*, op.cit., chapitres 5-8.

[8] Il va sans dire qu'une enquête sémantique complète sur le terme *pater*, dans l'ensemble des corpus ici rapidement examinés, nécessiterait une thèse à elle seule.

[9] Voir les deux volumes récents : Steven E. Jones, *Roberto Busa, S. J., and the Emergence of Humanities Computing: The Priest and the Punched Cards*, Londres / New York, Routledge, 2016 ; Roberto Busa, *One Origin of Digital Humanities: Fr Roberto Busa in His Own Words*, éd. par Julianne Nyhan et Marco Passarotti, New York, Springer, 2019. La référence au travail visionnaire de Roberto Busa est un passage obligé des articles employant les méthodes numériques depuis quelques années, devenant presque un topos. Ici, nous souhaitons souligner que sa vision se heurtait certes à des contraintes techniques, mais aussi et surtout aux difficultés intellectuelles d'alors.

[10] Nous nous permettons de renvoyer à Nicolas Perreaux, « De l'accumulation à l'exploitation ? Expériences et propositions pour l'indexation et l'utilisation des bases de données diplomatiques », dans Antonella Ambrosio, Sébastien Barret, Georg Vogeler (dir.), *Digital diplomatics. The computer as a tool for the diplomatist?*, Böhlau Verglag, Köln-Weimar-Wien, 2014, p. 187-210 (*Archiv für Diplomatik. Schriftgeschichte Siegel- und Wappenkunde, Beiheft* 14), où nous évoquions déjà ce problème.

[11] Alain Guerreau, *L'avenir d'un passé incertain. Quelle histoire du Moyen Âge au XXI*[e] *siècle ?,* Seuil, Paris, 2001, p. 191-237.

[12] Ces décomptes ne tiennent pas compte des *Acta Sanctorum*.

Dans cette perspective, le second objectif de cet article est de montrer quelques-unes des possibilités ouvertes par l'histoire du sens des mots et les analyses lexicales des grandes bases de données de textes anciens. La sémantique de la « paternité » médiévale paraît ainsi doublement intéressante : d'une part, comme nous l'avons dit, car nous hésitons encore sur le sens de termes qui lui sont relatifs (et leurs évolutions), d'autre part car ceux-ci sont extrêmement fréquents dans les textes – ce qui les rend d'autant plus délicat leur étude sans ordinateur.

---

[13] Jost Trier, *Der deutsche Wortschatz im Sinnbezirk des Verstandes : von den Anfängen bis zum Beginn des 13. Jahrhunderts*, Heidelberg, 1931 ; Lothar Schmidt (dir.), *Wortfeldforschung. Zur geschichte und theorie des sprachlichen feldes*, Darmstadt, Wissenschafliche Buchgesellschaft, 1973 ; Jost Trier, Anthony Van der Lee, Oskar Reichmann (dir.), *Aufsätze und Vorträge zur Wortfeldtheorie*, Den Haag, Mouton, 1973 ; Angela Castagnoli, *L'ipotesi del "campo semantico" di Jost Trier ed I suoi riflessi nella linguistica contemporanea*, Pise, Universita' degli studi di Pisa, 1987.

[14] Il s'agit d'un problème généralisé en médiévistique et en sémantique historique. Dès la fin du XIX^e siècle et plus encore dans les premières décennies du XX^e siècle, quelques pionniers ont vu les problèmes abstraits posés par la question de l'analyse diachronique et structurale du langage. Mais les obstacles techniques étaient alors insurmontables. En particulier, Jost Trier présenta à diverses reprises sa « méthode », mais elle était avant tout empirique et ne reposait sur aucune formalisation précise. De ce fait, les hypothèses proposées par ces théoriciens n'ont pu être explorées et son restées le plus souvent marginales, pour des raisons méthodologiques. L'obstacle disciplinaire a toutefois aussi joué un rôle dans ces blocages entre histoire, linguistique et statistique. Pour des raisons de partage scientifique, l'histoire s'est plutôt focalisé dès ses origines sur la dimension temporelle (en l'occurrence événementielle), tandis que la linguistique s'est massivement intéressé aux structures synchroniques – d'où les tentatives des linguistes pour démontrer la stabilité des langues, hors de tout cadre socio-historique. Les approches historico-sémantiques ont certes connu de grands succès scientifiques, mais elles sont restées le fait d'une minorité de chercheurs, le plus souvent âprement critiqués par les traditionnalistes (tant en histoire qu'en linguistique). Le divorce entre les disciplines n'aurait toutefois pas été totale si les historiens n'avaient pas eux-mêmes largement considérés jusqu'à récemment que les mots ne constituaient pas un objet central de l'histoire – privilégiant la logique fallacieuse selon laquelle *un* mot est égal à *une* traduction. Une logique qui éliminait *de facto* toute possibilité de réflexion sur la sémantique historique, et les liens entre le langage, les représentations et les structures sociales. Une tentative de réconciliation de ces pôles est lisible chez Ernst Cassirer, *La Philosophie des formes symboliques*, 3 volumes, Paris, Éditions de Minuit, 1972 (édition originale en 1923-1929), en particulier le tome 1.

[15] Pierre Guiraud, *Les caractères statistiques du vocabulaire*, Paris, PUF, 1954 ; id., *La sémantique*, Paris, PUF, 1955 (en référence à la note précédente, il est intéressant de noter que ce titre n'est pas mentionné dans la page Wikipedia de l'auteur, pourtant assez dense – le site a été consulté le 14 avril 2020) ; id., *Problèmes et méthodes de la statistique linguistique*, Dordrecht, D. Reidel Publishing Company, 1959 ; id., *Le Jargon de Villon ou le Gai Savoir de la coquille*, Paris, Gallimard, 1968 ; Georges Matoré, *Le vocabulaire et la société médiévale*, Paris, PUF, 1985.

[16] Reinhart Koselleck, *Vergangene Zukunft. Zur Semantik geschichtlicher Zeiten*, Francfort, Suhrkamp, 1979 ; Otto Brunner, Werner Conze, Reinhart Koselleck (dir.), *Geschichtliche Grundbegriffe*, Stuttgart, Klett-Cotta 1972-1997. L'historiographie allemande est certainement une des plus riches en matière de sémantique historique. Deux jalons récents : Moritz Wedell, *Zählen: Semantische und praxeologische Studien zum numerischen Wissen im Mittelatler*, Göttingen, Vandenhoeck & Ruprecht, 2011 ; Bernhard Jussen, « Historische Semantik aus der Sicht der Geschichtswissenschaft », *Jahrbuch für Germanistische Sprachgeschichte*, vol. 2, 2011, p. 51-61 ; Ludolf Kuchenbuch, *Reflexive Mediävistik: Textus - Opus - Feudalismus*, Francfort, Campus Verlag, 2012 ; Ernst Müller, et Falko Schmieder, *Begriffsgeschichte und historische Semantik*, Francfort, Suhrkamp, 2016.

[17] Une part non négligeable des outils statistiques employées ci-dessous sont le résultat de ses réflexions. Il développait certaines de ces idées dès les années 1980, dans Alain Guerreau, « Pourquoi (et comment) l'historien doit-il compter les mots ? », *Histoire & Mesure*, vol. 4, 1989, p. 81-105 ; plus récemment, voire id., « Pour un corpus de textes latins en ligne », *Bulletin du centre d'études médiévales d'Auxerre*, Collection CBMA – Les outils, http://journals.openedition.org/cem/11787, consulté le 14.04.2020.

## I. Une paternité ? Quelle paternité ?

### *I.1. La paternité n'est pas une structure universelle*

Quels caractères, relations, ou encore éléments définissent un « père » ou plutôt la « paternité » ? Pour nos sociétés du début du XXIe siècle, les réponses à ces questions sont *relativement* univoques[18]. D'après le *Trésor de la langue française*, le père est d'abord un « homme qui a engendré (sous-entendu biologiquement) » ou plus rarement « qui a adopté »[19]. Ce terme peut certes désigner le promoteur d'une idée – on dit par exemple de Ferdinand de Saussure qu'il est « le père de la linguistique structurale » –, mais le sens biologique reste prédominant dans le discours contemporain[20].

Cette situation semble relativement installée dès le XVIIIe siècle. L'*Encyclopédie* de Diderot et d'Alembert (1751-1772) possède différentes entrées pour le terme : « père (droit naturel », « père *naturel* », « père *légitime* », « père *putatif* », « père *adoptif* », « père (critique sacrée) », « père conscripts (*Hist. Rom.*) », et enfin « pères de l'Église (*Hist. ecclésiast.*) »[21]. L'organisation de l'article laisse peu de doutes : c'est le lien de filiation biologique et ses implications qui prime sur tout, avant la prétention juridique à la paternité, le recours à l'adoption, puis les sens liés au système ecclésial des siècles antérieurs (*i.e.* « critique sacrée »[22] et « pères de l'Église »), ou encore au monde antique (« père conscripts » ou *patres conscripti*). La paternité apparaît comme « la relation la plus étroite qu'il y ait dans la nature », car le père envisage ses enfants « sous deux rapports également intéressans (sic), & comme leurs héritiers, & comme leurs créatures ». La proportion des différents sens dans le texte de l'*Encyclopédie* est particulièrement instructive : à peine deux colonnes traitent de la paternité biologique, tandis que les sens ecclésiaux en occupent vingt-et-une[23]. Cette répartition n'est certainement pas imputable à l'importance pratique du second thème au XVIIIe siècle, les auteurs commençant d'abord par évoquer la paternité « naturelle »[24], mais plutôt au fait qu'une multiplicité de

---

[18] Nous n'ignorons pas les débats sur la paternité contemporaine et ses composantes abstraites (par exemple le paternalisme), ni les importants travaux des historiens, des sociologues, des juristes ou encore des anthropologues sur la question. Toutefois, nous pensons que ces débats portent plutôt sur *comment être (un bon) père ?*, que sur la question *qui est le père ?* L'évolution de la médecine et du contrat social au cours des dernières décennies tend à bouleverser ces lignes : Frédérique Granet, « Le père au regard du droit », dans Jean Delumeau, Daniel Roche (dir.), *Histoire des pères et de la paternité*, op.cit., p. 439-462 ; Brid Feathertone, *Contemporary Fathering: Theory, Research and Social Policy*, Portland, Policy Press, 2009 ; Thomas Johansson, Jesper Andreasson, *Fatherhood in Transition. Masculinity, Identity and Everyday Life*, Londres, Palgrave Macmillan, 2017.

[19] Bernard Quemada (dir.), *Trésor de la langue française*, volume 13, Paris, Gallimard, 1988 (version numérisée en ligne par l'ATILF : http://atilf.atilf.fr/, consultée le 14.04.2020).

[20] D'où l'obsession du père biologique (son identité ou sa personnalité) dans les productions cinématographiques contemporaines, de Star Wars à Harry Potter.

[21] *Encyclopédie, ou Dictionnaire raisonné des sciences, des arts et des métiers*, tome 12, Neufchastel, Samuel Faulche & Compagnie, 1765, p. 338-350.

[22] Où l'on trouve cette phrase : « *Dieu est nommé pere de tous les hommes, comme créateur & conservateur de toutes les créatures.* » (p. 339, col. 2).

[23] Ces multiples entrées furent toutefois toutes rédigées par le même auteur, le très prolifique Louis de Jaucourt [1704-1780]. Sur ce personnage, docteur en théologie et médecin, voir en dernier lieu Gilles Barroux, François Pépin (dir.), *Le chevalier de Jaucourt : l'homme aux dix-sept mille articles*, Paris, Société Diderot, 2015.

[24] Une recherche des principaux cooccurrents de « père » dans le corpus de l'*Encyclopédie* confirme d'ailleurs cette impression : fils, mère, enfant, famille, nom, femme, homme, apparaissent tous parmi les 20 termes les plus courants autour du lemme.

« pères » est évoquée dans la vision toute médiévale de l'article « pères de l'Église »[25]. Si le fait d'envisager les enfants comme des « créatures », et donc le père comme un créateur, constitue un héritage des représentations médiévales (quoi qu'ici fondamentalement transposé à la filiation), il semble donc que l'article présente deux visions concurrentes : celle, moderne, de la primauté de la filiation[26], et celle médiévale d'une multiplicité des pères, centrée sur les parentés spirituelles. Il n'est pas anodin que l'article, comme beaucoup d'autres d'ailleurs de l'*Encyclopédie*, ait été rédigé par Louis de Jaucourt, qui fut à la fois médecin et théologien. Ce dernier était ainsi parfaitement au fait tant de la logique chrétienne que des aspects dits « naturels » du champ socio-sémantique qu'il traitait ici. En juxtaposant mais aussi en hiérarchisant les différents sens du terme, Jaucourt offrait ainsi une forme de diachronie à son article.

Une enquête complémentaire dans plus de 10 000 romans du XIX^e siècle, disponibles sur le site Gallica, confirme nos impressions initiales[27]. Le lemme « père » désigne massivement l'être masculin qui a engendré, même si de rares cas alternatifs existent. Ces derniers concernent pour la plupart l'évocation de Dieu comme « Père » (le plus souvent avec une majuscule), mais aussi les personnages/auteurs cléricaux, parfois désignés par le substantif. L'évocation des Pères de l'Église y est plus rare que dans l'*Encyclopédie*, et joue dans ce corpus un rôle mineur. Une recherche plus fine sur un ensemble documentaire réduit, constitué principalement des textes de Balzac, Dumas, Flaubert, Hugo, Stendhal, Verne et Zola, donne des indications plus précises[28]. Pour les 3 800 mentions du lemme « père » dans ce corpus, les principaux cooccurrents sont « son », « mon », « votre », « mère », « fils », « ton » ou encore « enfant ». L'ensemble dessine un réseau très restreint et précis autour du père biologique, quand bien même il serait question de définir *qui est le père* d'un enfant ou d'évoquer une adoption[29].

Dans les sociétés non-occidentales ou précapitalistes, les réponses à ces questions sont nettement moins univoques[30]. Au sein du système médiéval en particulier, le lemme latin *pater* renvoie à une multitude de contextes et de personnages[31], qui restent d'ailleurs largement à définir. L'hypothèse n'est certes pas inédite – comme nous l'avons dit, un certain nombre de médiévistes ont montré que la parenté médiévale répondait à des logiques différentes de la nôtre depuis les années 1980, mais elle reste largement à explorer. Que signifiaient *pater* et *paternitas* dans l'Europe médiévale et à quel(s) ensemble(s) de représentations ces termes pouvaient-il se rattacher ?

---

[25] Cf. Alain Rauwel, « Paternité », dans Frédéric Gabriel, Dominique Iogna-Prat, Alain Rauwel (dir.), *Dictionnaire critique de l'Église*, à paraître – nous remercions vivement l'auteur de nous avoir donné accès à son article avant la publication.

[26] Sur l'émergence tardive de la pensée de l'« hérédité biologique », voir Maaike van der Lugt, Charles de Miramon, « Penser l'hérédité au Moyen Âge : une introduction », dans Maaike van der Lugt, Charles de Miramon (éd.), *L'hérédité entre Moyen Âge et Époque moderne. Perspectives historiques*, Florence, Sismel, 2008, p. 3-37.

[27] Nous remercions vivement Pierre-Carl Langlais de nous avoir transmis une copie de son important corpus, afin que nous puissions contrôler nos hypothèses.

[28] Soit une cinquantaine de volumes au total, datant de 1818 à 1885, pour environ 6,2 millions de mots. La liste est la suivante : Balzac : *Les Chouans ; La peau de chagrin ; L'illustre Gaudissart ; Eugénie Grandet ; Le colonel Chabert ; Le père Goriot ; La fille aux yeux d'or ; La duchesse de Langeais ; Petites misères de la vie conjugale* - Dumas : *Les trois mousquetaires ; Le Comte de Monte-Cristo (1-4) ; La dame de Monsoreau (1-3)* - Flaubert : *Madame Bovary ; Salambo ; L'éducation sentimentale ; La tentation de saint Antoine ; Trois contes ; Bouvard et Pécuchet* - Hugo : *Burg-Jargal ; Han d'Islande ; Le dernier jour d'un condamne ; Notre-Dame de Paris ; Les Miserables (1-5) ; Quatre-vingt-treize* - Stendhal : *Armance ; Le Rouge et le Noir ; Mémoires d'un touriste ; La Chartreuse de Parme ; Chroniques italiennes* - Verne : *Cinq semaines en ballon ; De la terre à la lune ; Vingt mille lieues sous les mers ; L'île mystérieuse ; Michel Strogoff ; Les cinq cents millions de la begum ; Les révoltés de la Bounty ; Robur le conquérant* - Zola : *Thérèse Raquin ; La curée ; La faute de l'abbé Mouret ; L'assomoir ; Au bonheur des dames ; Germinal*. Ce corpus constitué par Alain Guerreau a ici été lemmatisé via TreeTagger.

[29] La proportion des formes de « père » au pluriel sont d'ailleurs très rares dans ce corpus : nous y reviendrons.

[30] Maurice Godelier, *La production des Grands hommes*, Paris, Fayard, 1982 ; id., *Métamorphoses de la parenté*, Paris, 2004 ; Laurent Barry, *La parenté*, Paris, Gallimard, 2008. Voir aussi la remarque de Claude Lévi-Strauss : « L'idee que la filiation découle d'un lien biologique tend à l'emporter sur celle qui voit dans la filiation un lien social. Le droit anglais ignore même la notion de paternité sociale : le donneur de sperme pourrait légalement revendiquer l'enfant ou être tenu de pourvoir à ses besoins. », dans Claude Lévi-Strauss, *Nous sommes tous des cannibales*, Paris, Seuil, 2013, p. 91-92 ; ainsi qu'Aude-Marie Certin, « Überlegungen zu einer Geschichte der Vaterschaft in Westeuropa in der langen Dauer », *art.cit.*, p. 57-61.

### *I.2. Corpus et méthodes*

Contrairement à une opinion historienne assez répandue, les médiévistes sont loin de manquer de corpus numériques, et de textes en particulier[32]. Dans le cadre du présent article, différents ensembles documentaires ont été exploités. Le plus gros corpus de textes médiévaux reste à ce jour la *Patrologie Latine* (désormais PL) : elle contient environ 100 millions de mots, répartis dans près de 5 300 textes, soit près de 16 fois plus que tous les corpus latins de l'Antiquité païenne numérisés[33]. Malgré des biais sur lesquels nous ne pouvons nous attarder[34], la collection possède différents atouts : sa taille, l'importance des textes qui y sont imprimés, mais aussi le fait que ceux-ci soient ordonnés chronologiquement – ce qui permet de réaliser des enquêtes diachroniques.

---

[31] « Le *pater* et les *patres* évoqués par Dhuoda renvoient au *genitor* de son fils Guillaume, Bernard de Septimanie, mais aussi aux Pères de l'Église et à ceux qui suivent leurs préceptes. », dans Sylvie Joye, « La 'crise de la famille'. Débats contemporains et représentations médiévales à la lecture des sources du haut Moyen Âge occidental », *Mélanges de l'École française de Rome - Moyen Âge*, vol. 131-1, 2019, p. 55-65, ici p. 63.

[32] La déploration de la rareté documentaire est un *topos* de la médiévistique. Elle repose principalement sur un triple biais : 1) d'une part la surestimation de la taille du corpus antique païen (sans doute lié au prestige de l'Antiquité, cf. Jacques Le Goff, *Faut-il vraiment découper l'histoire en tranches ?*, Paris, Seuil, 2014), 2) d'autre part la sous-estimation du corpus médiéval – car les estimations globales demeurent rares. On peut en effet estimer qu'au moins un million de chartes subsistent en original ou en copie avant la fin du XIII^e^ siècle, à l'échelle européenne – tandis que les estimations concernant le nombre de volumes manuscrits médiévaux conservés, toutes typologies confondues, avoisinent aussi le million : Nicolas Perreaux, « L'écriture du monde (parties I et II) », Bulletin du Centre d'études médiévales d'Auxerre, vol. 19.2/20.1, 2015-2016, https://doi.org/10.4000/cem.14264 et https://doi.org/10.4000/cem.14452 (consultés le 14.04.2020) ; Eltjo Buringh, *Medieval Manuscript Production in the Latin West, Explorations with a Global Database*, Leiden, 2011 (*Global Economics History Series*, 6). Les métaphores historiennes ont aussi une part dans cette appréciation faussée du stock documentaire, car elle conduise à « romantiser » l'analyse documentaire. Sur ce point nous renvoyons à Joseph Morsel, « Les sources sont-elles « le pain de l'historien » », *Hypothèses*, vol. 7, 2004, p. 271-286 ; id., « Traces ? Quelles traces ? Réflexions pour une histoire non passéiste », *Revue historique*, vol. 80, 2016 (4), p. 813-868. Dans notre thèse, nous avions aussi souligné l'association des métaphores lumineuses et aquatiques, qui conduisent toujours à estimer le corpus médiéval sous l'angle de la déploration romantique (Nicolas Perreaux, *L'écriture du monde. Dynamique, perception, catégorisation du* mundus *au Moyen Âge (VII*e*-XIII*e* siècles). Recherches à partir des bases de données numérisées*, thèse de l'Université de Dijon, 2014, p. 434-443). 3) La dernière raison de la « déploration de la rareté documentaire » est sans doute la plus forte, mais aussi la plus complexe : il s'agit de l'approche historienne par la problématique a priori, qui conduit souvent à chercher dans la documentation des éléments qui s'y trouvent rarement, en délaissant ceux qui s'y trouvent en masse – parce qu'ils ne répondent pas à nos logiques contemporaines. En cela, nous pensons que l'approche corputielle change radicalement la donne, faisant précisément apparaître l'abondance de la documentation médiévale (certes relative face au déluge de données contemporain, mais néanmoins significative).

[33] Le corpus que nous avons pu constituer contient la presque totalité des textes latins païens entre le IVe siècle avant et le IIIe siècle après. Or, il ne représente que 6,9 millions de mots – soit grosso modo l'équivalent du corpus très sélectif de romans du XIXe siècle, précédemment employé, ou encore 5 fois les chartes de Cluny.

[34] En particulier : la présence d'éditions ne respectant pas certains critères scientifiques actuels, des attributions douteuses ou absentes (un tiers de la PL n'est pas attribuée), la variété des typologies documentaires (essentiellement de l'exégèse et de la théologie, mais aussi des textes normatifs, hagiographiques, diplomatiques, etc.), et la normalisation des graphies.

Le second plus vaste corpus numérique est celui des chartes. Pour la période allant du VIII^e^ au XIII^e^ siècle, il constitue même l'ensemble le plus dense, avec environ 75 millions de mots à ce jour. Ces documents sont généralement des textes courts (206 mots en médiane[35]) relatant des transferts de biens ou de droits (dons, ventes voire confirmations) ainsi que des accords, le plus souvent entre une personne et une institution, mais aussi plus rarement entre plusieurs institutions. Depuis 2009, nous nous sommes attachés à réunir tous les ensembles diplomatiques numérisés dans un corpus unique et standardisé, intitulé *Cartae Europae Medii Aevi* (désormais CEMA)[36]. Il réunit aujourd'hui plus de 225 000 chartes, interrogeables dans différents formats. Plus ponctuellement, nous avons eu recours à d'autres corpus : un ensemble de textes latins antiques, la Vulgate, les OpenMGH, le *Corpus Thomisticum*, et enfin une série de documents essentiellement théologiques du XV^e^ siècle[37].

Comment analyser ces bases ? Les médiévistes disposent d'une série d'outils de « lecture distante » (*Distant Reading*) à même de transformer profondément leurs perspectives[38]. Le latin étant une langue à flexion, les ensembles textuels précédemment évoqués ont tout d'abord été lemmatisé grâce aux paramètres médiolatins pour TreeTagger, développés par l'équipe de l'ANR Omnia[39]. Les corpus ont ainsi été préparés dans différents formats, à l'aide de scripts développés en Perl ou en Python. Parmi les outils d'analyse, nous avons retenu principalement trois logiciels : TXM, CQP-CWB et R. Les graphiques présentés proviennent essentiellement de bibliothèques de fonctions R : Cooc et Wordspace[40].

## II. Un modèle de paternité généralisée

### *II.1. La Trinité et l'affirmation du Dieu-père*

Au sein des corpus précités, on ne décompte pas moins de 301 500 mentions du substantif *pater* – ce qui le classe immédiatement dans la catégorie du lexique courant. Si ces occurrences sont plutôt régulières du II^e^ au XIII^e^ siècle dans la PL (fig. 1), avec peut-être une faible baisse tendancielle, on observe certaines évolutions fréquentielles sur le temps long, en particulier au regard des différentes typologies documentaires (fig. 2). C'est par ailleurs au sein d'un champ lexical plus large, que l'on peut constater des évolutions importantes.

---

[35] Cette mesure a été effectuée sur l'ensemble des CEMA (cf. ci-après). Elle porte sur les 209 000 chartes contenant au moins 10 mots (afin de retirer de l'analyse les actes tronqués). La moyenne se situe quant à elle à 280 mots.
[36] Présentation rapide de l'ensemble dans Nicolas Perreaux, « L'écriture du monde (parties I et II) », *op.cit.*
[37] Le corpus latin Antique proviennent pour l'essentiel de différentes bases de données qui ont été réunies (*All Texts Latin* et *Perseus* en particulier). Les OpenMGH sont une version pour le moment restreinte mais plus flexible des dMGH, c'est-à-dire une édition électronique des documents réunis dans la célèbre collection des *Monumenta Historica Germaniae*. Le *Corpus Thomisticum* a déjà été présenté il y a quelques pages : il contient les textes de Thomas d'Aquin et d'une partie de ses continuateurs. Quant à la base de latin du XV^e^ siècle, elle a été réunie par Alain Guerreau : elle contient 576 textes théologiques et narratifs de cette période, soit un peu plus de 5,7 millions de mots.
[38] En aucun cas cette approche ne dispense de la lecture des textes : Franco Moretti, *Graphs, Maps, Trees: Abstract Models for a Literary History*, London, New York, Verso, 2005 ; id., *Distant Reading*, Londres, Verso, 2013. L'intérêt principal des méthodes distantes n'est en outre certainement pas de « prouver » quelque chose (pas plus que les méthodes qualitatives d'ailleurs), mais de faire apparaître des éléments invisibles à l'œil nu, via un changement d'échelle, une approche globale des informations contenues dans les documents et une modélisation-formalisation de ces informations.
[39] Dirigée par Alain Guerreau. Certains des résultats de l'équipe sont visibles à l'adresse suivante : https://glossaria.eu (consulté le 14.04.2020). Le projet se développe actuellement sous la forme d'une nouvelle ANR, Velum, dirigée par Bruno Bon (IRHT).
[40] Respectivement développés par Alain Guerreau et Stefan Evert.

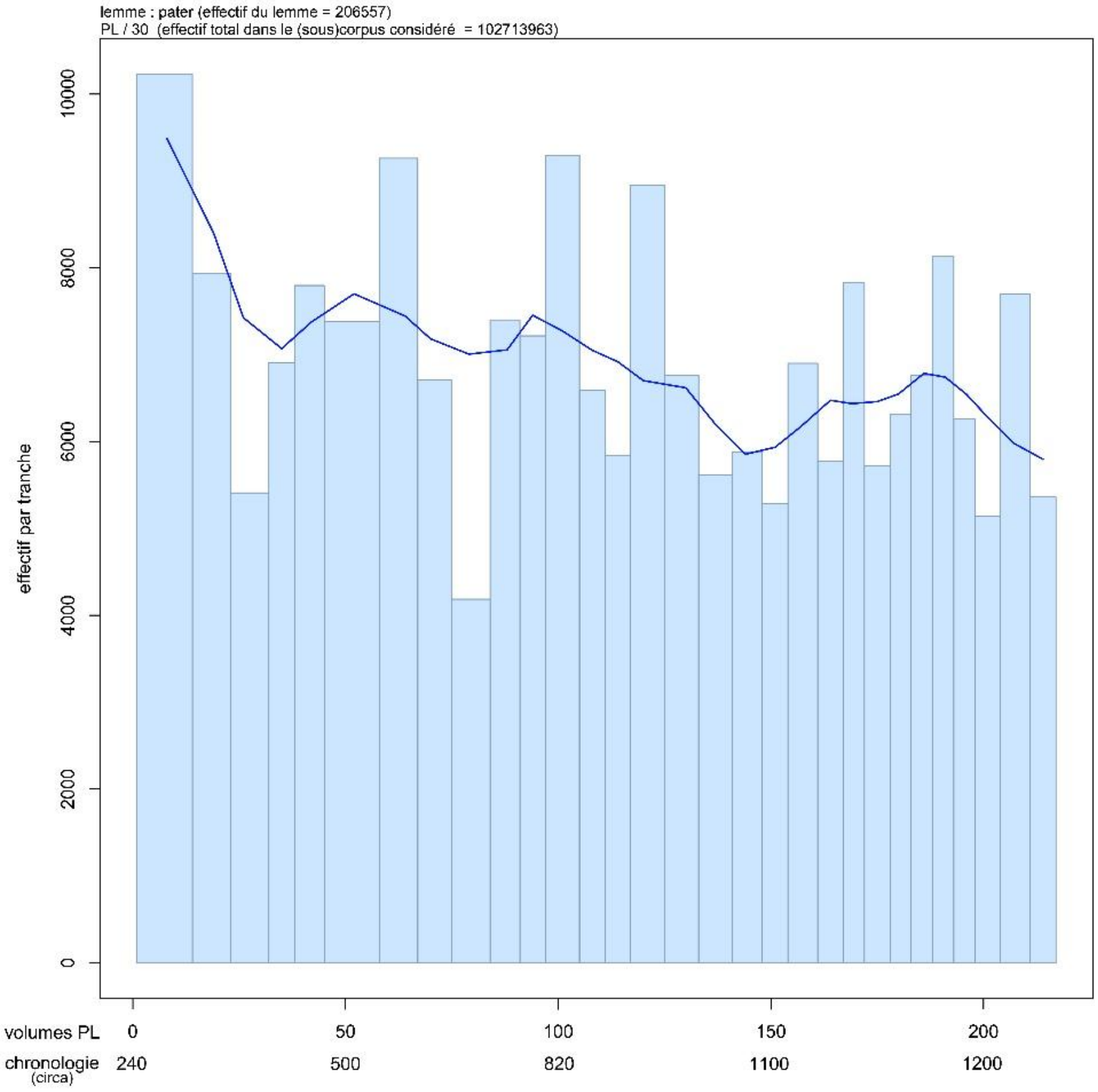

**Fig. 1 :** Évolution des mentions du lemme *pater* dans la PL (III^e-début du XIII^e s.).

| | *mater* | *pater* | *filius* | *filia* | *avus* | *proavus* | *abavus* | *atavus* | *tritavus* | *Somme* |
|---|---|---|---|---|---|---|---|---|---|---|
| **Antique** | 3244 | 8670 | 5354 | 1586 | 779 | 108 | 15 | 23 | 7 | *19786* |
| **Vulgate** | 323 | 1608 | 4689 | 1194 | 2 | 0 | 0 | 0 | 0 | *7816* |
| **CEMA** | 25425 | 47503 | 175945 | 21275 | 3993 | 384 | 23 | 103 | 0 | *274651* |
| **OpenMGH** | 4074 | 14699 | 18865 | 3365 | 709 | 113 | 20 | 42 | 9 | *41896* |
| **- (dont capitulaires)** | 74 | 318 | 411 | 65 | 96 | 2 | 0 | 1 | 0 | *967* |
| **PL (II^e-V^e s.)** | 8143 | 50273 | 63772 | 4130 | 492 | 73 | 1 | 22 | 0 | *126906* |
| **PL (V^e-VII^e s.)** | 3472 | 21676 | 20793 | 1931 | 370 | 42 | 6 | 22 | 12 | *48324* |
| **PL (VIII^e-IX^e s.)** | 7059 | 45484 | 54926 | 4271 | 672 | 66 | 13 | 29 | 5 | *112525* |
| **PL (X^e-mil. XI^e s.)** | 9357 | 40257 | 47466 | 5321 | 774 | 107 | 18 | 49 | 10 | *103359* |
| **PL (mil. XI^e s.-XIII^e s.)** | 9943 | 48410 | 64119 | 5447 | 607 | 72 | 12 | 21 | 0 | *128631* |
| ***Corpus Thomisticum*** | 2330 | 16850 | 20135 | 895 | 109 | 13 | 1 | 0 | 0 | *40333* |
| **Latin du XV^e s.** | 1685 | 6098 | 6504 | 835 | 207 | 25 | 0 | 5 | 0 | *15359* |
| *Somme* | *75055* | *301528* | *482568* | *50250* | *8714* | *1003* | *109* | *316* | *43* | *919586* |

**Fig. 2 :** Occurrences de *pater* et d'autres lemmes relatifs à la parenté, dans les corpus consultés.

*Pater* est certes très présent dès l'Antiquité païenne, avec 8 670 occurrences au sein d'un corpus de 6,9 millions de mots[41]. Mais un décompte complémentaire pour une série de termes désignant des membres de la parentèle ou relations de parenté montre que, pour la période Antique, l'importance du père est contrebalancée par de très nombreuses références à la figure maternelle (*mater*)[42]. On dénombre ainsi un peu plus de 2 occurrences de *pater* pour 1 de *mater* dans le latin païen. Or, ce ratio augmente fortement dès la Vulgate, où l'on trouve quatre fois plus de mentions de *pater* que de *mater*[43]. Enfin, dans une première partie de la *Patrologie latine* allant de Tertullien à Boèce[44], cette proportion est de 6,2 pour 1[45]. Il semble donc que le père gagne nettement en importance entre la fin de l'Antiquité et le très haut Moyen Âge, *a minima* dans cette série de textes. C'est d'ailleurs dans ce moment que s'affirme le lemme *filius*, et donc le couple *pater-filius*[46]. Parallèlement, nous avons pu noter que les lemmes désignant les ancêtres (*avus*, *proavus*, puis *abavus*, *atavus* et *tritavus-tetravus*[47]) jouent un rôle absolument mineur dans tous les textes médiévaux : au total, 10 fois moins que dans les textes antiques[48].

[41] Cela fait de *pater* le 73^e lemme le plus fréquent sur l'ensemble du corpus. Par comparaison, il se classe au rang 52 de la PL. Cette évolution montre que la promotion du lemme au Moyen Âge est réelle, mais qu'il était déjà omniprésent dans l'Antiquité. Concernant la paternité antique, les travaux sont trop nombreux pour que nous puissions y renvoyer ici en détail. Voir cependant le texte de Yann Thomas, « À Rome, pères citoyens et cité des pères, II^e s. av. J.-C. - II^e s. ap. J.-C. », dans Aline Rousselle, Giulia Sissa et Yan Thomas (éd.), *La Famille dans la Grèce antique et à Rome*, Paris, Éditions Complexe, 2005, p. 65-125.

Cette affirmation de la figure paternelle (*pater*), en lien avec le fils (*filius*), est évidemment liée au développement du christianisme et à la mise en place du système ecclésial, donc du dogme de la Trinité[49]. Dans cette perspective, l'importance des conciles de Nicée I [325] et de Constantinople I [381] ne peut être sous-estimée pour l'établissement de la figure médiévale du père[50]. La comparaison des principaux cooccurrents, autrement dit les termes apparaissant autour de *pater* dans une fenêtre de plus ou moins 5 mots, au sein du corpus des textes antiques et dans ceux de saint Augustin [† 430], montre des différences flagrantes (fig. 3)[51]. Pour l'Antiquité latine, ressortent les lemmes liés au système sénatorial, au grand-père, à la *familia* romaine, à la maisonnée et au système romain en général (*plebs*, *populus*, etc.). Pour saint Augustin se distinguent les lemmes relatifs à la Trinité (*spiritus*, *dominus*, *trinitas*), au Verbe (*verbum*), à la Création (*natura*, *homo*, *substantia*, *tempus*) et à la Chute (*peccatum*), ou encore aux couples esprit/chair (*spiritus*/*caro*) et créateur/créatures

---

42 Soit le rang 213 du corpus antique.

43 Soit 1 608 occurrences contre 323 (rang 51 vs. 242).

44 Ce dernier exclus, soit du tome 1 au tome 62.

45 Soit 50 273 occurrences contre 8 143.

46 Le premier cooccurrent de *pater* dans le corpus antique païen est en effet *mater*. Dans la première partie de la PL (qui correspond à une majeure partie des textes des Pères de l'Église), c'est *filius*.

47 Le lemme *bisavus* est absent de l'ensemble des textes consultés.

48 Où ils ne sont certes pas non plus omniprésents. Nous pensons toutefois que cette évolution est loin d'être anodine. Dans un article à paraître consacré au lemme *memoria*, nous avons par ailleurs pu montrer que les « ancêtres » jouaient un rôle mineur dans les discours mémoriels, et que leur part allait en diminuant au fil de la chronologie. Cette faiblesse de la profondeur généalogique avait été observée dans Régine Le Jan, *Famille et pouvoir dans le monde franc, VII^e^-X^e^ siècle [...]*, op.cit., p. 13.

49 Sur la mise en place de ce dogme et son importance cruciale dans le système de représentation médiéval, nous renvoyons à Jules Lebreton, *Histoire du dogme de la Trinité des origines au concile de Nicée*, 2 tomes, Paris, 1910-1928 ; Bertrand de Margerie, *La trinité chrétienne dans l'histoire*, Paris, 1975 ; Bernard Sesboüe, Bernard Meunier, *Dieu peut-il avoir un fils ? Le débat trinitaire au IV^e^ siècle*, Paris, 1993 ; Peter Widdicombe, *The Fatherhood of God from Origen to Athanasius*, Oxford, Oxford University Press, 1994 ; Alain Guerreau rédige actuellement un ouvrage dont les premiers chapitres sont consacrés à ces questions, chapitres auxquels nous avons pu accéder.

50 Paul Veyne, *Quand notre monde est devenu chrétien (312-394)*, Paris, Albin Michel, 2007 ; Peter Brown, *The Body and Society. Men, Women and Sexual Renunciation in Early Christianity*, New York, Columbia University press, 1988.

51 Si la sémantique du lemme *paterfamilias* évolue entre Antiquité et Europe médiévale, nous avons pu noter que la fréquence du lemme restait stable dans les deux corpus. En outre, elle est extrêmement faible : 123 mentions seulement dans le corpus Antique (soit 0,001782% du corpus) et 1873 dans la PL (0,001828% du corpus). Précisons que les formes dérivées de *paterfamilias* et *pater familias* ont été recherchées dans les deux cas, car les éditeurs optent pour l'une ou l'autre graphie. Nous sommes ainsi frappés par la différence qu'il existe entre l'importance du lemme dans les textes et son rôle historiographique : il y a presque 70 fois moins de *paterfamilias* que de *pater* dans le corpus Antique, et 109 fois moins dans la PL. Il est toutefois possible de considérer que *paterfamilias* ne renvoie directement pas aux documents anciens, mais à un concept historien, renvoyant quant à lui à une série d'idées et de notions. Dans cette perspective différente de la nôtre ici, nous renvoyons aux beaux développements du premier chapitre de Sylvie Joye, *L'autorité paternelle en Occident [...]*, op.cit., p. 60-128.

(*dominus*/*homo*). Il s'agit d'un basculement majeur pour le champ sémantique du père, qui passe par le développement de cette figure en tant que protagoniste de la Trinité.

| | Antique_nb | Antique_coef | Augustin_nb | Augustin_coef |
|---|---|---|---|---|
| *plebs* | 219 | 221994.93 | - | - |
| *res* | 214 | 117378.84 | - | - |
| *consul* | 135 | 115477.45 | - | - |
| *populus1* | 133 | 110572.57 | - | - |
| *familia* | 111 | 118326.4 | - | - |
| *senatus* | 110 | 96617.7 | - | - |
| *rex* | 108 | 90571.01 | - | - |
| *bellum* | 107 | 83609.06 | - | - |
| *uir* | 106 | 87819.51 | - | - |
| *nihil* | 100 | 75202.85 | - | - |
| *patria* | 100 | 100276.14 | - | - |
| *pars* | 96 | 70862.7 | - | - |
| *casa* | 94 | 71010.67 | - | - |
| *mors* | 91 | 84663.07 | - | - |
| *animus* | 87 | 67103.99 | - | - |
| *ius1* | 80 | 72307.2 | - | - |
| *puer* | 71 | 70317.71 | - | - |
| *honor* | 68 | 65411.84 | - | - |
| *memoria* | 66 | 68019.55 | - | - |
| *auus* | 65 | 71758.62 | - | - |
| *tribunus* | 63 | 63587.72 | - | - |
| *frater* | 141 | 136258.92 | 110 | 85194.56 |
| *filia* | 83 | 86076.25 | 79 | 68920.11 |
| *potestas* | 84 | 84822.53 | 84 | 69293.82 |
| *mater* | 221 | 213111.68 | 392 | 334286.57 |
| *nomen* | 101 | 86111.17 | 164 | 124007.78 |
| *filius* | 504 | 457765.86 | 3991 | 2502059.13 |
| *domus* | 84 | 74543.82 | 105 | 87530.4 |
| *deus* | 96 | 80958.63 | 1806 | 727551.85 |
| *regnum* | 73 | 71703.15 | 187 | 152730.69 |
| *abbas* | - | - | 75 | 69192.79 |
| *trinitas* | - | - | 78 | 69243.8 |
| *nemo* | - | - | 78 | 63663.39 |
| *opera* | - | - | 84 | 69442.83 |
| *ueritas* | - | - | 85 | 65117.37 |
| *natura* | - | - | 93 | 71498.28 |
| *sapientia* | - | - | 108 | 90298.35 |
| *terra* | - | - | 123 | 86189.08 |
| *tempus* | - | - | 124 | 91628.4 |
| *caro2* | - | - | 125 | 86902 |
| *peccatum* | - | - | 129 | 83681.59 |
| *forma* | - | - | 133 | 114599.83 |
| *substantia* | - | - | 143 | 123630.1 |
| *uoluntas* | - | - | 158 | 120281.71 |
| *uita* | - | - | 160 | 110637.26 |
| *dextera* | - | - | 162 | 144775.64 |
| *cela* | - | - | 219 | 198452.3 |
| *uerbum* | - | - | 350 | 231767.75 |
| *dominus* | - | - | 355 | 197757.01 |
| *homo* | - | - | 356 | 181081.99 |
| *spiritus* | - | - | 823 | 559540.37 |

**Fig. 3 :** Comparaison des principaux cooccurrents de *pater* chez les auteurs antiques païens et chez saint Augustin. En rose : les cooccurrents significatifs à *pater*, communs dans les deux corpus. Méthode : CoocC, coefficient de Dice[52].

[52] Aussi appelé « indice de Sørensen-Dice », celui-ci mesure la similarité entre deux séries statistiques. Il a été employé en statistique lexicale dès les années 1980. L'intérêt d'appliquer un coefficient aux mesures brutes est de

L'affirmation d'un Dieu-père et plus largement de la paternité divine au tournant des IVe-Ve siècles joue donc un rôle déterminant dans l'émergence d'un discours sur une « paternité » proprement médiévale. Si la Trinité n'est pas alors explicitement définie comme un système de parenté *stricto sensu*[53], elle propose en effet un modèle idéal de relation : d'une part le père et le fils sont consubstantiels (*consubstantialis*), bien que l'engendrement du Christ soit purement spirituel et que les deux soient *inseparabiliter* et co-éternels (*coeternus*)[54] ; d'autre part, Dieu est le créateur de toute chose, dont il possède la *paternitas*[55]. La promotion de la prière du Notre Père (*Pater noster*), texte à la fois biblique et liturgique qui évoque l'omnipotence du Dieu-père chrétien, roi (*adveniat regnum tuum*), nourricier (*panem nostrum quotidianum | da nobis hodie*) et salvateur (*sed libera nos a malo*), est aussi à considérer dans cette perspective[56]. Mis bout à bout, ces éléments laissent peu de doute : une certaine figure paternelle a connu une remarquable promotion lors de l'instauration sociale du système chrétien, promotion qui perdura tout au long des siècles de l'Europe médiévale. Ce développement induisit le renforcement de certaines personnes ou relations de parenté (en particulier *pater-filius*), au détriment d'autres (*mater*, *avus*, *proavus*, etc.), du moins dans un premier temps.

---

faire ressortir les cooccurrents *spécifiques* d'un terme pivot. Ainsi, si le lemme *ecclesia* apparaît comme cooccurrents de nombreux termes médiolatins, cela ne signifie pas que cette cooccurrence soit toujours sémantiquement significative. Pour cela, il faut que le lemme apparaisse plus fréquent dans les cooccurrents du pivot qu'il n'apparaît ailleurs. D'autres indices ou coefficients peuvent remplacer celui de Dice, qui a cependant montré son intérêt et sa robustesse.

53 Sylvie Joye indique que : « Le schéma trinitaire n'est donc guère utilisé pour décrire les relations entre père et fils, au sein de la cellule familiale. », dans *Ibid.*, p. 184, et encore : « En réalité, le débat trinitaire reposa bien peu sur la relation des Personnes de la Trinité en termes de parenté, bien que ce soient les Personnes du Père et du Fils qui aient mobilisé l'essentiel de la réflexion. C'est plutôt en terme de créateur et de créature que sont présentées les deux personnes dans les débats, et la relation de parenté n'est pas en tant que telle considérée comme un moteur de la réflexion. », dans *Id.*, p. 183. Il semble toutefois probable que le modèle trinitaire devint rapidement la matrice de la parenté médiévale et des relations sociales en général, non pas tant parce qu'elle se définit explicitement comme telle, mais parce que la relation des personnes de la Trinité est pensée comme idéale.

54 Les termes *consubstantialis*, *coeternus*, *consempiternus* ou encore *inseparabilis*/*inseparabiliter* apparaissent fréquemment lors des modélisations du champ sémantique de *pater*, à la fois chez Augustin, Grégoire de Tours, Bède, Raban Maur, Odon de Cluny, Pierre Damien, Bernard de Clairvaux ou encore Hugues de Saint-Victor. Pour des raisons de place bien compréhensibles, il est toutefois impossible de présenter la totalité des analyses réalisées. Ces éléments sémantiques impliquent cependant que les liens dans la Trinité existent sans la différence de générations qui précisément caractérise couramment le rapport père/fils dans notre système contemporain. Il s'agit d'un trait que l'on retrouve toutefois dans la paternité spirituelle médiévale, où un évêque peut devenir le père spirituel de son père charnel.

55 « Le Fils doit en effet être engendré, car sinon il ne serait pas fils ; mais il ne peut être créé, sans quoi il serait une créature et non pas divin au même titre que le Créateur. […] Une relation de paternité, fondée sur l'engendrement, s'inscrit ainsi au sein du noyau divin, entre les figures de la Trinité, différentes par leurs personnes mais égales par leur essence […]. », dans Jérôme Baschet, *La civilisation féodale. De l'An Mil à la colonisation de l'Amérique*, Champ Flammarion, Paris, 2006 (3e édition), p. 665. Sur la relation Créateur/créature, voir l'article fondamental d'Alain Guerreau, « Stabilità, via, visione : le creature e il Creatore nello spazio medievale », dans Enrico Castelnuovo, Giuseppe Sergi (dir.), *Arti e storia nel Medioevo, tome 3 : Del vedere : pubblici, forme e funzioni*, Torino, Einaudi, 2004, p. 167-197.

56 Les gloses/commentaires du *pater noster* sont légions dans l'Europe médiévale. Cf. Marc Philonenko, *Le Notre Père. De la Prière de Jésus à la prière des disciples*, Paris, Gallimard, 2001 ; Francesco Siri (éd.), *Le* pater noster *au XIIe siècle. Lectures et usages*, Turnhout, Brepols, 2015, où l'on trouvera l'ensemble de la bibliographie consacrée au *pater noster*.

### *II.2. Dans l'Europe médiévale : différentes paternités, une multitude de pères*

Au-delà du Dieu-père, qui était donc désigné comme *pater* au Moyen Âge ? Nous avancerons ici en utilisant à la fois des arguments qualitatifs et quantitatifs. La modélisation du champ sémantique du lemme dans le corpus des chartes (CEMA)[57] est instructive (fig. 4) :

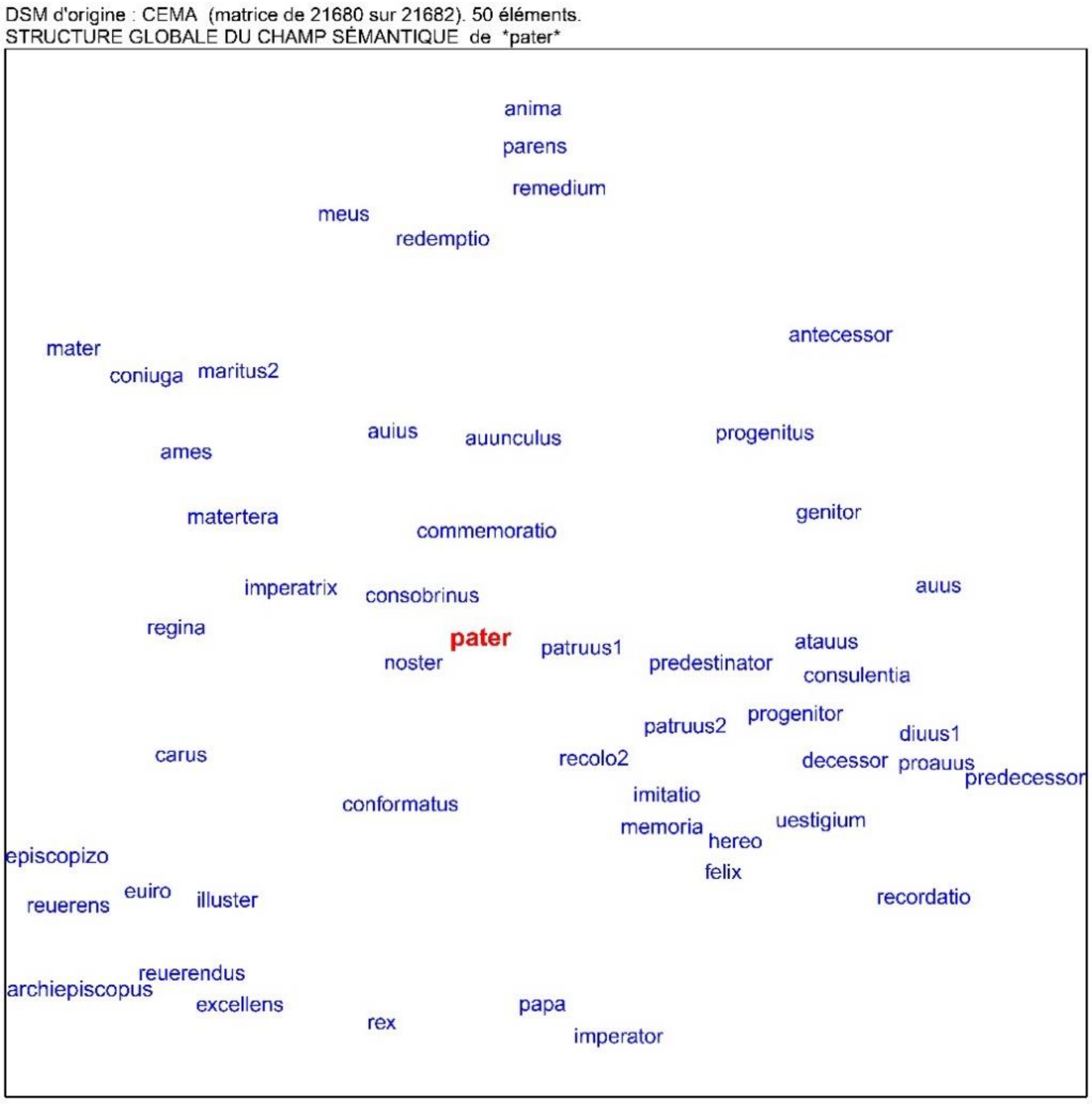

**Fig. 4 :** Champ sémantique de *pater* dans les chartes européennes des CEMA (50 termes).

Sur le premier axe factoriel (axe horizontal), la carte sémantique oppose à droite les morts, les ancêtres et le souvenir (*genitor*, *antecessor*, *progenitus*, *progenitor*, *avus*, *atavus*, *predecessor*, *memoria*, etc.), puis à gauche les personnages vivants (*mater*, *conjuga*, *maritus*, *avunculus*, *consobrinus*, *patruus*, *carus*, mais encore *regina-rex*, *imperatrix-imperator* ou encore *episcopizo*, *archiepiscopus* et *papa*). Sur l'axe 2 (axe vertical), elle oppose en haut la parenté charnelle (*mater*, *conjuga*, *maritus*), qui est présente dans les formules de don pour le remède des âmes (*anima*, *parens*, *remedium*, *meus*, *redemptio*)[58], et en bas à gauche la parenté spirituelle, « communautaire » ou « hiérarchiques » : l'évêque (*archiepiscopus*, *episcopizo*,

[57] Issues des packages R Wordspace et Cooc.

[58] Sur cette question, voir Eliana Magnani, « Transforming things and persons: The gift *pro anima* in the eleventh and twelfth centuries », dans Gadi Algazi, Valentin Groebner, Bernhard Jussen (dir.), *Negotiating the gift. Pre-modern figurations of exchange*, Göttingen, Vandenhoeck & Ruprecht, 2003 ; id., « Du don aux églises au don pour le salut de l'âme en Occident (IVe-XIe siècle) : le paradigme eucharistique », dans Nicole Bériou, Béatrice Caseau, Dominique Rigaux (dir.), *Pratiques de l'eucharistie dans les églises d'Orient et d'Occident*, Turnhout, Brepols, 2009, p. 1021-1042.

*reverens*), le pape (*papa*), le souverain (*rex, imperator*) – considéré comme le père de ses « sujets ». On comprend ainsi que le père médiéval constituait un trait d'union entre les vivants et les morts[59], mais aussi que le terme qualifiait une multitude de personnalités, qui allaient bien au-delà du père biologique[60]. Ces hypothèses quantitatives sont confirmées la lecture directe des documents, mais aussi des dictionnaires latins spécialisés.

Le long article du *Novum Glossarium* en particulier[61], indique qu'en matière de « parenté réelle »[62] *pater* peut certes désigner le père biologique, mais aussi un père par fosterage, les parents, un ancêtre ou des prédécesseurs – sans nécessairement que cela implique un lien généalogique direct. Il ne s'agit pas de nier le fait que les médiévaux savaient (la plupart du temps) *qui* était leur père biologique, mais d'avancer que cette situation était articulée à d'autres formes de paternités-filiations non seulement complémentaires, mais en fait essentielles. La paternité charnelle, plus ou moins restreinte[63], était ainsi articulée à une multitude de paternités spirituelles, qui allait des oncles aux abbés et aux évêques, dans des configurations sans cesse mouvantes. Les pères, pris comme un tout, permettaient d'articuler les relations sociales avec nombre de vivants, mais aussi de personnages défunts, avec différentes intensités et en fonction des circonstances. La paternité était ainsi plutôt un « type » de relation, qu'un état globalement stable aux contours définis, comme cela apparaît à l'inverse massivement dans les textes contemporains (fig. 5)[64].

---

[59] *Les vivants et les morts dans les sociétés médiévales*, Paris, Publications de la Sorbonne, 2018, où l'on trouvera une bibliographie très complète, et en particulier Michel Lauwers et Julien Loiseau, « Rapport introductif : l'historien (médiéviste) et les morts, Occident chrétien et pays d'Islam », p. 11-39.

[60] L'analyse factorielle peut se doubler d'une autre lecture, complémentaire : de nombreux termes à droite correspondent à l'ascendance (masculine), tandis qu'à gauche d'autres lemmes renvoient à l'alliance (féminine) ; en haut, on retrouve largement la question du salut par les œuvres, tandis qu'en bas apparaît le salut par l'Église et les autorités. Bien entendu, cette lecture redouble en partie celle proposée dans le corps du texte.

[61] *Novum Glossarium Mediae Latinitatis*, volume *Passibilis-Pazzu* (éd. Jacques Monfrin), Kopenhagen, E. Munksgaard, 1993, col. 651-668.

[62] Ce terme de « parenté réelle », proposé par les rédacteurs de l'article ne va pas sans poser de problème. Il suppose en effet que les parentés spirituelles et divines seraient « irréelles/symboliques ». Nous lui préférons l'expression de « parenté charnelle », proposée par Anita Guerreau-Jalabert. Le classement proposé par le dictionnaire peut aussi interroger : la paternité charnelle est en effet placée au départ de l'article, alors qu'elle constitue *de facto* une partie minoritaire des occurrences documentaires. Certes, la rédaction d'un article de ce type est toujours une affaire de découpage, plus ou moins arbitraire. Mais cette organisation pourrait laisser entendre que les parentés spirituelles/divines étaient plus ou moins factices, parce qu'elles découleraient d'un modèle biologique hiérarchiquement supérieur. Fidèles à la Bible, les étymologies de *pater* par les auteurs médiévaux commencent à l'inverse toutes par les paternités divine/spirituelle, avant d'aller vers les paternités charnelles (Isidore de Séville, Raban Maur, Alain de L'Isle, etc.). Plus généralement, la réintroduction des catégories « réelles » vs. « irréelles/symboliques » est un obstacle majeur dans l'étude des relations de parenté médiévales, puisqu'elle revient à retourner sans cesse au paradigme de la « famille ». Sur ces questions, voir en particulier Anita Guerreau-Jalabert, « *Spiritus* et *caritas*. Le baptême dans la société médiévale », dans Françoise Héritier-Augé, Élisabeth Copet-Rougier (éd.), *La parenté spirituelle*, Paris, Archives contemporaines, 1996, p. 133-203.

[63] Parce qu'elle ne se limitait pas à la paternité « biologique » au sens strict. Le *Novum Glossarium* indique bien que les « ancêtres » et « prédécesseurs » pouvaient en effet être qualifiés de *patres*, et il en va de même pour l'ensemble des parents biologiques (membres féminins compris ; même si ce cas est plus rare, l'article relève en effet un « *patrum meorum Godesteo et Godina* »). On lit par exemple chez Alain de L'Isle [† 1202/1203] : « *Pater ratione generis, unde Ioannes: « Patres vestri manducaverunt, etc. »* (Jean 6 :49), *id est praedecessores vestri.* », dans ALANUS AB INSULIS, *Distinctiones dictionum theologicalium*, PL 210, col. 685-1011, ici col. 894b. Dans les chartes des CEMA, l'association du lemme *pater* au pluriel (*patres, patrum, patribus*) et du lemme noster renvoie 283 occurrences, très largement issues des diplômes. Par exemple : « *patres nostri Pippinus videlicet et Karolus* », dans un acte de Lothaire II pour l'abbaye de Prüm, en 856 : Theodor Schieffer (éd.), *Die Urkunden Lothars I. und Lothars II.*, Berlin, Weidmann, 1966, n° 3 (MGH : DD Lo I / DD Lo II). On devrait donc en déduire que le mot *pater* pouvait désigner, peut-être par métonymie, les ascendants directs. Il s'agit par ailleurs de rappeler qu'en portugais, et ceci dès le Moyen Âge, « mes parents » (autrement dit le père et la mère) se dit « *meus pais* », *pai* au singulier signifiant par ailleurs « père ». Il en va de même avec l'expression espagnole « *mis padres* », ou encore en catalan.

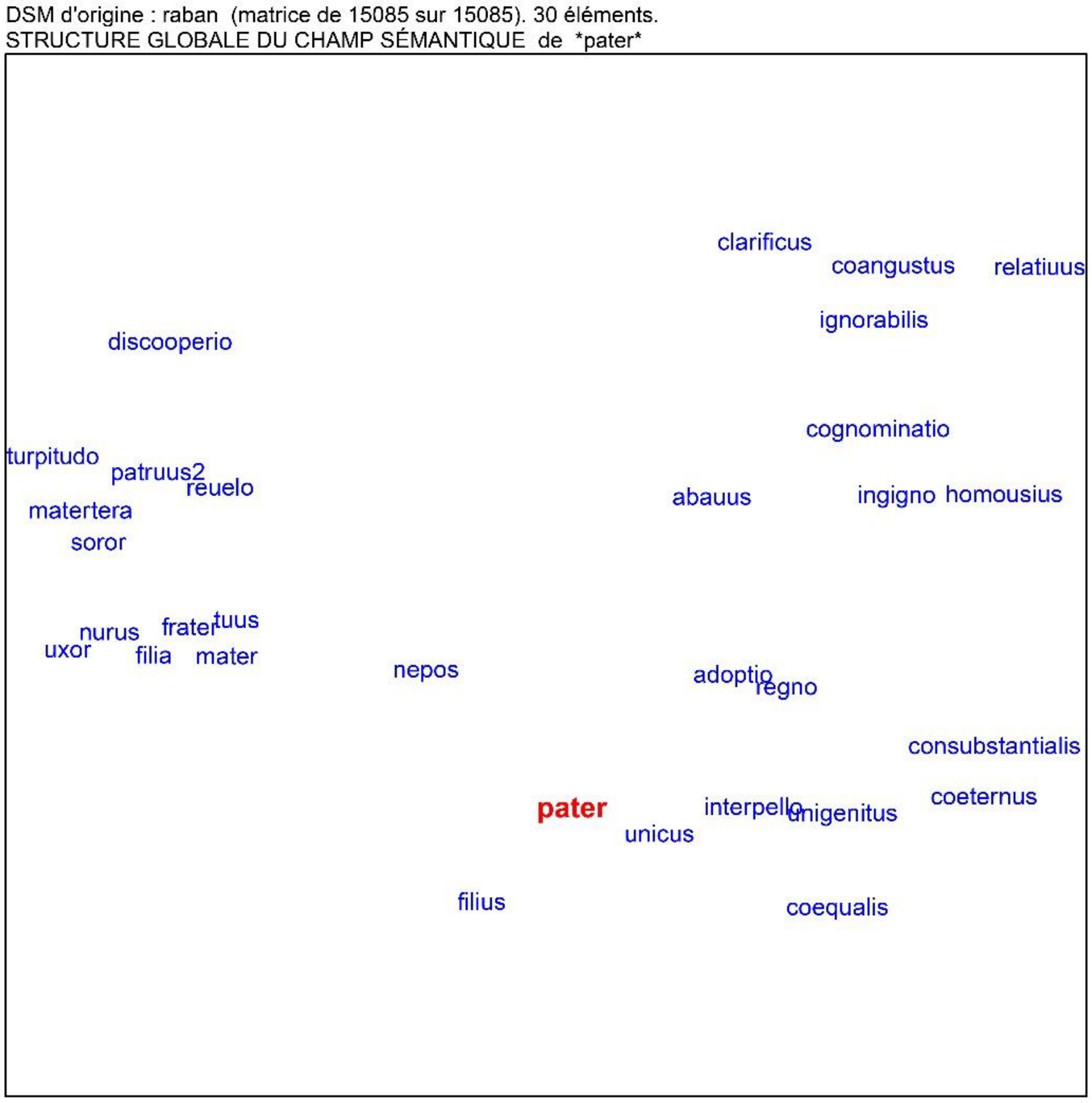

**Fig. 5 :** Champ sémantique de *pater* chez Raban Maur (PL, 30 termes). L'axe 1 (horizontal) oppose, à gauche, des termes relatifs dans ce contexte à la parenté charnelle (*uxor*, *filia*, *mater*, *frater*, etc.), tandis que la parenté spirituelle-divine se trouve à l'opposé, en bas à droite (*coequalis*, *coeternus*, etc.).

Ces deux paternités médiévales (charnelles et spirituelles) étaient toutefois complétées par le modèle paternel par excellence, déjà évoqué, celui de la paternité divine. L'auteur carolingien Raban Maur [† 856] consacre ainsi plusieurs paragraphes de son *De Universo* à préciser les différents sens du terme[65]. Mais pour lui, toute paternité dérive d'abord de celle de

[64] Si Hans Hummer dédicace son ouvrage sur la parenté à son « quatuor paternel » (« *To my four-fathers: Llyod Hummer | John McCulloh | Patrick Geary | Marc Kruman*. », dans *Visions of kinship in medieval Europe*, op.cit., p. V), c'est bien entendu parce qu'il a bien conscience de la multipaternité médiévale et qu'il joue sur cette alterité.

[65] « *Multis modis in Scripturis pater positus reperitur: quia aut naturaliter pater dicitur, ut Deus pater omnium, a quo sunt omnia, quia ab eo omnia originem habent; aut generatione, sicut nominatur ab eo, quod filium genuit, ut est illud: Abraham genuit Isaac, Isaac autem genuit Iacob (Matth. I) , etc.; aut posteritatem gentis ab eo ortae: sicut in Evangelio legitur, Iudaeos Domino dixisse: Pater noster Abraham est (Ioan. VIII) , et dives in inferno positus patrem vocavit Abraham (Luc. XVI) . Pater vero homo primus, Adam intelligitur: ut est illud Isaiae: Pater vester primus peccavit (Isa. XLIII) . Item pater, mundus sive diabolus est […]. Rursum patres apostoli, sive prophetae, ut in psalmo: Pro patribus tuis nati sunt tibi filii (Psal. XLIV). […] Aut veneratione, sicut monastica disciplina docet monachos: ut illum, qui super eos pastor constitutus est, abbatem et dominum vocent. Sed quaeritur, cum Dominus discipulis suis iusserit, dicens: Patrem nolite vocare vobis super terram, quia unus est*

Dieu : « *Deus pater omnium* », quand bien même celle-ci est d'abord adoptive[66]. Cette idée d'un Créateur, père de toutes choses, est évidemment présente dès la *Vulgate*, et en particulier dans l'Évangile de Matthieu (23:9) qui insiste sur la paternité première et totale de Dieu[67]. Lui seul pouvait véritablement être qualifié de *pater*. Ce passage biblique fut d'ailleurs abondamment commenté tout au long des siècles médiévaux, et l'on ne saurait sous-estimer sa portée[68]. Il contribua d'ailleurs sans doute à créer une hiérarchie des paternités, plutôt qu'une exclusivité de la paternité divine. Plus généralement, la conception d'un Dieu-père à partir duquel s'articulent toutes les autres paternités, était très répandue chez les auteurs médiévaux, depuis les Pères de l'Église jusqu'au XVe-XVIe siècle[69]. Ainsi, Nicolas de Cues [† 1464] insiste encore fortement sur la dimension « paternelle » du Dieu chrétien[70], la faisant entrer dans une série d'analogies extrêmement riches : le Créateur est à la Création ce que le roi est à son royaume, à l'Église ce que le *paterfamilias* est à sa maison (*domus*)[71]. Autrement dit, sa *paternitas* est absolue, première, même si encore une fois, cela n'exclut pas une « cascade » de pères spirituels et charnel, dont les paternités se redoublent selon une logique analogiste[72].

---

*Pater vester, qui in coelis est: nec vocemini magistri, quia magister vester unus est Christus (Matth. XXIII), quare adversum hoc praeceptum doctorem gentium Apostolus esse se dixerit: aut quomodo vulgato sermone, maxime in Palestinae et Aegypti monasteriis se invicem patres vocent? Quod sic solvitur, aliud esse natura patrem vel magistrum, aliud indulgentia. Nos si hominem patrem vocamus, honorem aetati deferimus, non auctorem nostrae ostendimus vitae: quia nec initium vitae ex eis habemus, sed transitum vitae per eos accepimus.* », dans Rabanus Maurus, *De Universo*, PL 111, col. 186b-186d.

66 La paternité de Dieu sur les hommes passe en effet par le baptême (cf. Anita Guerreau-Jalabert, « Qu'est-ce que l'*adoptio* dans la société médiévale ? », art. cité, p. 40-41 ; Joseph Morsel, « Dieu, l'homme, la femme et le pouvoir. Les fondements de l'ordre social d'après le 'Jeu d'Adam' », dans Monique Goullet (dir.), *Retour aux sources. Textes, études et documents d'histoire médiévale offerts à Michel Parisse*, Picard, Paris, 2004, p. 537-549, ici p. 543-544). Toutefois, il nous semble que ce mode de filiation spirituel n'est pas exclusif, puisque Dieu est déclaré père de toutes les créatures, y compris celles qui ne sont pas baptisées.

67 « *Et patrem nolite vocare vobis super terram unus enim est Pater vester qui in caelis est.* », Mat. 23:9. De même : « *Unus Deus et Pater omnium, qui est super omnes, et per omnia, et in omnibus nobis.* », Eph. 4:6. Le verset de Matthieu est l'un des plus fréquemment cité dans les études sur la paternité médiévale.

68 Mat. 23:9 est par exemple repris par saint Augustin [† 430], sous la forme suivante : « *Ne vobis dicatis patrem in terra; unus est enim pater vester Deus* », dans Augustinus Hipponensis, *Breviculus collationis cum Donatistis*, PL 43, col. 613-650, ici col. 628 (ce texte date du début du Ve siècle).

69 Karine Spiecker Stetina, *The Fatherhood of God in John Calvin's Thought*, Bletchley, Milton Keynes, 2016.

70 « *Que est autem via de patre familias ad creatorem ascendendi quasi de imagine ad exemplar, attendamus : In simplici patre familias multiplex paternitas coincidit. Nam est pater, quia genitor filii ; est pater, quia curam habet omnium et quia rector et senior, cui omnis honor debetur. Omnis autem paternitas, que in celo et in terra reperitur, est a Deo Patre. Instruimur autem per Filium Dei Jesum illum esse Patrem, quem Judei Deum nominant. Hic igitur Deus est ipsa paternitas absoluta que et Pater.* », dans Nicolai de Cusa, *De Aequalitate*, dans *Opera*, *Liber X*, Basilae, Henrici Petrina, 1565, p. 681. Le textes de Nicolas de Cues employés sont issus du corpus *Cusanus Portal*.

71 « *Alpha et Omega, in mundo ut rex in regno : Ubique regnat et imperat, in angelo ut decor in quantum veritas et sapor in quantum bonitas, in ecclesia sicut paterfamilias in domo, in electis ut liberator a malis, adiutor in bonis, in reprobis ut terror et horror, in fideli anima sicut rex in regno, fons in hortis, lux in tenebris, carbunculus in anulo. Item, Deus est eternus. Nullum ei competit tempus aut temporis mensura, sine principio et fine.* », dans Nicolai de Cusa, *Fides autem catholica*, dans *Opera omnia*, éd. par Rudolf Haubst, vol. XVI:1 (*Sermones* I, 1430-1441, *Fasciculus* I : *Sermones* I-IV), Hambourg, Felix Meiner, 1970, p. 71.

72 Sur la pensée analogiste, son fonctionnement et ses implications, voir Arthur Lovejoy, *The great chain of being: a study of the history of an idea*, Cambridge, Harvard University Press, 1936 ; Claude Lévi-Strauss, *La pensée sauvage*, Paris, Plon, 1962 ; Philippe Descola, *Par-delà nature et culture*, Paris, Gallimard, 2005 ; et plus spécifiquement en médiévistique : Joseph Morsel, « Dieu, l'homme, la femme et le pouvoir. Les fondements de l'ordre social d'après le 'Jeu d'Adam' », art. cité ; Anita Guerreau-Jalabert, « Occident médiéval et pensée analogique : le sens de *spiritus* et *caro* », dans Jean-Philippe Genet (dir.), *La légitimité implicite*, Paris-Rome, Publications de la Sorbonne-École française de Rome, 2015, p. 457-476 ; Amélie Didier, Alexane Girard, Gauthier Griffart, Héléna Lagréou, Arnaud Montreuil, Joseph Morsel, Carole Nestoret, Gilles Texier, « Similitude ? Quelle similitude ? Une recherche collective menée dans le cadre de la pépinière *Reproduction sociale au Moyen Âge* », *Les carnets du LaMOP*, mis en ligne le 15/11/2019, https://lamop.hypotheses.org/?p=5988 (consulté le 14.04.2020).

La supériorité de la paternité divine est évidemment à chercher dans le mode d'engendrement qui lui est en partie propre, bien qu'elle relève pleinement de la paternité spirituelle. La relation de Dieu à son fils le Christ est pensée comme une paternité parfaite, modèle de toute relation sociale, puisqu'elle n'implique aucun échange charnel : *pater* et *filius* sont unis par un amour parfait, parce que celui-ci est entièrement spirituel et indivisible[73]. Le père se situait ainsi au cœur des différentes formes de parenté médiévale : charnelle et spirituelle (donc aussi divine)[74]. Cette situation entraînait nécessairement une hiérarchisation des paternités : la paternité spirituelle-divine était idéale, car la substance du Père comme du Fils apparaissait « pure » et « simple »[75]. Autrement dit, cette forme de relation était a priori une paternité entre égaux, dans la consubstantialité. Son analogon séculier était la paternité spirituelle-terrestre, qui excluait a priori tout rapport charnel et entraînait une reproduction sans génération humaine : c'est la paternité des personnages ecclésiastiques, celle des saints, des papes, des évêques, des abbés, etc. Ces deux dimensions de la paternité, spirituelle-divine et spirituelle-terrestre, n'étaient toutefois pas dissociables : elles constituaient au contraire les deux faces d'une même pièce[76]. La paternité charnelle venait quant à elle en dernière position, car elle impliquait la reproduction sexuée, implication directe de la chute et du péché originel[77]. Dans cette perspective, il est probablement faux de penser qu'une figure comme Joseph, le père terrestre du Christ, constituait un père « affaibli » par son non-engendrement charnel[78]. Sa paternité, certes non-biologique et non-adoptive, était considérée comme parfaitement légitime : parce qu'elle était, comme toutes les autres d'ailleurs, « partagée »[79].

---

[73] Anita Guerreau-Jalabert, « Amour et amitié dans la société médiévale : jalons pour une analyse lexicale et sémantique », dans Laurent Jégou, Sylvie Joye, Thomas Lienhard, Jens Schneider (dir.), *Splendor Reginae: Passions, genre et famille. Mélanges en l'honneur de Régine Le Jan*, Turnhout, Brepols, 2015, p. 281-289.

[74] Jérôme Baschet, « Jeux de pères. La conversion de paternité dans quelques images médiévales », dans Chloé Maillet (dir.), *Parenté en images*, numéro d'*Images re-vues. Histoire, anthropologie et théorie de l'art*, vol. 9, 2011, https://journals.openedition.org/imagesrevues/1612

[75] Par exemple chez Pierre de Poitiers [† 1205/1206] : « *Item, Christus est substantia simplex, vel composita, utrumque videtur esse verum. Nam eadem est substantia quae et Pater. Cum ergo Pater sit substantia simplex, videtur Christus esse substantia simplex. Cum vero Christus sit homo verus et ita habeat et corpus et animam sicut et alii homines, videtur esse substantia corporea, et ita videtur habere partes sui animam et corpus.* », dans Petrus Pictaviensis, *Sententiae*, PL 211, col. 789-1280d, ici col. 1176d-1177a.

[76] Ce qui ne signifie pas que les clercs étaient incapables de distinguer la spécificité de la relation de paternité dans la Trinité. Simplement que la paternité spirituelle terrestre tirait précisément son efficacité de sa proximité avec un modèle : la paternité trinitaire, conçue comme une paternité spirituelle parfaite. L'association était d'autant plus essentielle que le couple charnel/spirituel constituait la matrice des représentations médiévales, et permettait d'affirmer la supériorité des clercs (et donc la supériorité de leur mode de reproduction) sur les laïcs – précisément car leur reproduction était assimilable (mais pas équivalente) à celle qui opérait dans la Trinité. Voir aussi les remarques de Joseph Morsel, qui vont plus loin encore sur l'emboîtement des différents niveaux sémantiques dans la parenté : « On peut ainsi considérer que l'usage monosémique de ces termes pour les parents charnels et, dirions-nous, analogique pour les parents spirituels, fait partie de la construction de la « famille » comme unité de base de la société (et la forme métonymique de la parenté). […] L'écart entre parenté charnelle/humaine et parenté divine/spirituelle ne pouvait dès lors apparaître en toute clarté que si les termes concernés étaient monosémiques. », dans Joseph Morsel, *Noblesse, parenté et reproduction sociale à la fin du Moyen Âge*, Paris, Picard, 2017, p. 119.

[77] Peter Brown, *Le renoncement à la chair. Virginité, célibat et continence dans le christianisme primitif*, Paris, Gallimard, 1995 ; Jérôme Baschet, « La parenté partagée : engendrement charnel et infusion de l'âme », dans Carla Casagrande, Silvana Vecchio (dir.), *Anima e corpo nella cultura medievale*, Florence, Sismel, 1999, p. 124-127, ainsi que la note 23, qui concerne la discussion entre Anita Guerreau-Jalabert et Jérôme Baschet sur ce point.

[78] Sur cette question, nous renvoyons à l'analyse de Paul Payan, *Joseph : une image de la paternité dans l'Occident médiéval*, op.cit., p. 280-291. Voir aussi Jean-Michel Sanchez, Jean-François Froger, Jean-Paul Dumontier, *Saint Joseph, image du Père*, Paris, Éditions grégoriennes, 2015.

[79] Jérôme Baschet, *Le sein du père […]*, op.cit.., chapitre VIII : « La paternité partagée » ; id., « La parenté partagée : engendrement charnel et infusion de l'âme », art. cité. Il n'est pas envisageable de développer ici des arguments sur la « chaîne généalogique » médiévale : voir Séverine Lepape, *Représenter la parenté du Christ et de la Vierge : l'iconographie de l'arbre de Jessé en France du Nord et en Angleterre, du XIII^e^ siècle au XVI^e^ siècle*, thèse de doctorat de l'EHESS, 2007.

### *II.3.* **Pater** *et* **dominus** *: la domination dans la paternité*

On comprend ainsi par extension que *pater* renvoyait à différents types de pères charnels, mais aussi et surtout spirituels (terrestres ou célestes). C'est d'ailleurs essentiellement dans ce sens que le *Novum Glossarium* donne les multiples significations du lemme[80], sans toutefois éclaircir la signification des définitions que le dictionnaire juxtapose. *Pater* qualifie celui qui donne le baptême ou encore le parrain[81], le fondateur d'un ordre, d'une règle monastique, d'un monastère ou d'une église[82]. Mais le lemme médiolatin désigne aussi les maîtres, les apôtres et les prophètes, les Pères de l'Église, les Pères du désert et les Pères des conciles[83]. On peut encore qualifier de *pater* le pape (pensons à l'expression « le Saint-Père »), les patriarches, les cardinaux, les évêques, les abbés[84], et surtout les saints[85]. En fait, tous les membres de l'Église jusqu'au simple prêtre pouvaient être appelés « père »[86]. Toujours par extension, le substantif peut désigner le diable (ce qui n'avait pas échappé à Freud[87]), ou le premier homme : Adam[88]. Enfin, certains laïcs peuvent eux aussi être qualifiés de pères spirituels : les souverains, les rois, les puissants en général (ducs, comtes, seigneurs divers et variés), ainsi que tous ceux impliqués dans une pratique de fosterage[89]. Les multiples significations de *pater* n'inviteraient-elles pas à penser qu'il existait des usages « réels » et d'autres « métaphoriques », ces derniers étant peu significatifs ? À rebours d'une partie de l'historiographie, nous pensons au contraire qu'il faut prendre l'ensemble de ces significations au sérieux, et tenter de les articuler.

---

[80] « *Pater* », dans *Novum Glossarium Mediae Latinitatis*, op.cit., col. 651-668.

[81] *Pater* peut en effet désigner soit le parrain au sens strict, soit le prêtre qui administre le sacrement. Par exemple chez Amalaire [† 850] : « *Illi qui baptizant, patres sunt baptizatorum.* », dans Amalarius, *De ecclesiasticis officiis*, PL 105, col. 985-1242d, ici col. 1056a – le passage est identifié par Bernhard Jussen, *Spiritual Kinship as Social Practice [...]*, op.cit., p. 281 (note 120). Autre exemple dans ce faux diplôme de Clovis pour Saint-Pierre-le-Vif de Sens (probablement de la fin du XI^e^-début du XII^e^ siècle), relatant le baptême de ce dernier : « *Anno tercio postquam baptismi sacramentum percepi per manus patris mei Remigii Remorum episcopi [...]* », dans *Die Urkunden der Merowinger*, éd. par Carlrichard Brühl, Theo Kölzer, 2 volumes, Hanovre, Hahnsche Buchhandlung, 2001, n° 4 (MGH – Diplomata, Merowinger) – même si dans ce cas précis, on pourrait se demander si *pater* s'applique au statut de l'évêque Rémi ou à son action. Sur cet acte, outre l'édition, nous renvoyons à Carlrichard Brühl, « Clovis chez les faussaires », dans Olivier Guyotjeannin (éd.), *Clovis chez les historiens*, Paris, 1996, p. 219-240, ici p. 226-231 (*Bibliothèques de l'École des chartes*, tome 154). Quant aux mentions de parrains-pères, elles sont abondantes, par exemple dans cet acte d'Otton I^er^ de juin 941 : « *comitum nostri ejusdem Geronis filio nostro autem spiritali filiolo videlicet Sigifrido quem sacri baptismatis fonte levauimus in comitatu prelibati patris ejus [...]* », dans *Die Urkunden Konrad I., Heinrich I. und Otto I*, éd. par Theodor Sickel, Hanovre, Hahn, 1879-1884, n° 40 (MGH – Diplomata, DD K I / DD H I / DD O I).

[82] C'est par exemple le cas de saint Benoît ou de saint Colomban (cf. « *Pater* », dans *Novum Glossarium Mediae Latinitatis*, op.cit., col. 657-658). Ainsi, l'expression « *sancti patris Benedicti* » apparaît 45 fois dans la PL. On pourrait toutefois se demander si *pater* qualifie ici le fait d'être un fondateur, ou plutôt d'être saint. Sur cette question, voir les notes qui suivent.

[83] Par exemple saint Augustin : « *sancti patris Augustini* ».

[84] Sur l'étymologie d'abbé (*abbas* / *ἀββᾶ*) signifiant « père » en araméen, l'historiographie est immense. Nous renvoyons à Jérôme Baschet, *Le sein du père [...]*, op.cit., p. 38. On pourrait aussi rappeler l'étymologie de « pape » (*papa* / *πάππας*) nous ramène elle aussi à la paternité. Il convient ici de signaler que la paternité « en Dieu » ou « en Christ » des évêques, papes, abbés, etc., relève certes de la parenté spirituelle, mais pas baptismale. Si c'est le baptême qui fait entrer dans la catégorie englobante des chrétiens, c'est l'élection (comme évêque, pape ou abbé) qui fait passer du statut de « frère » (par le baptême) à celui de « père ». Or, l'élection implique l'intervention du Saint-Esprit – ce qui justifie d'ailleurs d'autant plus de parler de parenté spirituelle. On signale d'ailleurs que les rois et empereurs peuvent eux aussi être qualifiés de *pater*, comme on le rappelle plus loin, ce qui renvoie probablement aussi à son caractère d'élu. Sur l'invocation-intervention du Saint-Esprit dans les élections épiscopales, voir Giovanni Caron, « Les élections épiscopales dans la doctrine et la pratique de l'Église », *Cahiers de civilisation médiévale*, vol. 11-44, 1968, p. 573-585, en particulier p. 577-578.

[85] Dans les CEMA, les cooccurrences associant directement (*i.e.* accolées) les lemmes *pater* et *sanctus* sont fréquentes : 913 au total. Nous avons pu en dénombrer 625 dans les OpenMGH, 4 045 dans la PL. La signification est claire : il était courant de qualifier un saint de « père ».

[86] Raban Maur ne dit pas autre chose dans le passage précité du *Du universo*.

[87] Dans des proportions non négligeables : celui-ci est fréquemment qualifié de « *pater mendacii* » (123 occurrences dans la PL), autrement dit « père du mensonge ». L'expression est intéressante, car si elle apparaît dès les Pères de l'Église (« *Diabolum dominus dixit patrem mendacii* », dans Augustinus Hipponensis, *In Ioannis evangelium tractatus CXXIV*, PL 35, col. 1379-1976, ici col. 1704). Elle se développe ensuite plus largement à partir du VIII^e^ siècle et au-delà. Dans le seul *Corpus Thomisticum*, l'expression apparaît ainsi une dizaine de fois. Plus rare dans les chartes, on la rencontre toutefois dans quelques occurrences, comme par exemple : « *Quoniam gesta mortalium facile labuntur ab animis gravedine carnis depressis et patre mendatii suggerente ab his qui ex parte ejus sunt falsis assertionibus [...]* », en 1178 dans un acte pour Notre-Dame de Léoncel : *Cartulaire de*

Ce relevé trop rapide permet en effet de formuler une hypothèse : toute personne dominante dans la société médiévale, autrement dit un seigneur (*dominus*), pouvait être qualifiée de *pater* dans un contexte social déterminé, qui était beaucoup plus étendu que celui de la paternité biologique contemporaine. Le pape, les évêques, les abbés et encore une fois les saints étaient régulièrement qualifiés de « seigneur et père » (« *domino et patri* »). Une simple comparaison avec les textes antiques païens permet de confirmer cette impression : il existe seulement 45 cooccurrents associant *pater* et *dominus* dans le corpus non-chrétien consulté, contre 9 778 cooccurrences dans les chartes des CEMA – soit une association environ 41 fois plus forte[90]. On constate par ailleurs, à travers une analyse de l'évolution des cooccurrences, que les associations entre *pater* et *dominus* ne cessent de se renforcer entre le IX^e^ et le XIII^e^ siècle, avec une accentuation plus nette aux XII^e^-XIII^e^ siècles, toujours dans les chartes (fig. 6). Quel sens donner à ces observations ? Une possible explication est que cette situation est fondée sur la proximité sémantique entre *dominus* (qui désigne le Seigneur du ciel, mais aussi

---

*l'abbaye de Notre-Dame de Léoncel au diocèse de Die, ordre de Cîteaux*, éd. par Ulysse Chevalier, Montélimar, 1869, n° XXVIII. Plus généralement, les associations directes de *pater* et de *diabolus* (souvent sous la forme « *patre diabolo* ») se rencontrent 449 fois dans la PL – le plus souvent en référence au verset Jean 8:44 : « *Vos ex patre diabolo estis […]* ». Sur ce dernier passage, voir Émile PUECH, « Le Diable, homicide, menteur et père du mensonge en Jean 8,44 », *Revue Biblique*, vol. 112, 2005, p. 215-252. Les liens entre paternité et Diable ont été peu relevés par les médiévistes. Voir cependant la géniale intuition de Freud, où ce dernier évoque de diable « comme substitut du père » : Sigmund FREUD, « Une névrose démoniaque au XVII^e^ siècle », *La Revue française de psychanalyse*, tome 1 (fascicules 1, 2 et 3), Paris, Doin, 1927 (première publication en allemand en 1923 – repris dans *Essais de psychanalyse appliquée*, Paris, Éditions Gallimard, 1933).

[88] Toujours chez Raban Maur : « *Pater vero homo primus, Adam intelligitur: ut est illud Isaiae: Pater vester primus peccavit* (Isa. 43) ». La paternité première et charnelle d'Adam est toutefois renvoie bien entendu à la Bible, elle est aussi affirmée très tôt, par exemple chez l'évêque de Brescia Philastre [† vers 387], dans son *De Hœresibus* : « *Unde et in hac coniunctione sanctae adorandaeque Trinitatis gaudemus, in qua ante Adam pater noster, Noe, Abraham, Moses, et prophetae, sacerdotes, iudices, et apostoli pariter, et evangelistae praedicantes meruerunt consequi angelicam dignitatem.* », dans Philastrius, *De Hæresibus*, PL 12, col. 1111-1302a, ici col. 1292c. L'association de l'anthroponyme « Adam » et du lemme *pater*, à plus ou moins cinq mots de distance, est présente près de 200 fois dans la PL. Pour une chronologie plus tardive, les occurrences sont tout aussi nombreuses, par exemple chez Guillaume Perault [† 1271], qui reprend ici des passages de saint Augustin (*Sermo de disciplina Christiana*, III) : « *Quarto, fraternitas spiritualis, de qua Augustinus de Doct. Chr.: omnes quidem fratres secundum quod homines sumus, quanto magis secundum quod Christiani sumus: ad id quod homo es, unus pater fuit Adam, una mater fuit Eua; ad id quod Christianus unus pater est Deus, una mater est Ecclesia.* », dans Guilelmus Peraltus, *De eruditione principum*, livre II, chapitre XII, Parme, 1864 (dans le *Corpus Thomisticum*). Plus tardivement encore, chez Marsile Ficin [† 1499] : « *Pater primus scilicet Adam peccavit, quondo preceptum divinum neglexit.* », dans Ficino Marsilio, *De christiana religione liber*, Paris, 1559, fol. 100v. Sur Adam et sa postérité médiévale, voir Irène Rosier-Catach, Gianluca Briguglia (dir.), *Adam, la nature humaine, avant et après. Épistémologie de la Chute*, Éditions de la Sorbonne, Paris, 2016, en particulier l'article de Sylvain Piron, « Ève au fuseau, Adam jardinier », p. 283-315, qui revient sur cette dimension paternelle ; Joseph Morsel, « Dieu, l'homme, la femme et le pouvoir. Les fondements de l'ordre social d'après le 'Jeu d'Adam' », art. cité, p. 543-544.

[89] Sur ces associations, le mémoire inédit de Sylvie Joye est particulièrement riche : *L'autorité paternelle en Occident*, op.cit., en particulier le chapitre III, « L'imagine du père : atout et piège politique », p. 202-281.

[90] Parmi les expressions fréquentes employant les deux lemmes, on lit par exemple « *venerabilis patris (et/ac) domini* », « *sanctissimi/reverendi in Christo patris et domini* », « *patrem nostrum dominum* », etc. Dans les textes théologiques de la PL, on relève 10 732 cooccurrences de *pater* et *dominus*, à une distance de plus ou moins cinq mots.

n'importe quel seigneur terrestre) et *pater* (qui renvoie à Dieu le Père, mais aussi à tous dominants). Ainsi, le rapprochement entre les deux termes ne signifie pas seulement que Dieu ne cessait de s'affirmer comme un « père ultime », mais aussi que de cette relation découlait un progressif renforcement entre le fait d'être seigneur (Dieu étant aussi le « seigneur ultime ») et d'être un père spirituel. De cette extension, qui passe à la fois par le développement de la logique seigneuriale proprement médiévale et par le développement de certaines propriétés spirituelles attribuées aux seigneurs terrestres, il découle une situation spécifique et dynamique[91]. Tout se passe en effet comme si, dans l'Europe médiévale, toute personne se trouvait être dans une situation de multi-paternité à la fois charnelle et spirituelle (y compris divine) – parce qu'il était à la fois le *filius* d'un père biologique[92], de différents seigneurs et de personnages saints et divins –, mais aussi que cette relation dépendait de la logique de domination qui sous-tendait l'ensemble de la structure des rapports sociaux et de production. Le développement des cooccurrences entre *pater* et *dominus* montre en outre que cette logique n'a fait que s'accroitre.

[91] Que l'on devienne père par l'élection est rigoureusement congruent avec cette hypothèse d'un lien entre paternité et *dominium*, puisque l'élection médiévale est précisément conçue au sein d'un système de représentations où le pouvoir descend (il vient de Dieu), donc où l'élection n'est nullement une forme de délégation par les votants – mais plutôt une forme de reconnaissance anticipée d'un résultat voulu par Dieu. Parallèlement, Jacques Le Goff interprétait dès 1977 le rituel symbolique de la vassalité en lien à la fois avec la *potestas* et avec la paternité (celle du *senior* – dans Jacques Le Goff, « Le rituel symbolique de la vassalité », *Pour un autre Moyen Age*, Paris, Gallimard, 1977, p. 349-420, en particulier p. 355 et 373, où il articule *manus*, *potestas* et *pater*, puis p. 380-382, 415, où il revient sur les liens entre ce rituel et la parenté/paternité : « Dans le cas de l'investiture féodale c'est, me semble-t-il, dans la sphère du sacré parental que se meut la symbolique. »). De ce fait, il est probable que le rapport père-fils soit aussi à examiner comme analogon du rapport vieux-jeune, qui est lui aussi un rapport de domination, à articuler au rapport *dominus-homo*.

[92] Le père jouant un rôle fondamental dans la reproduction charnelle, car la semence maternelle (le sang) était le plus souvent considérée comme « passive ». Sur ce thème particulièrement riche, voir Maaike van der Lugt, *Le Ver, le démon et la vierge. Les théories médiévales de la génération extraordinaire*, Paris, Les Belles Lettres (*L'Âne d'Or*, 20), 2004 ; Charles de Miramon, Maaike van der Lugt, « Sang, hérédité et parenté au Moyen Âge : modèle biologique et modèle social. Albert le Grand et Balde », *Annales de démographie historique*, n° 137, 2019, p. 21-48.

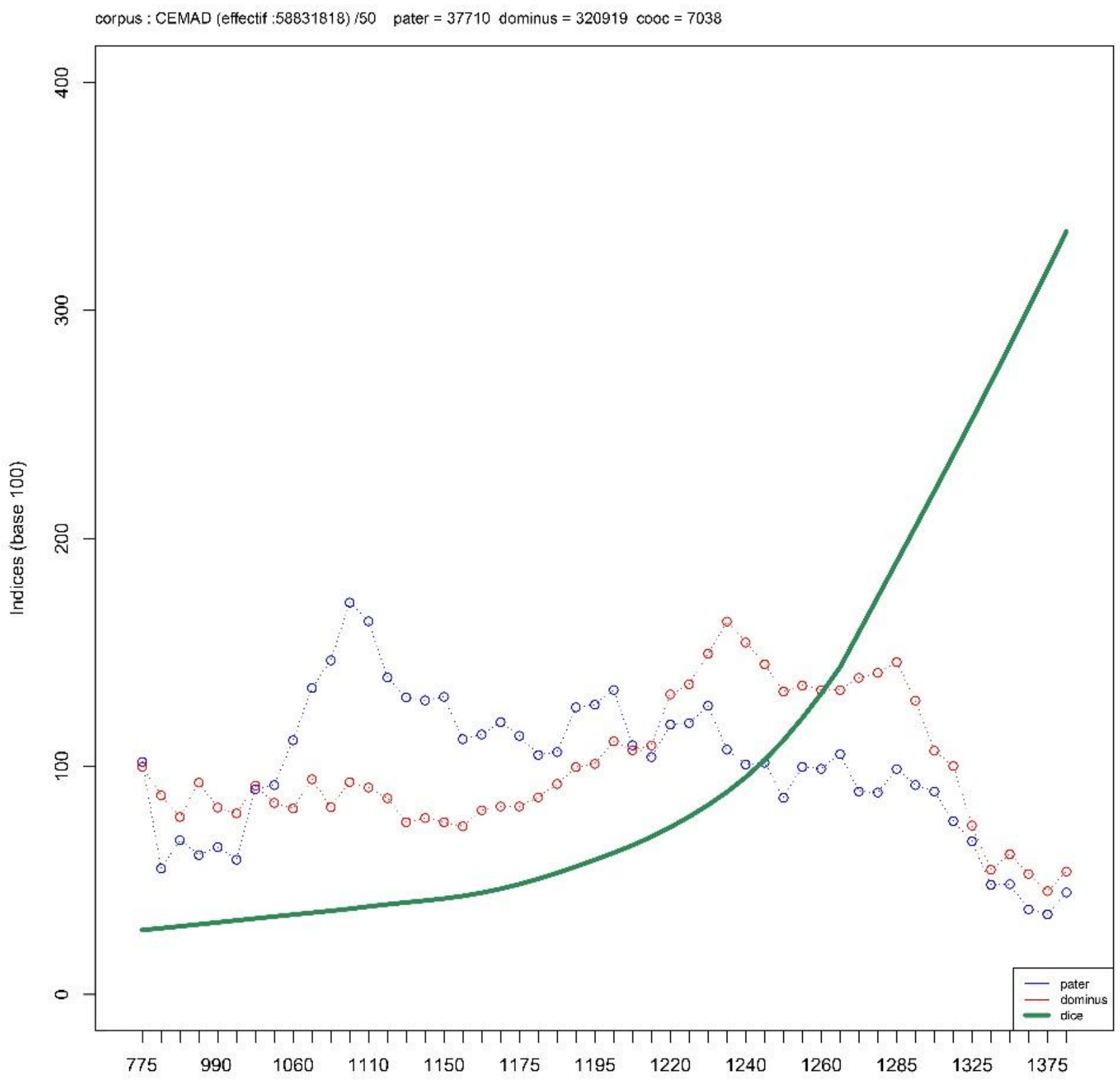

**Fig. 6 :** Évolution des cooccurrences des lemmes *pater* et *dominus*, dans les chartes des CEMA (VIII^e^-XIV^e^ siècles). En vert : mesure de l'association, selon le coefficient de Dice.

Ces observations permettent de proposer une hypothèse théorique : si les lemmes *dominus-servus* forment une paire fondamentale du système de représentation de l'Europe médiévale, et que *pater-filius* en constituent une autre[93], l'alliance de *pater* et *dominus* devrait entraîner l'existence de certains liens entre *filius* et de *servus*[94].

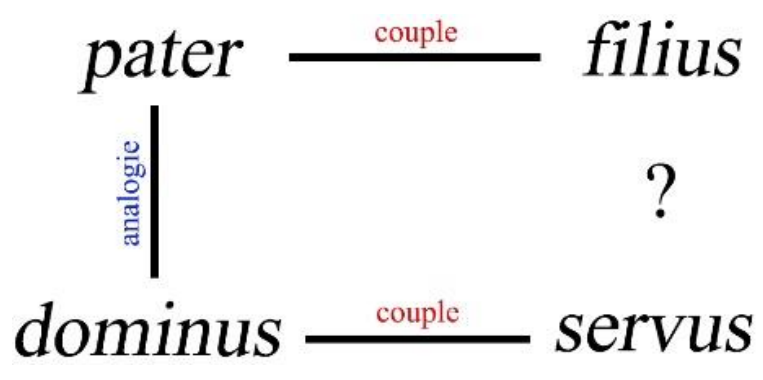

Or, la lecture des textes confirme en effet la présence de certains « passages » entre les termes[95]. Ainsi Thomas d'Aquin rapproche dans sa *Somme théologique* [1266-1273] la relation

[93] Cf. nos remarques en II.1.

[94] Certes, ces relations n'ont rien de mécanique et ne constituent pas une paire conceptuelle au même titre que *Deus-homo*, *dominus-homo*, *dominus-servus*, ou encore *spiritus-caro*. Il s'agit simplement ici de souligner que, dans certaines configurations, *filius* et *servus* se trouvent non pas sur un même plan conceptuel, mais articulés à des paires opposables qui sont elles même reliées par analogie. Il ne s'agit en aucun cas d'une relation d'égalité, mais plutôt d'une congruence conceptuelle. La stricte équivalence de *filius* et de *servus* aurait d'ailleurs posé différents problèmes conceptuels dans le système de représentation, le Christ (*filius* par excellence) ne pouvant être considéré comme *servus* (de la même façon qu'*homo* ne renvoyait pas automatiquement à *servus*).

[95] D'autres liens avaient été relevés autour d'*uxor*/*femina* : Eliana Magnani, « *Uxor* et *femina*. Enquête sur la désignation des femmes dans les documents diplomatiques bourguignons (IX^e^-XI^e^ siècle) », dans Jean-Paul Renard (dir.), *La place et le rôle des femmes dans l'histoire de Cluny : en hommage à Ermengarde de Blesle, mère de Guillaume le Pieux*, Brioude, Créer, 2013, p. 125-138 ; Id., « Le genre d'*Ego* ou les « stratégies de la

d'une femme (*uxor*) à son époux (*maritus*), de celle d'un serviteur (*servus*) à son seigneur (*dominus*), mais encore de celle d'un fils (*filius*) à son père (*pater*)[96]. Plus globalement, les lemmes *filius* et *servus* sont cooccurrents plus de 4 800 fois dans la PL[97], et la force de cette association quadruple quasiment entre les premiers et les derniers tomes du corpus (IIIe-début du XIIIe siècle).

Pour résumer, nous avons pu observer qu'une majorité des occurrences de *pater* renvoyaient à la paternité spirituelle ou divine. Ce fait ne peut pas être écarté au prétexte que ces mentions formeraient un ensemble « symbolique » ou « métaphorique », déconnecté des pratiques sociales. Les saints, papes, évêques, abbés et seigneurs étaient bel et bien qualifiés de *pater*, et cette désignation impliquait des relations pratiques : suppliques, soutiens, encadrements, mais aussi devoirs, ponctions, polarisations, etc. Derrière cette typologie bien différente de la nôtre, se trouvait l'association directe ou indirecte du titre de S/seigneur (*dominus*) et de celui de père (*pater*), redoublée dans certains cas par l'association discursive entre le serviteur (*servus*) et le fils (*filius*). Ce repérage rapide permet de faire remarquer que si la multi-paternité était la règle dans l'Europe médiévale (par le jeu même de la sémantique de *pater*, qui englobait une liste très longue, mais aussi variable, en fonction des configurations, de dominants), la « multi-seigneurialité » l'était tout autant[98]. Dans ces conditions, on peut faire l'hypothèse que l'ordre de la paternité reflétait l'ordre social au sens large.

Le système analogique médiévale, à la fois flexible mais aussi robuste, permettait ces circulations qui justifiaient la domination : le père était un créateur, un nourricier et un protecteur, il fallait suivre ses lois et, en bon fils, donner ou verser les rentes[99]. C'était encore

---

différence ». Esquisse de champ sémantique (IXe-XIe siècle) », dans Laurent Jégou, Sylvie Joye, Thomas Lienhard, Jens Schneider (dir.), *Splendor Reginae. Passions, genre et famille [...]*, op.cit., p. 179-195.

[96] « *Sicut uxor potest frangere fidem marito, ita etiam servus domino, et filius patri. Sed ad investigandam injuriam servi in dominum, vel filii in patrem, non est institutum in lege aliquod sacrificium.* », dans Thomas Aquinus, *Summa Theologica*, Rome, 1892, Quaestio 105, Articulus 4.

[97] C'est certes un peu moins que la cooccurrence de *pater* et *dominus*, qui se rencontrent quant à elle plus de 10 700 fois. Néanmoins, 1) les données sont comparables, 2) le système analogiste médiévale ne fonctionne pas dans la stricte équivalence (i.e. *dominus* = *pater*, donc *servus* = *filius*), mais plutôt dans les glissements de sens, autour des couples sémantiques fondamentaux. Le cas est ici d'autant plus typique que *filius* renvoie bien entendu fréquemment au Christ, lieu même Seigneur (*dominus*) par excellence.

[98] La multitude des droits exercés par différents seigneurs sur un même lieu, parfois sur une même parcelle, générait en effet un maillage de relations sociales, où s'imbriquaient dominants et dominés. À une échelle plus vaste, on dépendait toujours d'un évêque, du pape, mais encore d'un prince/roi/empereur, lui-même qualifié de *dominus* et donc potentiellement de *pater*. On peut arguer que cette situation, loin d'être un chaos désorganisé, renforçait la polarisation et donc l'efficacité du système médiéval, en concourant à la fixation spatio-temporelle des dominés (*servus*, *filius*). Les travaux sur cette perspective essentielle sont nombreux mais généralement disséminé sous la forme de remarques ponctuelles. Voir néanmoins : Alain Guerreau, « Fief, féodalité, féodalisme. Enjeux sociaux et réflexion historienne », *Annales ESC*, vol. 45:1, 1990, p. 137-166 ; id., « Seigneurie », dans André Vauchez (dir.), *Dictionnaire encyclopédique du Moyen Âge*, Paris, Éditions du Cerf, 1997, p. 1415-1416 ; id., « Féodalité », dans Jacques Le Goff, Jean-Claude Schmitt (dir.), *Dictionnaire raisonné de l'Occident médiéval*, op.cit., p. 387-406 ; Hélène Débax, *La féodalité languedocienne, XIe-XIIe siècles. Serments, hommages et fiefs dans le Languedoc des Trencavel*, Toulouse, Presses universitaires du Midi, 2003 ; id., *La seigneurie collective. Pairs, pariers, paratge : les coseigneurs du XIe au XIIIe siècle*, Rennes, Presses universitaires de Rennes, 2012 ; Joseph Morsel, *L'artistocratie médiévale*, Paris, Armand Collin, 2004, p. 176-178.

[99] Roland Viader, « La dîme dans l'Europe des féodalités. Rapport introductif », dans Roland Viader (dir.), *La dîme dans l'Europe médiévale et moderne*, Toulouse, Presses universitaires du Mirail, 2010, p. 7-36 ; Michel Lauwers (dir.), *La dîme, l'Église et la société féodale*, Turnhout, Brepols, 2012. Dans ce dernier volume, Valentina Tonneato (« Dime et construction de la communauté chrétienne, des Pères de l'Église aux Carolingiens (IVe-VIIIe siècle) », p. 65-86, ici p. 70) mentionne un passage fort intéressant des sermons de Césaire d'Arles : « *Quare non accipiam partem de substantia tua, qui tibi praemia praeparavit aeterna? Quare non accipiat decimum, qui contulit totum ? Contra terrenum patrimonium Deus offert caelum [...] tunc vobis respondebit pater, dominus et amicus, cum quo fecistis caeleste commercium*. », dans *Sermo* 31:5, éd. G. Morin et trad. M.-J. Delage, *Sermons au peuple*, tome 2, Paris, 1978, p. 138 (*Sources chrétiennes*, 243). La rhétorique classique de Césaire consiste à

plus clair dans le cas de l'Église, où quasiment tous les membres apparaissaient comme des seigneurs dans une configuration donnée, en particulier face aux laïcs. Cette tendance paraît d'ailleurs se renforcer, ainsi que le montre l'augmentation très nette de la fréquence des quelques 7 000 cooccurrences de *dominus* et de *pater* dans les chartes consultées, entre le VII^e^ et le XIII^e^ siècle.

### *II.4. Principales évolutions sémantiques de* **pater**

Le renforcement des liens entre *pater* et *dominus*, fondamental à nos yeux, n'est cependant pas isolé. Il s'inscrit dans une dynamique sémantique plus large, dont il s'agit désormais de brosser les grands traits. C'est comme souvent dans les chartes que nous avons pu relever les évolutions les plus nettes[100]. Elles concernent en particulier le développement des liens entre paternité et épiscopat, entre paternité et sainteté, et enfin une forme de « défamiliarisation » du champ paternel[101]. Un graphique divisant l'ensemble du corpus en 10 tranches chronologiques (P1-P10) permet de repérer les cooccurrents de *pater* évoluant le plus fortement[102]. Cette méthode a été employée, en complément d'analyse statistique sur des couples précis (par exemple le couple *pater-dominus* déjà évoqué) et d'un contrôle serré sur les actes eux-mêmes.

---

inviter à donner un dixième au Seigneur (en citant ensuite Math 25:34-37), qui est à la fois « *pater, dominus et amicus* », parce que celui-ci promet non seulement des récompenses célestes, mais aussi parce qu'il est le créateur de tous.

100 Dans la PL, le champ sémantique de *pater* est globalement plus stable. Cette divergence, que l'on n'observe pas pour tous les termes, pourrait s'expliquer par la prégnance du « père » biblique dans les textes théologiques.

101 Nous dérivons ici le terme « défamiliarisation » du concept de « déparentalisation du social », proposé dans Joseph Morsel, *L'Histoire (du Moyen Âge) est un sport de combat*, Paris, LaMOP, 2007, chap. V. On pourrait répondre à cela que ce phénomène est plutôt, a contrario, une « hyper-familiarisation », c'est-à-dire une extension de la logique parentélaire à l'ensemble de la société, à travers la spiritualisation de l'ensemble des relations sociales. Extension que l'on retrouve aussi bien dans *pater* que dans *frater*.

102 Il s'agit de la fonction CoocE2() de la bibliothèque Cooc, développée par Alain Guerreau. L'intérêt de diviser le corpus en dix parties contenant un nombre de mots équivalent (et non pas selon des périodes chronologiques d'égales durées) s'explique par la nature des distributions lexicales. Les mesures de cooccurrents nécessitant l'usage de coefficients, il s'agit en effet de raisonner sur des échantillons de taille équivalente, afin d'éviter que les caractères statistiques du lexique ne changent (ce qui fausserait irrémédiablement les comparaisons). Par ailleurs, nous n'avons retenu comme cooccurrents que les substantifs et les qualificatifs. Nous n'avons pas détecté d'évolution signifiante en ce qui concerne les verbes cooccurrents – ce qui est assez fréquent lorsqu'on retient un substantif comme pivot (ici *pater*). D'autres recherches seraient toutefois nécessaires pour confirmer cette hypothèse.

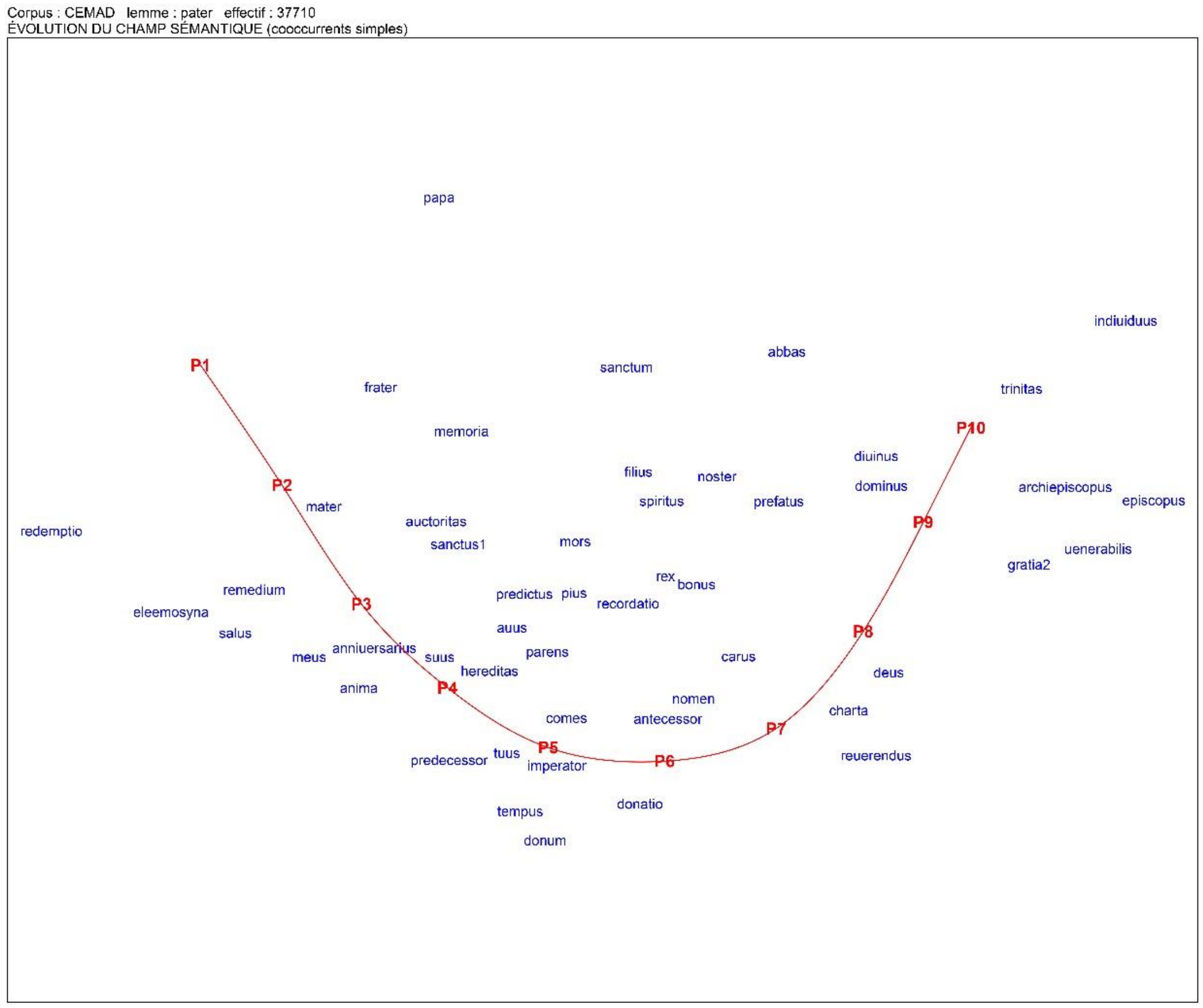

**Fig. 7 :** Évolutions des cooccurrents de *pater* dans les CEMA, divisés en 10 tranches chronologiques de taille équivalente (P1-P10, soit du haut Moyen Âge au XIVe siècle).

Parmi les occurrences les plus anciennes de *pater* dans les chartes, une part non négligeable concerne la parentèle charnelle. Elles apparaissent en particulier dans les mentions de dons pour le salut des âmes, fréquentes au VIIIe-XIe siècle[103]. Elles sont visibles sur l'analyse à gauche, à travers les cooccurrents *redemptio*, *eleemosyna*, *remedium*, *salus*, *anima* (fig. 7). Dans ces discours diplomatiques se révèlent aussi certains membres de la parentèle restreinte : *mater*, *frater*, *meus*[104]. Assez rapidement, d'autres énoncés prennent le pas sur les précédents : ils évoquent la *memoria* des *parens*, *antecessor*, *predecessor*, ces termes pouvant désigner à la fois des pères charnels et spirituels. Ces lemmes, plus propres aux Xe-XIe siècles, sont fréquents dans les diplômes (*imperator*, *rex*) et actes de l'aristocratie (*comes*). À la toute fin de la chronologie, nous constatons toutefois un glissement assez net : les cooccurrents spécifiques des XIIe-XIIIe siècles sont plutôt *dominus*, *abbas*, *divinus*, *trinitas*, *archiepiscopus*, *episcopus*, *venerabilis*, *gratia*, voire *deus*. Le champ sémantique de *pater* dans les textes diplomatiques glisse ainsi progressivement du domaine de la parenté charnelle à celui de la parenté spirituelle. Tout se passe comme si nous observions une forme d'extension du champ paternel – puisque l'évolution n'est pas une sortie de la parenté, mais plutôt une focalisation sur les formes non-charnelles de cette dernière[105].

---

[103] Eliana Magnani, « Transforming things and persons: The gift *pro anima* in the eleventh and twelfth centuries », art. cité ; id., « Du don aux églises au don pour le salut de l'âme en Occident (IVe-XIe siècle) : le paradigme eucharistique », art. cité.

[104] Mais aussi d'autres termes comme uxor, qui voient leurs cooccurrences avec *pater* chuter à partir du XIIe siècle – même s'ils n'apparaissent pas nécessairement sur le graphique de synthèse.

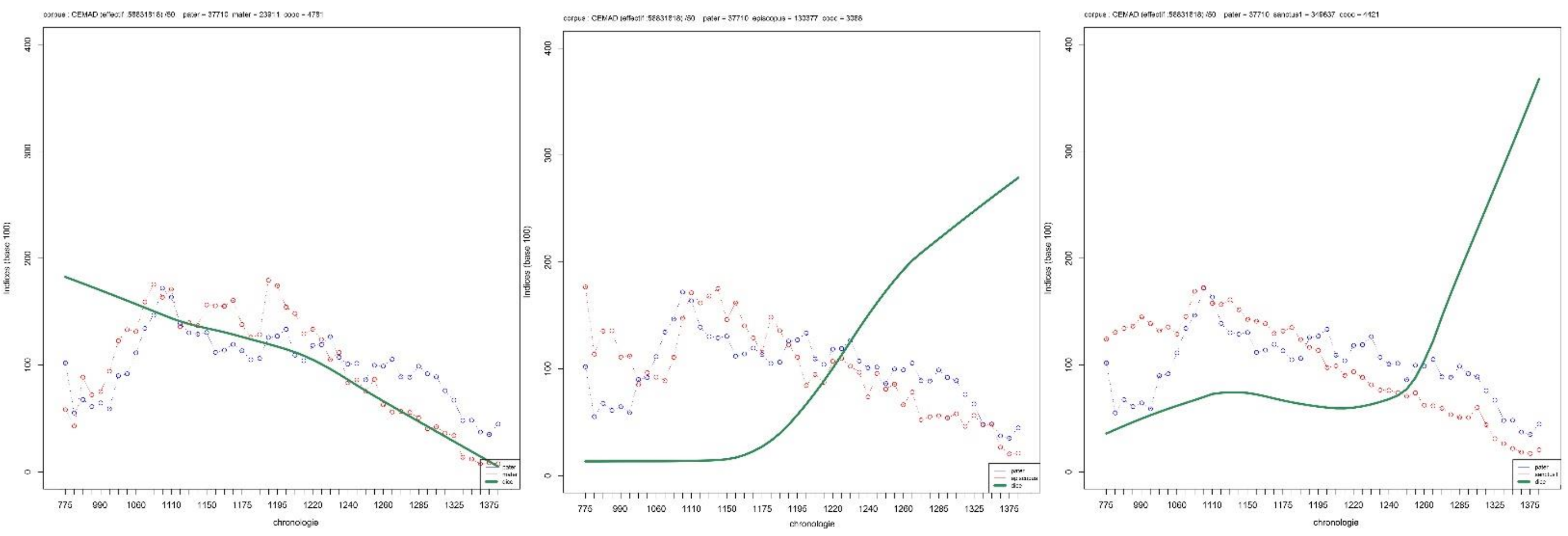

**Fig. 8a, b et c :** Évolution des cooccurrences des lemmes a) *pater* et *mater*, b) *pater* et *episcopus*, c) *pater* et *sanctus*, dans les chartes des CEMA (VIII^e^-XIV^e^ siècles). En vert : mesure de l'association, selon le coefficient de Dice.

Parmi les éléments les plus tranchés, l'affirmation de la paternalité épiscopale est nette[106]. Dans nos analyses, les associations entre *pater* et *episcopus* s'envolent en effet dans la seconde moitié du XII^e^ siècle, tout en continuant d'évoluer ensuite à la hausse (cf. fig. 8b)[107]. La tendance est très proche dans le cas de *pater* et *sanctus*, à ceci près qu'elle s'affirme d'abord lentement, dès les VIII^e^-IX^e^ siècles, avec un premier pic au XI^e^ siècle[108]. C'est pourtant dans la seconde moitié du XIII^e^ siècle que la cooccurrence des deux termes se développe fortement.

Parallèlement, se déroule une baisse très nette des associations entre *pater* et le qualificatif *omnipotens*[109]. Cette dynamique est certes liée à l'évolution de ce dernier terme lui-même, qui connaît une chute importante dans les textes diplomatiques entre le VII^e^ et le XI^e^ siècle. Mais nous pouvons sans doute aller plus loin et relier cette tendance à celles précédemment observées : la plus faible insistante sur la toute-puissance de la paternité divine, d'abord principalement contrebalancée par la paternité charnelle, semble en effet s'associer aux autres formes de paternités spirituelles précédemment évoquées[110]. Entre les pôles charnels et divins, se développent en effet les paternités spirituelles, non seulement baptismales mais aussi épiscopales, abbatiales et saintes. Cette tendance conduit à notre sens à l'accentuation de la multi-paternité, du moins à l'extension de la perception de cette dernière. Or, nous avons aussi constaté que dans la PL les associations entre *pater* et *celum* chutaient progressivement, tandis que celles entre *pater* et *mundus* augmentaient fortement[111]. Cette « descente du P/père sur terre », entre le Xᵉ et le XIIIᵉ siècle, est probablement à articuler avec la multiplication des *patres* intermédiaires entre le Seigneur céleste et le père charnel lors de cette période[112].

---

[105] Voir Joseph Morsel, *L'Histoire (du Moyen Âge) est un sport de combat*, op.cit., chap. V.

[106] Sur le développement de l'autorité épiscopale dans l'Occident médiéval, nous renvoyons à Geneviève Bührer-Thierry, *Évêques et pouvoir dans le royaume de Germanie. Les Églises de Bavière et de Souabe 876-973*, Paris, Picard, 1997 ; Steffen Patzold, *Episcopus. Wissen über Bischöfe im Frankenreich des späten 8. bis frühen 10. Jahrhunderts*, Ostfildern, Thorbecke, 2008 ; Laurent Jégou, *L'évêque, juge de paix : l'autorité épiscopale et le règlement des conflits entre Loire et Elbe (milieu VIII^e^-milieu XI^e^ siècle)*, Turnhout, Brepols, 2011.

[107] 3 088 cooccurrences ont été détectées pour cette association, seulement dans les actes datés des CEMA (les actes non-datés ont été exclus pour réaliser la courbe d'évolution chronologique).

[108] Si d'autres investigations sont nécessaires, le développement du thème de la paternité des saints, dans les chartes, ne nous paraît pas anodin. Il pourrait être à mettre en lien avec l'encellulement des X^e^-XI^e^ siècle, et avec la fixation des communautés observable lors de cette période – le saint étant bien entendu l'opérateur essentiel de la polarisation ecclésiale. Nous renvoyons ici à Robert Fossier, *Enfance de l'Europe, X^e^-XII^e^ siècles : aspects économiques et sociaux*, 2 volumes, Paris, Presses universitaires de France, 1982 ; Alain Guerreau, « Structure et évolution des représentations de l'espace dans le haut Moyen Âge occidental », dans *Uomo e spazio nell'alto medioevo, Spolète, Presso la sede del centro*, 2003, p. 91-115 (*Settimane di Studio*, 50).

[109] Cette dynamique est aussi sensible dans la PL.

Dans cette dynamique, il convient toutefois de remarquer que nous n'avons pas observé d'évolution des cooccurrences entre *pater* et *filia*, qui sont d'ailleurs extrêmement rares tout au long de la chronologie. On pourrait certes arguer que cette tendance est liée à l'omniprésence des discours sur la parenté divine, et donc la relation de Dieu et du Christ. Mais la même logique semble à l'œuvre dans les chartes, si bien que l'on pourrait dire que s'affirmer être le « père » d'une fille demeure chose assez rare dans l'Europe médiévale latine.

En définitive, nous avons ainsi constaté une tendance de *pater* à passer, dans les chartes, du domaine de la parenté charnelle à celui de la parenté spirituelle[113] : autrement dit, une forme de spiritualisation des figures paternelles[114]. Parallèlement, il semble que la multi-paternité s'affirme, même si le terme *pater* reste privilégié pour décrire les relations père-fils (et/ou dominants-dominés).

## III. Qu'est-ce que la *paternitas* ? Origine et diffusion d'un concept chrétien

### *III.1. Dans les premiers siècles de l'Église*

Ces considérations nous amènent à nous interroger sur les termes désignant la « paternité » médiévale, en tant que relation à proprement parler : comment les *personnes* se définissaient-elles face à ces pères multiples ? Les historiens usent en effet du concept de « paternité » afin d'évoquer les liens entre père et fils à toutes les époques, agissant comme s'il s'agissait d'un concept universel. Tout comme *pater*, le lemme *paternitas* possède pourtant une histoire et un sens, qui n'ont à notre connaissance jamais été étudiés. Or, une première recherche montre que ce dernier est totalement absent du corpus latin païen : il s'agit d'un cas assez peu fréquent et qui doit être souligné[115]. Le terme apparaît en effet dans le *Nouveau Testament*, avec une unique occurrence dans l'Épître aux Éphésiens [seconde moitié du Ier siècle] : « *ex quo omnis paternitas in caelis et in terra nominatur* » (Ep. 3:15) [116]. Cette mention est importante pour l'analyse, car elle insiste sur le caractère divin de la *paternitas*[117]. Dieu est considéré comme le Père de toutes créatures, au ciel comme sur terre[118].

[110] Dans la PL, la principale tandence nous paraît être un glissement sémantique depuis *pater* comme membre céleste de la Trinité (*omnipotens*, *regnum*, *persona*, *celum*, *eternus*, *gloria*, *eternus*), dont les théologiens discutent la *natura* (*substantia*), à un *pater* plus incarné, terrestre (*mundus*) et dominant (*dominus*).
[111] Cf. la note précédente.
[112] Dans le corpus des romans du XIXᵉ siècle examiné, la proportion des mentions de « pères | Pères » (au pluriel donc) représente seulement 3,33% de la totalité du lemme « père ». Elle est beaucoup plus élevée dans les corpus médiévaux (*patres*, *patrum*, *patribus*) : 8,17% dans les CEMA (alors que les chartes sont souvent considérées comme des documents « pragmatiques ») et presque 25% dans les textes théologiques ou narratifs (22,62% dans la PL, 24,1 dans les OpenMGH).
[113] Cela va sans dire, sans que les logiques familiales et de filiation d'effacent totalement dans la paternité. Elles s'intègrent simplement de plus en plus nettement dans un ensemble plus global de relations sociales.
[114] Nous avions constaté une évolution comparable en étudiant la sémantique de *memoria*. Dans son important livre, Jérôme Baschet relevait déjà une « amplification de la parenté spirituelle » partant de la paternité : id., *Le sein du père [...]*, op.cit., chapitre VIII, partie III : « La hiérarchie des pères ». Sur ce mouvement, voir l'article d'Anita Guerreau-Jalabert, « L'Arbre de Jessé et l'ordre chrétien de la parenté », dans Dominique Iogna-Prat, Éric Palazzo, Daniel Russo (dir.), *Marie. Le culte de la Vierge dans la société médiévale,* Paris, Beauchesne, 1996, p. 137-170.

Or, cette formule biblique va profondément marquer les auteurs chrétiens. Sur les 110 occurrences de *paternitas* dans la PL pour la période allant de Tertullien à Boèce (ce dernier exclus), 35 renvoient explicitement au verset des Éphésiens. La première est attribuée au pape Marcelin [296-304 pour son pontificat], qui l'emploierait pour la première fois dans une lettre datée de 299, mais il s'agit en fait d'un texte issu du corpus pseudo-Isidorien [830-850][119]. On trouve cependant, pour le IVe siècle, des occurrences chez Marius Victorinus[120], Hilaire de Poitiers[121], Eusèbe de Verceil[122], le pseudo-Ambroise[123] ou encore Palladius de Ratiaria[124] – probablement en lien avec l'arianisme et plus largement avec la définition du dogme trinitaire[125], donc la relation entre *pater* et *filius*. Nous pensons toutefois que ce sont Jérôme et Augustin qui contribuent définitivement à sceller l'importance du verset, avec respectivement 17 et 18 occurrences chez les deux auteurs[126].

---

[115] Une liste des inventions latines des Pères, dans leurs textes ou dans la Vulgate, serait sans doute très instructive. Nous travaillons à une contribution dans ce sens.

[116] L'Épître était autrefois généralement attribuée à saint Paul [† c. 67-68], mais les chercheurs l'attribuent aujourd'hui plus régulièrement à l'un de ses disciples, probablement dans les années 80 du Ier siècle après. Concernant ce texte et sa critique, voir en premier lieu : Martin Kitchen, *Ephesians*, London-New York, Routledge, 1994 (en particulier p. 4-7 pour l'attribution) ; Peter T. O'Brien, *The Letter to the Ephesians*, Grand Rapids, Eerdmans, 1999 (chapitre I pour les questions d'attribution) ; Bart D. Ehrman, *The New Testament: A Historical Introduction to the Early Christian Writings*, New York, Oxford, 2000 (2nd édition), p. 350-354 ; Donald A. Hagner, *The New Testament: A Historical and Theological Introduction*, Grand Rapids, Baker Academic, 2012, p. 585-604 ; Pierre Boucaud, « *Corpus Paulinum*. L'exégèse grecque et latine des Épîtres au premier millénaire », *Revue de l'histoire des religions*, vol. 3, 2013, p. 299-332. Pour une mise au point actuelle et globale sur le *Nouveau Testament*, voir Hugh Houghton, *The Latin New Testament. A Guide to its Early History, Texts, and Manuscripts*, Oxford, Oxford University Press, 2016.

[117] Le terme grec est « *πατριὰ* ». Une littérature importante existe sur la traduction de ce terme, qui est souvent rendu par « famille/*family* » dans les travaux contemporains (par exemple dans Charles John Ellicott, *St. Paul's Epistle to the Ephesians: with a critical and grammatical commentary, and a revised translation*, Andover, W.F. Draper, 1863, p. 38 ; Frederick Fyvie Bruce, *The Epistles to the Colossians, to Philemon, and to the Ephesians*, Grand Rapids, Eerdemans, 1984, p. 324-325, qui critique à l'inverse ce choix de traduction ; le Bailly propose quant à lui « descendance, lignée, particulièrement du côté du père », puis « race, famille » et enfin « tribu, caste »). Ce sens « familial » ne nous retient pas vraiment ici, dans la mesure où il n'a jamais été envisagé par les auteurs médiévaux. Tout au contraire, avant même la Vulgate, *πατριὰ* est rendu en latin par *paternitas* : c'est ce que confirme à la fois la consultation du corpus des *Vetus Latina* (Hermann Josef Frede, *Epistula ad Ephesios*, Freiburg, Herder, 1962-1964 (*Vetus Latina*, 24:1)) et l'analyse directe des mentions (PL et CLCLT – ce dernier corpus ayant été employé systématiquement pour contrôler les dernières éditions). Toutes les mentions latines du verset antérieures à Jérôme contiennent en effet déjà le lemme. La chose n'est d'ailleurs pas totalement étonnante, dans la mesure où les anciennes traductions latines des Épîtres furent très largement reprises dans la Vulgate. Elle est toutefois significative, car *πατριὰ* renvoyait probablement à un sens plus restreint que *paternitas* (cf. Markus Barth, *Ephesians. Introduction, Translation, and Commentary on Chapters 1-3*, New York, Garden City, 1974, p. 382). La traduction latine pourrait donc correspondre à une extension de la paternité divine, présentée comme universelle. Sur ces mentions antérieures à la Vulgate, voir les développements qui suivent. Nous remercions vivement Marie Frey Rébeillé-Borgell pour ses conseils en matière de Bible ancienne.

[118] Pour Frederick Fyvie Bruce, Dieu est l'« archétype » du père, parce que la paternité que toutes les « paternités dérivent de la sienne » (*The Epistles to the Colossians, to Philemon, and to the Ephesians*, op.cit., p. 324).

[119] « *Hujus rei gratia, ut ait Apostolus, flecto genua mea ad Patrem Domini nostri Jesu Christi, ex quo omnis paternitas in celis et in terra nominatur.* », dans Marcellinus Papa, *Ad orientales episcopos*, PL 7, col. 1085-1092c, ici col. 1092b. On pourra plutôt se reporter à *Decretales pseudo-Isidorianae et capitula Angilramni*, éd. par Paul Hinschius, Leipzig, B. Tauchnitz, 1863, p. 221-223 (aussi dans PL 130, col. 7-1177, ici col. 218b). Différentes nouvelles éditions des fameuses *Fausses décrétales*, dont celle des MGH, toujours en préparation, confirment la présence de ce texte dans une forme proche au sein de la collection. Pour une introduction à ce corpus, voir Horst Fuhrmann, « The Pseudo-Isidorian Forgeries », dans Wilfried Hartmann, Kenneth Pennington (éd.), *Papal Letters in the Early Middle Ages. History of Medieval Canon Law*, Washington, Catholic University of America Press, 2001, p. 135–195 ; Clara Harder, *Pseudoisidor und das Papsttum: Funktion und Bedeutung des apostolischen*, Köln-Weimar, Bölhau, 2014 ; Eric Knibbs, « Ebo of Reims, Pseudo-Isidore and the Date of the False Decretals », *Speculum*, vol. 92, 2017, p. 144–183.

Le lemme s'installe ainsi progressivement dans le latin des Pères puis au haut Moyen Âge. Au-delà de la missive faussement attribuée à Marcellin, les papes jouent probablement un rôle important dans sa diffusion, en particulier à travers les lettres pontificales[127]. On constate en effet la présence du lemme dans une missive adressée au pape Libère [352-366], puis dans des lettres de l'antipape Félix II [355-365], d'Innocent Ier [401-417] et de Zosime [417-418][128].

---

[120] « *Paulus ad Ephesios: huius rei gratia flecto genua mea ad patrem domini nostri Iesu Christi, ex quo omnis paternitas in caelis et in terra nominatur [...]* », dans Marius Victorinus, *Adversus Arium*, PL 8, col. 1039-1138b, ici col. 1040d ; ou encore dans id., *In epistolam Pauli ad Ephesios*, PL 8, col. 1235-1294d, ici col. 1268b.

[121] « *Credimus in unum Deum patrem omnipotentem, institutorem et creatorem omnium, ex quo omnis creatura paternitas in coelo et in terra nominatur.* », dans Hilarius Pictaviensis, *Liber de synodis seu fide orientalium*, PL 10, col. 479-546b, ici col. 506c et encore col. 509b.

[122] « *Ab ipso ergo est omnis paternitas in coelo, et in terra. Filium quoque de substantia Patris sine initio ante saecula natum, nec tamen factum esse fatemur. Quia nec Pater sine Filio, nec Filius aliquando exstitit sine Patre, et tamen non sicut Filius de Patre, ita Pater de Filio, quia non Pater a Filio, sed Filius a Patre generationem accepit. Filius ergo Deus de Patre. Pater autem Deus, sed non de Filio. Pater quidem Filii, non Deus de Filio. Ille autem Filius Patris, et Deus de Patre, aequalis tamen per omnia Filius Deo Patri, quia non esse coepit aliquando, nec desiit . . . Filius unius cum Patre substantiae creditur, sic quod et homousion Patri dicitur, hoc est eiusdem cum Patre substantiae: OMO enim Graece unum, USIA vero substantia dicitur, quod utrumque coniunctum sonat unam substantiam. Non enim de nihilo, neque de alia substantia, sed de Patris utero, id est, de Substantia eiusdem Filius genitus, vel naturae credendus est.* », dans Eusebius Vercellensis, *De Trinitate confessio*, PL 12, col. 959a-968b, ici col. 959a-960a.

[123] Ambrosiaster, *Commentaria in Epistolam ad Corinthios Secundam*, PL 17, col. 275-338b, ici col. 297b ; ID., *Commentaria in Epistolam ad Ephesios*, PL 17, col. 371-404a, ici col. 383d ; id., *Commentaria in Epistolam ad Romanos*, PL 17, col. 45-184a, ici col. 52a.

[124] « *Talem namque ac tantum pro sua singulari omnipotentia pater genuit filium qui in ostensione genitricis potentiae posset cuncta propria craeare virtute sicuti et apostolus nunc differentiae memor servato gradu unum auctorem et unum opificem rettulit cunctorum dicens unus deus pater ex quo omnia et nos in ipso et unus dominus Ihesus cristus per quem omnia et nos per ipsum nunc de solius patris prestantia unus deus et pater omnium qui super omnia et per omnes et in omnibus nobis item ut deus domini nostri ihesu cristi pater gloriae item deus et pater domini nostri Ihesu cristi item genua flecto ad patrem domini nostri ihesu cristi ex quo omnis paternitas in caelis et in terris nominatur quomodo uos etiam tres sempiternos dixistis cum de uno scribtum sit sempiterna quoque eius uirtus et diuinitas.* », dans Palladius Ratiarensis, *Fragment d'une apologie des condamnées d'Aquilée*, éd. par Roger Gryson, *Scolies ariennes sur le concile d'Aquilée*, Paris, Le Cerf, 1980, p. 314 (*Sources chrétiennes*, 267). Le verset Eph 3:15 ne pouvait pas, bien entendu, ne pas attirer l'attention des Ariens – tant il permettait pour eux d'appuyer la supériorité du Père sur le Christ.

[125] Voir la note précédente.

[126] Jérôme utilise ainsi 15 fois le lemme dans son seul commentaire sur l'Épître aux Éphésiens, ce qui montre à quel point le verset l'a retenu (Hieronymus Stridonensis, *Commentaria in Epistolam ad Ephesios*, PL 26, col. 15-218d). Les deux autres occurrences de l'auteur sont aussi des mentions / commentaires du verset (id., *Commentaria in Naum*, PL 25, col. 1231-1272d, ici col. 1251a ; id., *Epistola Pauli ad Ephesios*, PL 29, col. 777-784c, ici col. 780d). Augustin procède en revanche à une utilisation plus pratique du terme, puisque 7 occurrences de *paternitas* sont chez lui issues de lettres – même s'il est très probable qu'il existe ici un biais documentaire (Augustinus Hipponensis, *Epistolae*, PL 33, col. 61-1094, ici col. 451, 565, 761, 975, 1012, 1069). La différence entre les deux auteurs est d'autant plus nette que si Jérôme commente le verset Ep. 3:15 dans son ensemble, Augustin va plus loin et réemploie le terme dans d'autres contextes.

[127] Peut-être en imitation des lettres attribuées à Augustin ? Voir les notes qui précèdent.

[128] L'analyse de la tradition de ces textes est redoutable, y compris pour les spécialistes – dont nous ne sommes évidemment pas. Malgré ces difficultés « d'archéologie textuelle », nous pensons que ces mentions sont révélatrices d'une première tendance. L'occurrence pour Libère concerne encore la controverse Arienne : « *Vestrae beatissimae paternitatis iura penes Deum sunt manifesta, dum praedicationibus scilicet apostolicis et*

Malheureusement, il est souvent bien difficile de déterminer l'authenticité de ces documents, même s'ils concernent pour une bonne part la controverse sur l'Arianisme – et doivent être considérés d'une part dans l'immense effort documentaire produit pour y répondre, d'autre part au prisme des enjeux christologiques de la querelle[129]. Cette réflexion sur la *paternitas*, dans les textes des IVe et Ve siècles est cependant tout sauf anodine, dans les lettres mais aussi dans les commentaires bibliques : elle annonce le développement du thème de la paternité divine puis terrestre dans l'Europe des VIe-XIIIe siècles. Dans tous les cas, nous pouvons dire que la *paternitas* constitue une invention proprement patristique et médiévale, c'est-à-dire profondément ecclésiale – parce qu'elle est à proprement parler à situer dans l'Église et la production de ses discours.

### *III.2. Dynamique de* **paternitas** *chez les auteurs médiévaux*

Alors que le terme est inconnu en latin ancien, il se développe donc très fortement entre le IIIe et le XIIIe siècle, s'étendant progressivement hors des lettres, vers l'exégèse, la théologie en général, puis tous les autres genres textuels : conciles, chartes, capitulaires, sermons, encyclopédies, etc. Il est certes moins fréquent que *pater*, avec respectivement environ 2 000 et 1 700 occurrences dans la PL et les CEMA, mais son développement est encore une fois proprement médiéval. Des auteurs majeurs comme Isidore de Séville, Bède le Vénérable, Boniface de Mayence, Alcuin, Raban Maur, Haymon d'Auxerre, Smaragde de Saint-Mihiel, Loup de Ferrières, Paschase Radbert, Jean Scot Erigène, Abbon de Fleury, Anselme de Cantorbéry, Yves de Chartres, Pierre le Vénérable, Adam Scot, Alain de L'Isle, Bernard de Clairvaux, ou encore Richard de Saint-Victor, l'emploient abondamment[130]. Son usage semble cependant réservé à des contextes particuliers, qui marquent son importance – nous y reviendrons. Encore rare dans les textes des Pères, on note que *paternitas* se développe tout au long de la chronologie considérée, avec deux accélérations : dans la seconde moitié du IXe siècle puis au XIIe et surtout XIIIe siècle (fig. 9). Dans ce dernier cas, nous pouvons imaginer un rôle de l'évolution de la typologie documentaire au sein de la PL, avec le développement

---

*doctrinis, verae fidei cultura, universa repleta sit terra.* », dans Liberius Papa, *Epistolae,* PL 8, col. 1395-1408c, ici col. 1403c. Migne tient son édition de *In quatuor tomos conciliorum omnium, tum generalium, tum provincialium atque particularium [...]*, éd. par Lorenz Sauer, tome 1, Cologne, Gerwinum Calenium et haeredes Johannis Quentelii, 1567, p. 438. Il semble toutefois que cette occurrence proviennent du corpus pseudo-Isidorien, et nous ne pouvons donc pas nous prononcer sur son authenticité (*op.cit.*, PL 130, col. 633d ; sur le corpus pseudo-Isidorien, voir nos remarques précédentes). Concernant Félix II, on trouve cette mention : « *Ad salutationem ergo vestrae sanctae et honorandae paternitatis vice nostra [...]* », dans Felix II, *Epistolae et decreta*, PL 13, col. 11-28d, ici col. 16a-16b ; *In quatuor tomos conciliorum omnium [...]*, op.cit., p. 443 ; *Divi Clementis opera*, éd. par Johannes Sichardt, Paris, Jean Roigny, 1544, fol. 124r. Ces éditeurs n'indiquent pas si la lettre est fausse, mais on la retrouve à nouveau dans le corpus pseudo-Isidorien (*op.cit.*, col. 643a). Concernant la lettre de Zosime (datée de 417), la tradition éditoriale hésite toutefois entre *paternitas* (version donnée par *Annales ecclesiastici*, éd. par Cesare Baronio, vol. 7, 417:20 – p. 90 de la réédition de 1866) et *fraternitas* – ce qui est intéressant en soi : « *tamen ne fraternitatis/paternitatis vestrae de adventu ac discussione praedicti diutius penderet exspectatio [...]* ». Outre les *Annales ecclesiastici*, la lettre est éditée dans PL 20, col. 649-654, ici col. 650a (qui mentionne les différentes versions), ainsi que dans *Epistulae imperatorum pontificum*, éd. par Otto Guenther, vol. 1, Vindobonae, F. Tempsky, 1895, p. 99 (lettre n° 45).

[129] Voir la note précédente.

[130] C'est particulièrement le cas dans leurs lettres. On dénombre par exemple 34 mentions chez Boniface de Mayence : Bonifacius Moguntinus, *Epistolae*, PL 89, col. 687-804b ; 21 dans les lettres d'Alcuin : Alcuinus, *Epistolae*, PL 100, col. 139-512b ; 18 dans celles d'Anselme de Cantorbéry : Anselmus Cantuariensis, *Epistolae*, PL 158, col. 1059-1208a ; ou encore 16 chez Bernard de Clairvaux : Bernardus Claraevallensis, *Epistolae*, PL 182, col. 1191-1196. Outre l'exégèse du verset Eph 3:15, le terme revient fréquemment dans les traités théologiques : sept fois par exemple dans le *De tripartito tabernaculo* d'Adam Scott (PL 198, col. 609-792c) ou encore 10 fois dans *Theologicae regulae* d'Alain de l'Isle (PL 210, col. 617-684c).

des lettres carolingiennes et les registres d’Innocent III[131]. Mais cela n’explique pas nécessairement l’ensemble du mouvement. Une hypothèse de travail serait que la diffusion du thème est liée plus généralement au renforcement de la hiérarchisation ecclésiale et à l’accroissement de la domination épiscopale – puisque tous les plus hauts personnages de l’Église se voient graduellement attribuer la *paternitas*[132].

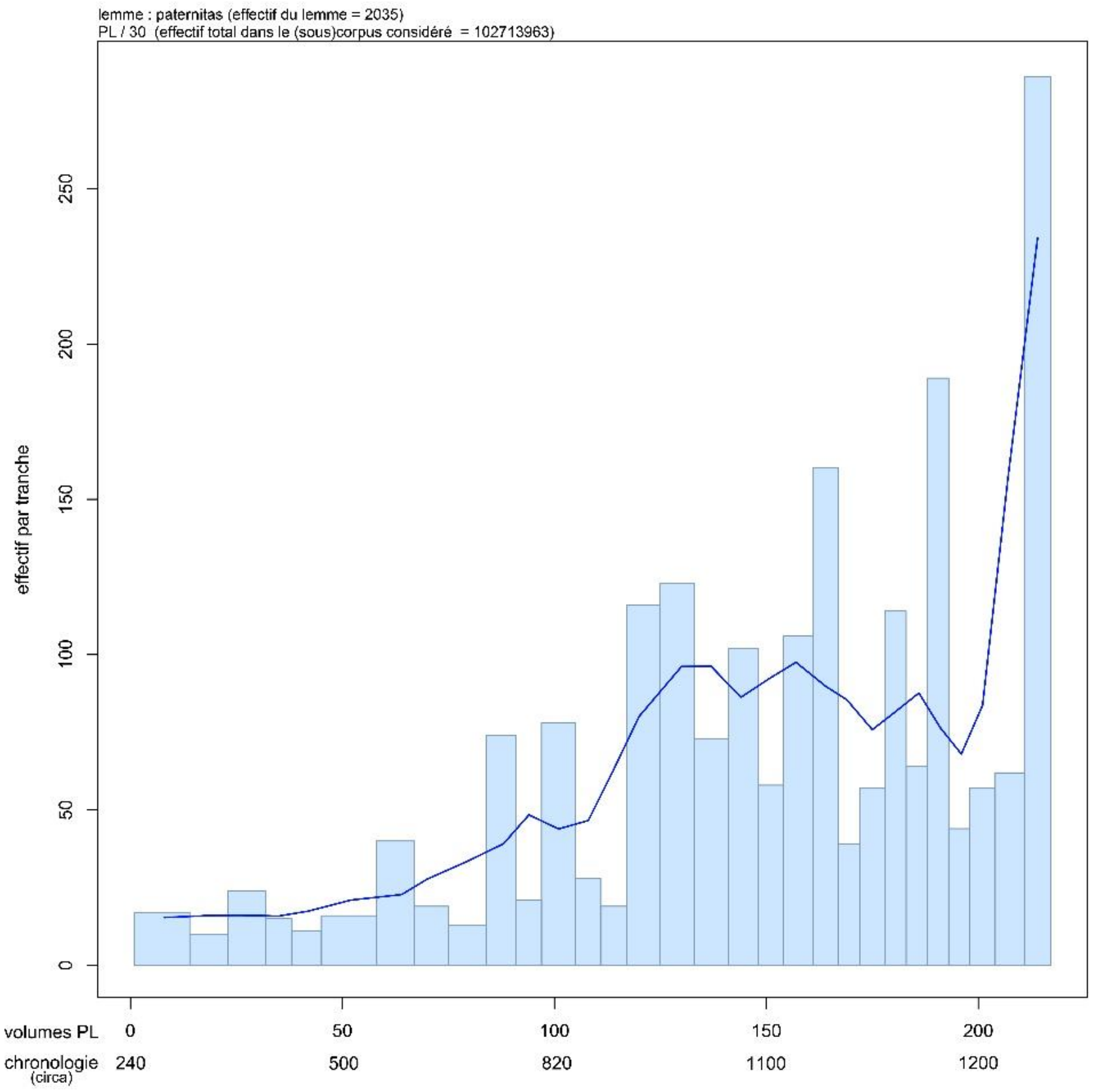

**Fig. 9 :** Évolution des occurrences de *paternitas* dans la PL (IIIe-XIIIe siècle).

Cette situation contraste fortement avec la grande rareté des mentions de *maternitas*[133], même si elle-même est saisie et conceptualisée à travers la christologie, et donc la parenté spirituelle. Nous n’en dénombrons pas plus d’une vingtaine dans la PL et seulement 5 dans les chartes des CEMA[134]. Les occurrences des chartes sont d’ailleurs toutes issues d’un même

---

[131] PL 214-217.

[132] Voir les paragraphes qui suivent.

[133] Ce qui ne signifie évidemment pas que le thème de la « maternité médiévale » est infructueux, bien au contraire – simplement que *maternitas* a un sens qui ne se confond pas avec notre « maternité », et plus encore que le couple *maternitas*/*paternitas* n’a qu’un rapport latéral avec le couple contemporain maternité/paternité. Sur cette question, nous renvoyons à Dominique Iogna-Prat, Éric Palazzo, Daniel Russo (dir.), *Marie. Le culte de la Vierge dans la société médiévale*, op.cit. ; pour une compréhension extensive, Sylvie Barnay, *La Vierge. Femme au visage divin*, Paris, Gallimard, 2000 ; Eléonore Fournié, Séverine Lepape (dir.), *L’Immaculée Conception de la Vierge : histoire et représentations figurées du Moyen Âge à la Contre-Réforme*, numéro spécial de *L’Atelier du Centre de Recherches Historiques*, n° 10, 2012 ; Anita Guerreau-Jalabert, « Maternité », dans Frédéric Gabriel, Dominique Iogna-Prat, Alain Rauwel (dir.), *Dictionnaire critique de l’Église*, op.cit.

[134] Une exploration plus attentive montre que ce nombre devrait être encore plus réduit pour la PL. Les bases de données font par exemple apparaître une mention chez Ildefonse de Tolède [† 667], dans son traité sur la Vierge (*De perpetua virginitate S. Mariae*, PL 96, col. 53-110a, ici col. 64b). Or, un retour au texte édité montre que le

corpus et d'une même édition : celle des privilèges et lettres du pape Calixte II pour l'abbaye Saint-Philibert de Tournus, d'ailleurs reprises par Migne[135]. Dans la PL, la plus ancienne occurrence du lemme provient du fameux commentaire du Pseudo-Denys l'Aréopagite par Jean Scott Érigène [† c. 876], dans lequel il oppose *paternitas* et *maternitas*[136]. Il n'est d'ailleurs pas anodin que le terme apparaisse pour la première fois dans la traduction d'un texte grec[137].

Par la suite, les mentions n'arrivent pas avant la fin du XI^e^ et le début du XII^e^ siècle, avec Jean de Wurzbourg, Guibert de Nogent, Geoffroi de Vendôme[138] et surtout le pape Calixte II, déjà évoqué. Ces rares mentions concernent soit la filiation ecclésiale dans le droit fil de la filiation du Fils qui la fonde, soit la « maternité » de la Vierge, toujours dans le fil de la filiation du Fils, les deux fils pouvant (et devant) se croiser sans cesse[139]. C'est à partir des années 1140-1160, puis vers les années 1200 et de manière décisive à partir des années 1230, que le thème de la *maternitas* se propage, probablement en lien avec le développement du culte marial et d'une réflexion conséquente sur la maternité, virginale ou non[140]. Au XIII^e^ siècle toujours, le *Corpus Thomisticum* renvoie ainsi 15 occurrences du lemme, et une enquête complémentaire dans le CLCLT donne cette fois un total plus important de 143 mentions – dont près de 95% sont postérieures à 1200[141]. Reste que le contraste est saisissant : il y a près de 200 fois moins de mentions de *maternitas* que de *paternitas* dans les CEMA et la PL cumulés. Cette

---

lemme apparaît en fait dans un titre, sans aucun doute ajouté par un transcripteur plus tardif, médiéval ou contemporain : « *Caput III. Sanctam Mariam ex natione et stirpe Iudaeorum esse, ex fide autem et honore, laude et amore, Christianorum. Tandem aliquot prophetarum oraculis probat virginitatem summam oportere cum maternitatis gloria copulari* ». Ce genre de pièges plaide encore une fois pour l'articulation permanente du qualitatif et du quantitatif.

[135] *Nouvelle Histoire de l'abbaye royale et collégiale de Saint-Filibert et de la ville de Tournus*, éd. par Pierre Juenin, Dijon, A. de Fay, 1733, p. 142-145 ; Calixtus II, *Epistolae et privilegia*, PL 163, col. 1093-1338a, ici col. 1234d, 1235b, 1236a, 1236d, 1237b et 1313d.

[136] « *Et hoc non est divina intelligere volentium proprium, sed sonos leves accipientium, et hos usque aures transeuntes extrinsecus continentium, et nolentium scire, quid quidem qualis dictio significet, quomodove eam oporteat et per alias aequepotentes et manifestiores dictiones declarare, patientiumque elementis, et literis non intellectis, et syllabis et dictionibus incognitis et non ingredientibus in animae suae intellectuale, sed foris circa labia et auditus suos percrepitantibus, ac si non liceat, quattuor numerum per bis duo significare, aut simplam lineam per rectam lineam, aut maternitatem per paternitatem, aut aliis quibusdam multis orationis partibus idipsum significantibus.* », dans Ioannes Scotus Erigena, *Versio operum S. Dionysii*, PL 122, col. 1023-1194c, ici col. 1135a ; voir l'édition récente du *Commentaire de la Hiérarchie céleste du Pseudo-Denys (865-870),* éd. par Jeanne Barbet, Turnhout, Brepols, 1975 (*Corpus Christianorum, Continuatio mediaevalis*, 31).

[137] Nous rejoignons donc ici les remarques de Sylvie Joye : « Ce rapprochement de Marie avec la promotion de la généalogie carolingienne reste cependant une exception, même si elle a un grand succès et cette thématique mariale ne se diffuse pas communément au sein du modèle familial, les Carolingiens et leurs évêques étant plutôt de grands défenseurs de la figure paternelle par ailleurs, symbole de la stabilité sociale. », dans *ibid., L'autorité paternelle en Occident [...],* op.cit., p. 188.

[138] Ioannes Wirziburgensis, *Descriptio Terrae Sanctae*, PL 155, col. 1053-1090c, ici col. 1078a ; Guibertus S. Mariae de Novigento, *De laude S. Mariae*, PL 156, col. 537-577, ici col. 548a ; Goffridus Vindocinensis, *Epistolae*, PL 157, col. 33-212c, ici col. 39a.

[139] « *Benedicta maternitas tua, benedicta verba tua, benedicta pudicitia tua, benedicta humilitas tua, benedicta plenitudo tua, benedicta integritas tua.* », dans *Tractatus ad laudem gloriosae Virginis Matris,* PL 182, col. 1133-1148c (*Opuscula duo* – texte autrefois attribué à saint Bernard), ici col. 1145b.

[140] Dominique Iogna-Prat, Éric Palazzo, Daniel Russo (dir.), *Marie. Le culte de la Vierge dans la société médiévale,* op.cit. ; Caroline Walker Bynum, « Jesus as Mother and Abbot as Mother: Some Themes in Twelfth-Century Cistercian Writing », *Harvard Theological Review*, vol. 70:3-4, p. 257-284 ; id., *Jesus as Mother: Studies in the Spirituality of the High Middle Ages*, Berkeley, University of California Press, 1982.

[141] Ce qui explique le faible nombre d'occurrences dans la PL, qui s'arrête en 1216 avec Innocent III. Parmi les auteurs employant plusieurs fois *maternitas* dans leurs textes, on trouve saint Bonaventure [† 1274], Pierre de Jean Olivi [† 1298], Jean Duns Scot [† 1308], Raymond Lulle [† 1315], Jean de Gerson [† 1429], Bernardin de Sienne [† 1444], Nicolas de Cues, Denys le Chartreux [† 1471] ou encore plus tardivement Laurent de Brindisi [† 1619].

différente est à conserver à l'esprit lorsque nous tenterons d'établir la sémantique de *paternitas*, dans les paragraphes qui suivent.

Dans les chartes, l'augmentation des usages de *paternitas* est plus tardive que dans les lettres et l'exégèse, mais tout de même très nette au XIIIe siècle. On observe tout d'abord un nombre assez élevé de mentions du lemme pour la période allant du VIIe au début du Xe siècle. Les occurrences chutent ensuite fortement, entre 950 et 1150, pour remonter vers 1200. Cette évolution peut probablement être corrélée aux changements dans la typologie diplomatique, en particulier à la proportion d'actes pontificaux, de diplômes et/ou de lettres dans le haut Moyen Âge[142]. *Paternitas* apparaissant en particulier dans les marques de déférence (*vestra paternitas*), il est logique qu'il soit plutôt présent lors de cette période[143].

Les premiers actes émanant de chancelleries de grands laïcs à en faire usage sont ceux de Charlemagne, dans les dernières décennies du VIIIe siècle, alors que ce dernier fait référence au pape[144]. Elles sont suivies de mentions dans les documents royaux anglo-saxons et dans d'autres actes solennels, qui indiquent le registre lexical élevé du lemme[145]. Cette logique se poursuit entre 850 et 1150, où malgré la chute des mentions, on observe que *paternitas* renvoie toujours à une personne de haut rang, le plus souvent ecclésial[146]. C'est lors de cette période que la typologie des personnages dont on invoque la *paternitas* commence à évoluer plus nettement, en particulier vers les évêques[147]. L'augmentation du XIIIe siècle en diplomatique

---

[142] Une enquête spécifique serait sans doute utile pour étayer cette hypothèse, mais elle nécessiterait la présence de métadonnées extensives sur l'ensemble du corpus. On peut toutefois espérer que les méthodes issues du *Machine Learning* (apprentissage supervisé) permettront de renseigner intégralement les auteurs des actes inclus dans les CEMA d'ici quelques mois. Nous nous permettons de renvoyer à Nicolas Perreaux, « De l'accumulation à l'exploitation ? […] », *op.cit.* ; id., « Les actes pontificaux dans la masse documentaire », dans Rolf Grosse, Laurent Morelle, Olivier Guyotjeannin (dir.), *Les actes pontificaux. Un trésor à exploiter*, De Gruyter, Göttingen, à paraître.

[143] Car la proportion de diplômes et de bulles est très élevée dans le haut Moyen Âge. L'arrivée en masse des actes de petits laïcs et de petits seigneurs au cours des Xe-XIIe siècle change cette fortement cette proportion.

[144] « *Deinde dicendum erit Dominus noster filius vester hec parva munuscula paternitati vestre […]* », dans un acte pour Adrien Ier, en 785 (Artem n° 4512) : *Capitularia regum Francorum. 1*, éd. Alfred Boretius et al., Hannover 1883 (MGH, *Capitularia regum Francorum* 1), n° 111 (p. 226). L'acte concerne des instructions données par le souverain à ses *missi dominici* auprès du pape. On trouve aussi cette mention dans un diplôme pour l'église de Paris, daté des années 774-800 (Artem n° 2001 – qui indique l'acte comme « douteux », ce que ne fait pas l'édition des MGH) : « *Igitur cognoscat omnium secutura posteritas presulumque nobis succedentium pia paternitas, quia petierunt a nostra exiguitate fratres nostri sanctę matris ęcclesię […]* », dans *Die Urkunden Pippins, Karlmanns und Karls des Großen*, éd. par Engelbert Mühlbacher, Alfons Dopsch, Johan Lechner, Michael Tangl, Hanovre 1906 (MGH, *Diplomata, Die Urkunden der Karolinger. Erster Band*), n° 193 (p. 259).

[145] « *Notum sit paternitati tue quia sicut primitus a sancta Romana et apostolica sede beatissimo papa Gregorio dirigente […]* », dans un acte d'Æthelheard, archevêque de Canterbury, en 798, dans *Charters of Christ Church Canterbury*, éd. par Susan Kelly, Nicholas Brooks, Oxford, Oxford University Press, 2013 (*Anglo-Saxon Charters* 17) – Sawyer n° 1258. Resterait à savoir qui rédigent quoi dans ces actes : les auteurs diplomatiques, ou les destinataires.

[146] Par exemple dans la lettre de mars 863 du comte Girart de Roussillon au pape Nicolas Ier, concernant les abbayes de Pothières et de Vézelay qu'il vient de fonder : « *per deum vestram reverentissimam paternitatem* », dans *Monumenta Vizeliacensia : textes relatifs à l'histoire de l'abbaye de Vézelay*, éd. par Robert Burchard Constantijn Huygens, 1 volume et supplément, Turnhout, Brepols, 1976-1980, acte n° 2 (p. 249-254) – le lemme *paternitas* apparaît 8 fois dans l'acte, toujours en référence au pape. La bulle-réponse de Nicolas Ier (865) contient là aussi le terme : « *successores nostri accipientes pie paternitatis suffragium eidem monasterio […]* », dans *Cartulaire général de l'Yonne*, éd. par Maximilien Quantin, Auxerre, Perriquet, 1854-1860, tome 1, n° XLIV.

[147] « *Nos igitur huiusmodi petitionem suscipientes nostrae etiam atque ipsius miseriae condolentes omni que carentem dolo cognoscentes migrandi facultatem liberalissime indulsimus eminentiam vestram ad quos pervenerit obnixe exposcentes ut in sacrosancto vestrae paternitatis gremio eum suscipientes ministerii sui officium infra vestram diocesim celebrare sinatis.* », dans un acte de Jean, évêque de Cambrai, en 880 – dans *Formulae*

correspond toutefois à un autre phénomène, qui pourrait être un transfert du lemme depuis les écrits théologiques, les diplômes et les bulles, vers des actes de la pratique moins importants.

Cette approche fréquentielle, d'abord dans les documents théologiques puis dans les chartes, peut enfin être complété par une analyse rapide de corpus plus tardifs. Au sein du *Corpus Thomisticum* apparaissent par exemple 577 occurrences de *paternitas*. Proportionnellement, ce score est sensiblement le même, voir un peu plus élevé, que celui des derniers tomes de la PL[148]. On peut donc supposer que l'usage du lemme continue d'augmenter, même si d'autres enquêtes seraient à mener afin d'appuyer cette hypothèse de travail. Plus généralement, le développement accéléré de *paternitas* à partir du XII^e^-XIII^e^ siècle participe au mouvement de création/diffusion conceptuel très riche de ce siècle[149]. On note en effet qu'émerge alors lentement un autre terme qui formera bientôt une paire conceptuelle avec *paternitas* : *filiatio* (fig. 10). Comme dans le cas du premier, le lemme n'existe pas dans l'Antiquité. Mais contrairement à *paternitas*, *filiatio* n'apparaît pas dans la Vulgate : il progresse lentement au fil des siècles médiévaux, avec une augmentation plus nette des mentions au XII^e^-XV^e^ siècles. Chez Thomas d'Aquin, il apparaît ainsi parmi les cooccurrents les plus significatifs de *paternitas*[150]. Cette tendance confirme que l'Europe médiévale voit se développer une réflexion spécifique sur la « paternité » en tant que concept, et bientôt autour de son double conceptuel, la *filiatio* – l'ensemble désignant en premier lieu la relation divine du Christ et du Père, modèle de relation paternelle. L'observation n'est pas sans conséquence : la *paternitas* apparaît ainsi comme une pensée proprement médiévale, qui ne concerne certes pas l'hérédité biologique (tout au contraire), mais qui constitue une étape de la pensée occidentale des relations entre père et enfants, créateur et créatures.

---

*Merowingici et Karolini aevi*, éd. par Karl Zeumer, Hannover, 1886, n° 15 (MGH, *Leges* VI, *Formulae Merowingici et Karolini aevi* 1), aussi *Diplomata Belgica* n° 3802 ; « *Isarnus Gratiapolitanense sedis episcopus quumquidem Adalbert et sorore sua Guidtrud nostram expetierunt paternitatem quod et fecerunt.* », dans un acte de l'évêque de Grenoble Isarn (976) – dans *Cartulaires de l'église cathédrale de Grenoble dits cartulaires de Saint-Hugues*, éd. par Jules Marion, Paris, Imprimerie impériale, 1869, n°16 ; « *Successorum itaque nostrorum reverendam deprecamur paternitatem ut huius nostre institutionis testamentum ita inviolabiliter observent […]* », dans un acte de l'évêque d'Autun Walon (918), pour son chapitre cathédral – Artem n° 796 ; « *Unde etiam suggesserunt reuerenti paternitati nostre […]* », dans un don d'église par l'évêque d'Autun Rotmond (milieu du X^e^ siècle) – dans *Cartulaire de l'Eglise d'Autun*, éd. par Anatole De Charmasse, Paris, Durand, 1865, n° XLIX.

148 Le contrôle a été effectué sur les volumes 201(Arnoul de Lisieux - Guillaume de Tyr) à 217 (Innocent III).

149 En analysant par exemple l'évolution de *reformatio*, nous avons pu montrer que le passage du verbe *reformo* vers ce « qualificatif substantivé » s'accélérait fortement au tournant des XII^e^ et XIII^e^ siècles.

150 Dans les 577 mentions de *paternitas* dans le *Corpus Thomisticum*, *filiatio* apparaît comme cooccurrent 152 fois (à une distance de plus ou moins 6 mots), ce qui est assez considérable. Il arrive d'ailleurs en seconde position, juste après *pater*. Par ailleurs, *paternitas* tout comme *filiatio* sont définis comme des *relatio* à Dieu par Thomas d'Aquin : « *Ad tertium dicendum, quod filiatio est relatio eius quod est a principio; paternitas autem est relatio principii. Tota autem Trinitas est principium creature. Unde magis potest trahi nomen paternitatis ad alias personas respectu creature, quam filiationis nomen.* », dans Thomas Aquinas, *Scriptum super Sententiis*, Parme, 1858, Liber 3, Questio 2, Quaestiuncula 2. Sur ces questions, nous renvoyons à Luc-Thomas Somme, *Fils adoptifs de Dieu par Jésus Christ : la filiation divine par adoption dans la théologie de saint Thomas d'Aquin*, Paris, Vrin, 1997.

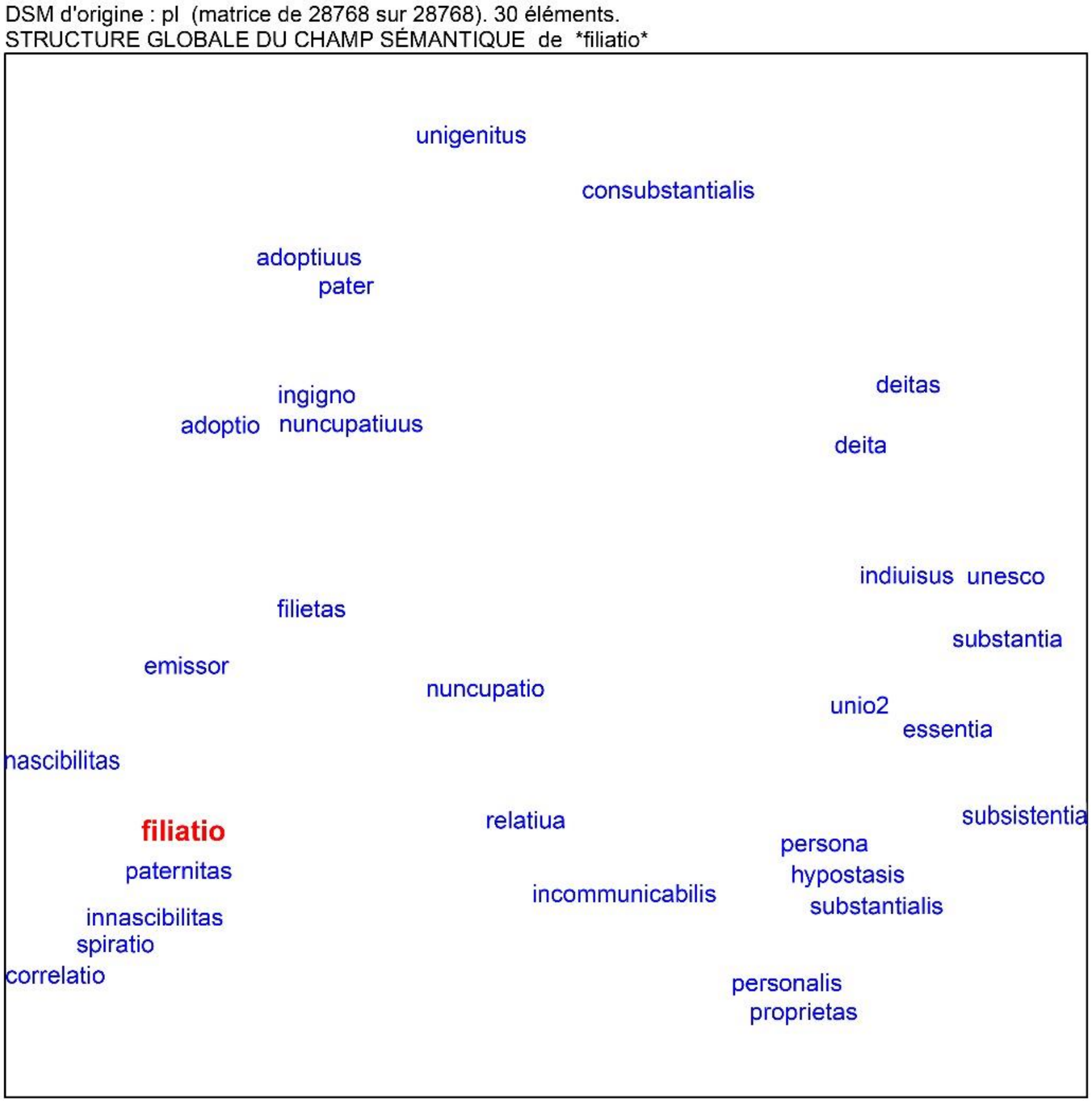

**Fig. 10 :** Champ sémantique de *filiatio* dans l'ensemble de la PL. Nous observons la présence de *pater*, mais aussi de *paternitas* (au plus près de *filiatio*), dans les éléments significatifs.

Notre analyse serait toutefois incomplète si nous ne nous attardions pas sur le qualificatif *paternus* et le substantif *paternum*, qui désignent selon les dictionnaires « ce qui paternel » et plus particulièrement (dans le cas du substantif) l'« héritage »[151]. *Paternus* est en effet présent plus de 6 500 fois dans la PL et 1 900 dans les chartes des CEMA[152]. Une visualisation de son champ lexical confirme deux groupes sémantiques assez distincts (fig. 11) : d'une part, à gauche, ce qui est relatif à la transmission (*hereditarius*, *heres*, *successio*), en lien avec la provenance d'ego (*parens*, *patricus*, *avitus*, *maternus*-*maternum*), d'autre part, à droite, des termes relatifs à la « qualité » des relations paternelles (*affectio*-*affectus*, *compassio*, *fraternus*, *dilectio*, *filialis*, *devotio*, *pietas*, *pius*, *benevolens*, *affectuosus*, *benignus*, *benignitas*, *paternitas*)[153]. Sommes-nous là dans l'ordre des « affects », au sens où l'entendent Damien

[151] Michel Parisse (dir.), *Lexique Latin-Français : Antiquité et Moyen Age*, Paris, Picard, 2006. En pratique, le lemmatiseur confond parfois les deux termes, même si des différences importantes dans les modélisations sémantiques indiquent que, globalement, les lemmes sont effectivement différenciés. Le *Novum Glossarium* ne donne en outre qu'une seule entrée, *paternus*, indiquant que le qualificatif peut parfois être employé comme substantif (définitions « II », dans « *Paternus* », dans *Novum Glossarium Mediae Latinitatis*, op.cit., col. 674-677, ici col. 676-677). Le corpus numérique *Corpus Corporum* (http://mlat.uzh.ch/, site accédé le 14.04.2020), contrairement au lemmatiseur du *Novum Glossarium*, ne donne d'ailleurs qu'un seul lemme : *paternus*.

[152] Si l'on additionne les termes lemmatisés comme *paternus* et *paternum*, nous obtenons respectivement 9 218 et 3 473 mentions dans la PL et les CEMA (actes datés seulement).

[153] *Paternum* apparaît en outre comme un terme très proche sémantiquement de *paternus*.

Boquet et Piroska Nagy[154] ? Ces différents termes indiquent en tout cas que *paternus* relève encore une fois d'une sémantique plus large que la paternité contemporaine, et s'insère dans la logique des relations sociales médiévales : la *caritas*[155]. Nous constatons là encore que la dimension charnelle n'est pas la clé de lecture sémantique majoritaire (encore moins une lecture biologique)[156], mais au contraire que la transmission des patrimoines (*successio*, *heres*, *hereditarius*) s'insère dans une logique paternelle généralisée, en lien avec la théorie médiévale du lien social. Les mentions de *paternus* sont d'ailleurs nettement plus fréquentes dans les corpus théologiques que dans les textes dits de la pratique[157]. En cela, nous pouvons dire que *paternus* (et le putatif *paternum*) qualifie(nt) tout autant les relations charnelles que les relations spirituelles ou divines[158] – ce qui nous ramène à la multi-paternité et à l'articulation des couples *dominus-servus*, *pater-filius*.

---

[154] Damien Boquet, *L'ordre de l'affect au Moyen Âge. Autour de l'anthropologie affective d'Aelred de Rievaulx*, Caen, Publications du CRAHM, 2005 ; Piroska Nagy, Damien Boquet, *Sensible Moyen Âge. Une histoire des émotions dans l'Occident médiéval*, Paris, Le Seuil, 2015.

[155] Le terme n'apparaît pas directement sur le graphique, sans doute parce que nous avons limité à 30 lemmes l'analyse sémantique, mais il est sous-jacent : « *Quid est autem virtus pietatis ? Caritas in Deum et proximum* », explique saint Augustin dans son *Sermo* 229 V (cité par Anita Guerreau-Jalabert et Bruno Bon, « *Pietas [...]* », op.cit., p. 84, qui montre que certains termes (apparaissant sur les graphiques) sont très proches de *caritas* : *pietas*, *dilectio*, etc.) ; voir aussi François Dolbeau, « Nouveaux sermons de saint Augustin pour la conversion des païens et des donatistes (III) », *Revue des Études Augustiniennes*, vol. 38, 1992, p. 50-79, ici p. 75 : « *Quae est virtus pietatis ? Caritas.* »). Sur la signification de la *caritas* médiévale, nous renvoyons à Anita Guerreau-Jalabert, « *Spiritus* et *caritas*. Le baptême dans la société médiévale », art. cité.

[156] Là encore, on pourrait remarquer que l'article *paternus* du *Novum Glossarium* commence par les relations biologiques/familiales/lignagères (avec en particulier des exemples issus des diplômes carolingiens et post-carolingiens ; les définitions sont d'abord : « appartenant ou ayant appartenu au père », « fait par le père », « transmis par le père, en parlant d'un bien, d'un héritage [ou] d'un trait de caractère », « en parlant de la lignée ou de la famille »), comme si celles-ci constituaient l'élément le plus significatif (« *Paternus* », dans *Novum Glossarium Mediae Latinitatis*, op.cit.). Les éléments concernant la paternité spirituelle arrivent ensuite (définitions 5-8), avec cette fois des citations issues des textes théologiques.

[157] Encore une fois, avec 9 218 occurrences dans la PL et 3 473 dans les CEMA (actes datés seulement), celles-ci sont presque deux fois plus présentes dans le premier corpus.

[158] Les définitions classées en « B » dans le *Novum Glossarium* (« ancestral, qui vient des ancêtres ») sont tout à fait intéressantes, car elles confirment l'hypothèse de la multi-paternité médiévale, évoquant tour à tour les « ancêtres » au sens général, Adam et les Pères de l'Église.

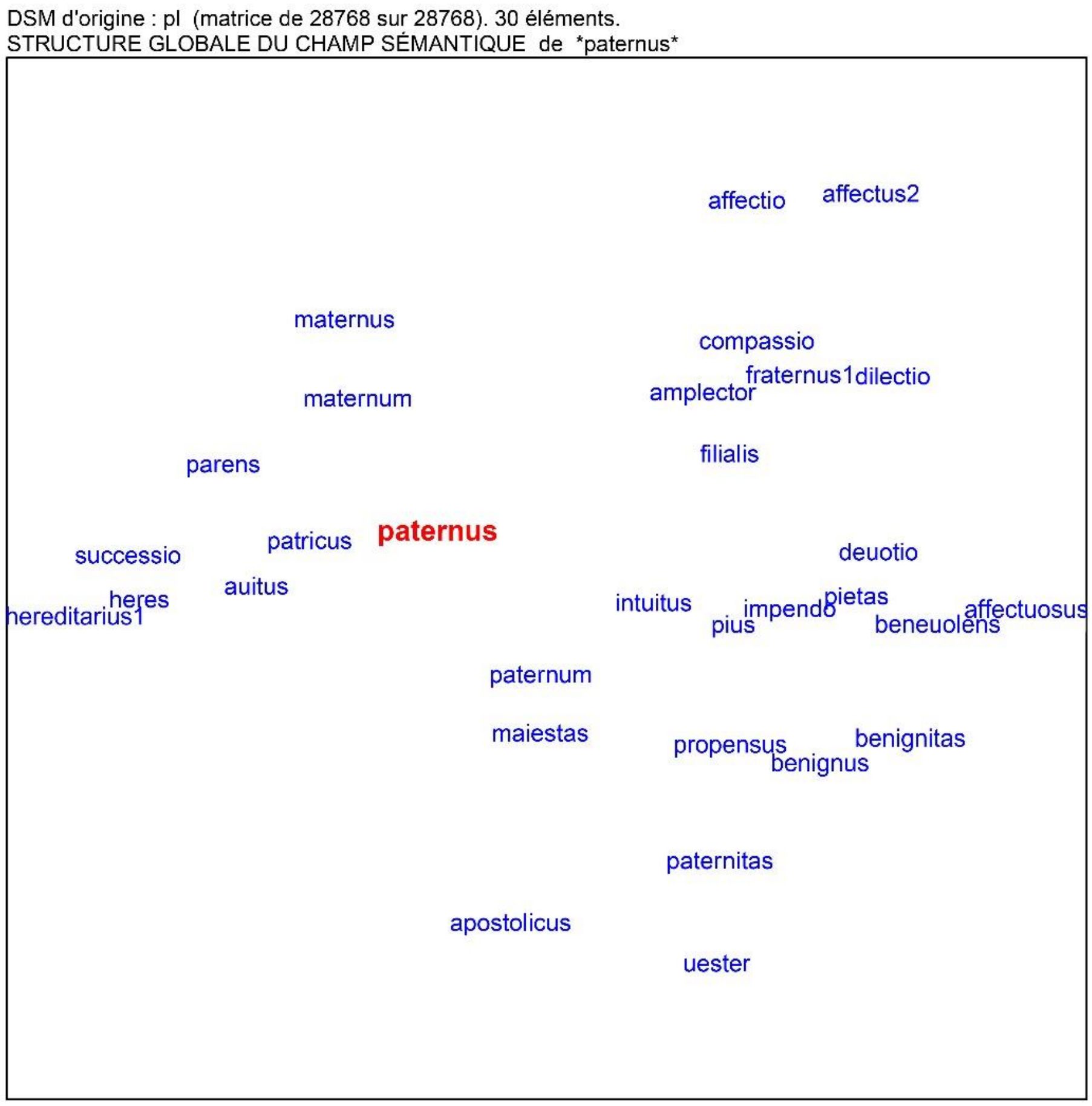

**Fig. 11 :** Champ sémantique de *paternus* dans l'ensemble de la PL.

### *III.3. La* **paternitas** *comme paradigme de la parenté spirituelle ?*

Que signifie en définitive la *partenitas* ? La lecture qualitative des textes et les analyses numériques montrent que le terme évoque avant tout une qualité, dont la force vient du fait qu'elle renvoie à l'acte créateur divin. Toute « paternité » dérivant de cette action originelle, le qualificatif prend un sens éminemment positif. C'est ce qui explique que le substantif est fréquemment employé dans les suppliques.

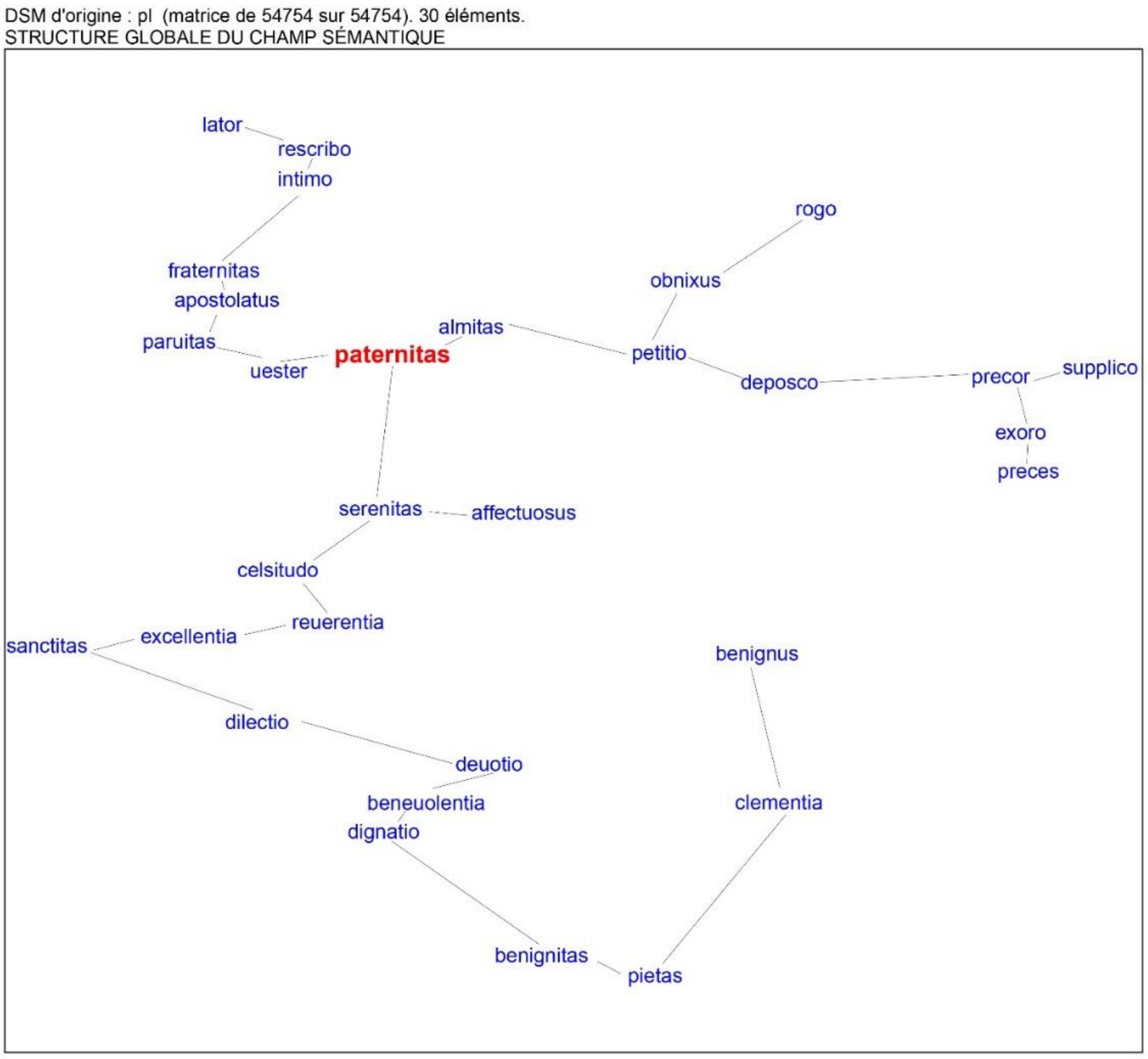

**Fig. 12 :** Champ sémantique de *paternitas* dans l'ensemble de la PL.

Lors d'une analyse du champ sémantique de *paternitas* (fig. 12), c'est ce lexique de la « demande » qui apparaît en haut à droite (*rogo*, *petitio*, *precor*, *supplico*, *exoro*, *preces*). On s'adresse à une personne en invoquant sa *paternitas*, d'où certains cooccurrents qui émergent en haut à gauche de l'analyse : *vester*, à nouveau, puisqu'on en appelle à la « paternité » d'un dominant[159]. Le vouvoiement renvoie bien entendu à des questions de hiérarchie et de domination[160] : tandis que *vester* est directement associé plus de 1 220 fois à *paternitas* dans les CEMA, c'est le cas pour *tuus* dans seulement 17 mentions. La situation est d'ailleurs inversée avec *fraternitas*, où l'association avec *tuus* renvoie 1 041 occurrences, contre 627 avec *vester* : une observation logique, car nous sommes alors dans le registre de l'égalité. Le discours sur la paternalité engendre donc une autodiminution de celui qui appelle une faveur : le suppliciant insiste sur sa *parvitas*, tout en requérant la *fraternitas*[161]. En bas de l'analyse apparaissent ainsi diverses qualités morales, concernant en particulier le lien social (*caritas*), dont certaines jouent un rôle absolument fondamental dans le système médiéval (*serenitas*, *sanctitas*, *devotio*, *clementia* et même *pietas*)[162]. Ces éléments montrent que la *paternitas* est une relation de père à fils très valorisée, car hautement spirituelle[163]. Ainsi, le lemme n'est pratiquement jamais lié au registre charnel, et ne présente aucun lien par exemple avec *caro*, *genitor* ou encore *proles*[164].

[159] Par exemple : « *Notum sit igitur vestre paternitati* », à l'adresse de l'abbé de Cluny Hugues de Semur, vers 1070, dans *Recueil des chartes de l'abbaye de Cluny*, éd. par Alexandre Bernard et Alexandre Bruel, Tome 4 : 1027-1090, Paris, Imprimerie nationale, 1888, p. 540-541, n° 3430 ; « *Excusatam igitur habeat gravedinem immo impossibilitatem meam vestra paternitas et compatiatur pedibus vestris prostrato seni discretionis apostolice mansueta caritas.* », à l'adresse du pape Urbain II, en 1095, dans *Cartulaire de l'église d'Angoulême*, éd. par Jean Nanglard, Angoulême, G. Chassaignac, 1900, n° 18. On trouve presque 400 associations de ce type dans les seuls CEMA.

Peut-on mesurer une évolution dans cette structure sémantique, mise en place à partir d'une formule biblique ? Qui possède en définitive cette fameuse *paternitas*, outre Dieu lui-même ? La première extension de la *paternitas* divine concerne comme nous l'avons montré la papauté, qui s'approprie le terme assez rapidement, dans les lettres (et pseudo-lettres) pontificales des Ve-VIIe siècles. Rapidement néanmoins, le qualificatif se voit rapproché de la figure épiscopale[165]. Ce dernier est en effet chargé de la reproduction spirituelle du système, qui nous semble précisément être désignée par le lemme *paternitas*. Car enfin, qu'est-ce qu'une paternité strictement spirituelle, si ce n'est une reproduction sociale ? Bien qu'irrémédiablement lié étymologiquement à *pater*, les abbés se verront eux aussi progressivement associés à *paternitas*, principalement à partir du XIe-XIIIe siècles[166]. Cette « descente hiérarchique » (du pape vers les évêques, puis des évêques vers les abbés) n'est certes pas linéaire et exclusive, mais elle indique globalement une évolution dans l'usage du terme, parallèle à sa diffusion quantitative. Autrement dit : plus le lemme est courant, moins il s'applique à des personnalités hiérarchiquement importantes – ce qui en soi n'est pas illogique[167]. Cela n'empêche pas en outre la *paternitas* d'être de plus en plus liée au pôle spirituel : au fil des siècles, nous observons un renforcement des cooccurrences entre le lemme et *spiritus*, mais aussi *gratia* ou encore *caritas*.

---

160 Les travaux sur la question demeurent relativement rares. Voir cependant les travaux de Philippe Wolff, « Premières recherches sur l'apparition du vouvoiement en latin médiéval », *Comptes rendus des séances de l'Académie des Inscriptions et Belles-Lettres*, vol. 130:2, 1986, p. 370-383 ; Id., « Les sociétés allemandes et le vouvoiement au Moyen Âge », dans *Histoire et société. Mélanges offerts à Georges Duby*, volume 1 : Le couple, l'ami et le prochain, Aix-en-Provence, Publications de l'Université de Provence, 1992, p. 157-166 ; Id., *Vous. Une histoire internationale du vouvoiement*, Toulouse, Signes du monde, 1994.

161 Les passages associant directement *paternitas* et *fraternitas* sont relativement rares (à plus ou moins cinq mots de distance : 5 occurrences dans les CEMA, 17 dans la PL et 3 dans les OpenMGH). Mais c'est essentiellement parce que les termes sont, dans les contextes diplomatiques/épistolaires, où ils apparaissent, permutables. Nous trouvons ainsi de nombreuses mentions de « *vestre fraternitati* », « *fraternitas vestra* », « *fraternitatem vestram* », « *fraternitas tue/tua* », etc. La fréquence de *fraternitas* dans les chartes est d'ailleurs assez comparable à celle de *paternitas*, avec 3 101 mentions.

162 Bruno Bon, Anita Guerreau-Jalabert, « *Pietas*. Réflexions sur l'analyse sémantique et le traitement lexicographique d'un vocable médiéval », art. cité.

163 Ce qui pose aussi la question de l'Esprit et le fait entrer dans ce mouvement, puisqu'il relie le Père et le Fils.

164 Il serait bien entendu possible de relever, comme toujours, quelques exemples contradictoires, mais ils sont ici particulièrement rares. Voir ceux donnés dans l'article du *Novum Glossarium*, op.cit., col. 671-674. Concernant proles, nous renvoyons à Esther Dehoux, « Une simple affaire de famille ? Usage et portée du mot *proles* dans la France de l'Ouest (IXe-XIIe siècle) », *Bulletin de la Société Archéologique du Finistère*, vol. 140, 2012, p. 241-266.

165 Un exemple parmi d'autres : dans les MGH, on dénombre 7 occurrences de *paternitas* dans les capitulaires (toutes concernant les évêques et/ou le pape), tandis que le corpus des conciles (qui est seulement deux fois plus important que celui des capitulaires) compte 77 occurrences à lui seul.

166 Ce qui montre par ailleurs que *paternitas* ne désigne pas une paternité systématique ou simple.

167 Dans une étude à paraître sur *sigillum*, il nous a semblé qu'un même mécanisme était à l'œuvre.

## Conclusion

Si elle ne pouvait être que lacunaire, l'enquête présentée ici a permis de relever différents éléments inédits, tout en confirmant d'autres hypothèses déjà proposées par l'historiographie. Nos trois principales observations peuvent être synthétisées de la façon suivante :

1. Le champ sémantique de *pater* et ses dérivés ont connu une rupture presque totale au tournant de l'Antiquité et du très haut Moyen Âge. Cette évolution fut intrinséquement liée à la mise en place du système chrétien, en particulier du dogme de la Trinité, qui proposa dès les Pères une nouvelle définition des rapports père-fils. Associée au mariage chrétien, aux théories de procréations médiévales, au refus de l'adoption – et donc à la maximisation progressive des relations d'alliance –, cette rupture ne peut pas être sous-estimée, car les pères jouaient un rôle absolument central dans les relations sociales médiévales : étroitement associés au titre de seigneur (*dominus*), ils apparaissaient à la fois comme des créateurs, des protecteurs et des nourriciers. Chaque individu s'insérait alors dans un réseau de relations père-fils très dense et diversifié, dans lequel il était à la fois le fils d'un père charnel, mais aussi celui de Dieu et de nombreux pères spirituels – sans que ces relations spirituelles constituent des rapports lattéraux, « symboliques » ou amoindris, bien au contraire. Plus près de nous, la modernité représente une autre rupture radicale pour le champ de la paternité : au cours des XVII$^{e}$-XIX$^{e}$ siècles, nous avons pu observer un net ressèrement des sens de « père ». Cette période délaisse largement la multi-paternité (reléguée au rang de curiosité ecclésiale, par exemple dans l'*Encyclopédie*) et insiste sur la filiation biologique, qui devient le paradigme central de la paternité[168].

2. Ces observations confirment l'impossibilité stricte d'une analyse de la parenté médiévale qui ne reconnaisse pas un rôle pivot à la parenté spirituelle. Les cas de *pater* mais aussi de *paternitas* sont à ce titre absolument édifiants. Si l'on divise le champ sémantique entre une putative paternité « biologique » et des paternités symboliques, secondaires, on détruit irrémédiablement le sens même des représentations médiévales et toute possibilité de reconstruire la structure sociale. L'hypothèse la plus plausible est à l'inverse que les sens charnels et spirituels de la paternité étaient étroitement hiérarchisés et articulés dans la parenté. À la paternité charnelle répondait une multitude de paternités spirituelles, allant de la paternité baptismale à la paternité divine, en passant par les paternités ecclésiales et seigneuriales (donc royales)[169]. Or, le sens et l'orde de ces strates formaient précisément une structure qui répondait à la logique globale et idéelle de la société. Dans cette perspective, l'association de *pater* et de *dominus*, qui trouvait son fondement dans le fait que Dieu le Père était aussi le Seigneur, ne peut pas être sous-estimée.

3. L'importance relative croissante de la parenté spirituelle au fur et à mesure des siècles examinés semble enfin être indéniable. Le cas de *paternitas* est particulièrement net : néologisme biblique, d'abord réservée à Dieu lui-même, cette qualité (une paternité spirituelle parfaite, associée au vouvoiement) s'est progressivement étendue aux papes, aux évêques, aux abbés, et *in fine* à différents personnages clés de l'institution ecclésiale. Cette tendance s'inscrit en parallèle de la multiplication des pères, phénomène déjà observé par différents chercheurs[170], et qui va aussi dans le sens d'une extension de la paternité spirituelle à tous les dominants. C'est encore dans ce cadre qu'il faut sans doute comprendre le développement des associations entre *dominus* et *pater*.

---

[168] Outre l'adoption, l'autre champ dominant semble être celui de l'invention/des découvertes (eg. Lavoisier comme « père de la chimie moderne »), qui s'insère d'ailleurs dans de nombreux cadres contemporains : le droit d'auteur, l'industrie, les sciences, etc.

[169] Jérôme Baschet, *Le sein du père. Abraham et la paternité dans l'Occident médiéval*, op.cit.

Ce schéma ne saurait évidemment se substituer à une étude complète des relations de paternité et du rôle structurel du père dans l'Europe médiévale[171]. La raison en est simple : cette étude nécessiterait de réintégrer *pater* au sein du lexique de la parenté dans son ensemble. Au-delà du couple *pater-filius*, de nouvelles enquêtes sur *mater*, *filia*, mais aussi *frater* et sur le lexique adelphique en général nous paraissent indispensables. Les remarques trop rapides que nous avons proposées autour de *paternitas-fraternitas* pourraient en particulier faire l'objet d'études approfondies. Alors que tout chrétien médiéval s'insérait dans un réseau « vertical » de paternités multiples, à la fois structurant et hiérarchisant, il s'intégrait aussi dans un réseau horizontal de parentés spirituelles tout aussi riche : celles dessinées par la fraternité de tous les chrétiens, dont l'origine était précisément l'entrée en chrétienté par le batpême, et donc la reconnaissance simultanée de la paternité divine et de la paternité baptismale.

Dans cette perspective, il y a fort à parier que de belles découvertes attendent les historiens qui accepteront de s'emparer des outils numériques et de la fouille de données, de relever le défi des corpus et de la statistique lexicale. La sémantique historique apparaît comme l'un des domaines les plus prométeurs pour aborder les systèmes de représentations des sociétés anciennes, et partant de là leurs structures sociales. En révélant des éléments invisibles à l'œil nu, elle permet de jeter un regard inédit sur leur pensée, en limitant fortement le rôle de nos présupposés dans l'analyse. C'est dans cette perspective que pourraient s'ouvrir des enquêtes originales sur la parenté médiévale, à la croisée de l'histoire, de l'anthropologie, de la linguistique et de la fouille de données.

Nicolas Perreaux (LaMOP)

[170] Au-delà de notre chronologie, en particulier à partir des XV$^{e}$-XVI$^{e}$ siècles, Aude-Marie Certin observe d'importantes évolutions dans l'articulation du couple charnel-spirituel à la figure paternelle. Voir Aude-Marie Certin, « Überlegungen zu einer Geschichte der Vaterschaft in Westeuropa in der langen Dauer », *art.cit.*, p. 69-74. Ce qu'elle constate très justement ressemble fort, selon nous, à une spiritualisation au moins partielle de la figure du père charnel – *a minima* dans le protestantisme.

[171] On notera en particulier que la question des liens entre paternité et mariage a volontairement été exclue de la présente analyse ; sans compter les liens entre paternité et baptême, ou encore entre paternité et procréation, qui ont été à peine effleurés ici.